\pdfoutput=1

\documentclass[11pt,twoside,a4paper,cmspaper,final,collab]{cms-tdr}
\def\svnVersion{298271}\def\cmsCernNo{2013-037}\def\cmsCernDate{\today}\def\cmsMessage{Published in the Journal of High Energy Physics as \href{http://dx.doi.org/10.1007/JHEP06(2015)116}{\doi{10.1007/JHEP06(2015)116.}}}

\begin{document}\cmsNoteHeader{SUS-14-001}

\hyphenation{had-ron-i-za-tion}
\hyphenation{cal-or-i-me-ter}
\hyphenation{de-vices}
\RCS$Revision: 298268 $
\RCS$HeadURL: svn+ssh://svn.cern.ch/reps/tdr2/papers/SUS-14-001/trunk/SUS-14-001.tex $
\RCS$Id: SUS-14-001.tex 298268 2015-07-28 12:33:29Z rlucas $
\newlength\cmsFigWidth
\ifthenelse{\boolean{cms@external}}{\setlength\cmsFigWidth{0.85\columnwidth}}{\setlength\cmsFigWidth{0.4\textwidth}}
\ifthenelse{\boolean{cms@external}}{\providecommand{\cmsLeft}{top}}{\providecommand{\cmsLeft}{left}}
\ifthenelse{\boolean{cms@external}}{\providecommand{\cmsRight}{bottom}}{\providecommand{\cmsRight}{right}}
\newcolumntype{,}{D{,}{\pm}{-1}}
\newcommand{\susy}{SUSY\xspace}
\newcommand{\ptvecmissmu}{\ensuremath{{\vec p}_{\mathrm{T}}^{\text{miss},\mu}}\xspace}
\newcommand{\jet}{\ensuremath{{j}}\xspace}
\newcommand{\tauh}{\ensuremath{\tau_\mathrm{h}}\xspace}
\newcommand{\wpj}{\ensuremath{\PW\,\text{+jets}}\xspace}
\newcommand{\njets}{\ensuremath{N_{\text{jets}}}\xspace}
\newcommand{\nbjets}{\ensuremath{N_{\text{b jets}}}\xspace}
\newcommand{\MCT}{\ensuremath{ M_{\mathrm{CT}}}\xspace}
\newcommand{\MTt}{\ensuremath{M^{\text{3-jet}}_{\mathrm{T}}}\xspace}
\newcommand{\MTb}{\ensuremath{M^{\text{R-sys}}_{\mathrm{T}}}\xspace}
\newcommand{\pTopv}{\ensuremath{\vec{p}^{\text{3-jet}}}\xspace}
\newcommand{\pTopT}{\ensuremath{p_{\mathrm{T}}^{\text{3-jet}}}\xspace}
\newcommand{\pTopFourVec}{\ensuremath{\text{P}^{\text{3-jet}}}\xspace}
\newcommand{\pRsystemFourVec}{\ensuremath{\text{P}^{\text{R-sys}}}\xspace}

\newcommand{\zll}{\ensuremath{\Z\to\ell^+\ell^-}\xspace}
\newcommand{\znunu}{\ensuremath{\Z \to \PGn \PAGn}\xspace}
\newcommand{\zmumu}{\ensuremath{\Z\to\Pgmp\Pgmm}\xspace}
\newcommand{\znunubr}{\ensuremath{\Z(\PGn\PAGn)}\xspace}
\newcommand{\wlnubr}{\ensuremath{\PW(\ell\PGn)}\xspace}
\newcommand{\zellellbr}{\ensuremath{\Z(\ell^+\ell^-)}\xspace}
\newcommand{\ttwobb}{\ensuremath{\PSQb\PASQb\to\bbbar\PSGczDo\PSGczDo}\xspace}
\newcommand{\ttwocc}{\ensuremath{\PSQt\PASQt\to\ccbar\PSGczDo\PSGczDo}\xspace}
\newcommand{\ttwott}{\ensuremath{\PSQt\PASQt\to\ttbar\PSGczDo\PSGczDo}\xspace}
\newcommand{\ttwotb}{\ensuremath{\PSQt\PASQt\to\PQt\PSGczDo\PAQb\PSGcpmDo\to\PQt\PAQb\PW^\pm\PSGczDo\PSGczDo}\xspace}
\newcommand{\ttbarW}{\ensuremath{{\ttbar\PW}}\xspace}
\newcommand{\MTTwo}{\ensuremath{M_{\mathrm{T2}}}\xspace}
\newcommand{\wmunubr}{\ensuremath{\PW(\mu\nu)}\xspace}
\newcommand{\METmu}{\ensuremath{p^{\text{miss},\mu}_\mathrm{T}}\xspace}
\newcommand{\zmumubr}{\ensuremath{\Z (\mu \mu)}\xspace}
\newcommand{\ttbarZ}{\ensuremath{{\ttbar\Z}}\xspace}
\newcommand{\RQCD}{\ensuremath{R_\mathrm{QCD}}\xspace}
\newcommand{\RmumuDataMC}{\ensuremath{R^{\mu\mu}_\mathrm{data/MC}}\xspace}
\newcommand{\DeltaPhi}{\ensuremath{\Delta\phi}\xspace}
\providecommand{\MT}{\ensuremath{M_\mathrm{T}}\xspace}
\providecommand{\ptmiss}{\ensuremath{p_\mathrm{T}^\text{miss}}\xspace}
\newcommand{\fulllumi}{19.4}

\cmsNoteHeader{SUS-14-001}
\title{Searches for third-generation squark production in fully hadronic final states in proton-proton collisions at $\sqrt{s}=8$\TeV}

\date{\today}

\abstract{
  Searches for third-generation squarks in fully hadronic final states are presented using data samples corresponding to integrated luminosities of
  19.4 or 19.7\fbinv, collected at a centre-of-mass energy of 8\TeV with the CMS detector at the LHC.
  Three mutually exclusive searches are presented, each optimized for a different decay topology.
  They include a multijet search requiring one fully reconstructed top quark,
  a dijet search requiring one or two jets originating from b quarks,
  and a monojet search.
  No excesses above the standard model expectations are seen,
  and limits are set on top and bottom squark production in the context of simplified models of supersymmetry.
}

\hypersetup{%
pdfauthor={CMS Collaboration},%
pdftitle={Searches for third generation squark production in fully hadronic final states in proton-proton collisions at sqrt(s)=8 TeV},%
pdfsubject={CMS},%
pdfkeywords={CMS, physics, SUSY, third generation}}

\maketitle

\section{Introduction}

The standard model (SM) of particle physics has proven to be remarkably robust.
Nonetheless, the SM has well-known shortcomings, such as an instability in the calculation of the Higgs boson mass known as the fine-tuning (or hierarchy) problem~\cite{finetune2,finetune3,finetune1,finetune4,finetune6}.
The discovery of a Higgs boson with a mass of about 125\GeV~\cite{Chatrchyan:2012ufa,Aad:2012tfa,HiggsLegacyCMS} at the CERN LHC has reinforced the acuteness of this problem.
These shortcomings suggest that the SM is merely a low-energy approximation of a deeper, more complete theory.
Supersymmetry (\susy)~\cite{SUSY0,SUSY1,SUSY2,SUSY3,SUSY4,Martin:SUSYprimer,bib:SUSYLang} is a widely considered extension of the SM that introduces an additional symmetry of nature between fermions and bosons.
A new supersymmetric particle (sparticle) is proposed for each SM particle,
with the same mass and quantum numbers but with a spin that differs by a half-integer unit.
For example, squarks are the SUSY partners of quarks.
Supersymmetric models contain extended Higgs sectors.
The SUSY partners of the Higgs bosons are higgsinos.
Neutral (charged) higgsinos mix with the SUSY partners of the neutral (charged) electroweak gauge bosons to form neutralinos \PSGcz (charginos \PSGcpm).
Divergent quantum corrections to the Higgs boson mass due to virtual SM particles are cancelled by corresponding
contributions from virtual sparticles~\cite{Barbieri:1987fn, deCarlos1993320, natSUSY2, Barbieri199676}, thus resolving the fine-tuning problem.

The symmetry proposed by SUSY cannot be exact, as no sparticles have yet been observed.
However, the stabilising features of \susy can survive with a modest amount of fine tuning if sparticles are not much heavier than their SM counterparts.
For third-generation particles in particular, the mass difference between a particle and its corresponding sparticle
should not be too large, in order for SUSY to provide a so-called ``natural'' solution~\cite{finetune5,natSUSY3,natSUSY4,lightstop} to the fine-tuning problem.
Thus the SUSY partners of top and bottom quarks, the top and bottom squarks \PSQt and \PSQb, respectively,
might have masses below or around the TeV scale and be accessible at the LHC.
In \susy models with $R$-parity~\cite{DMrparity} conservation, top and bottom squarks can be pair produced,
with each top or bottom squark initiating a decay chain in which the end products are SM particles and a stable lightest supersymmetric particle (LSP).
In many \susy scenarios the LSP is the lightest neutralino \PSGczDo,
which is weakly interacting and will escape detection,
leading to a distinctive experimental signature of large momentum imbalance in the plane perpendicular to the beam axis.

This paper presents three complementary searches for the direct production of either a pair of top squarks ($\PSQt\PASQt$) or bottom squarks ($\PSQb\PASQb$) decaying to fully hadronic final states with large transverse momentum imbalance.
The searches are based on proton-proton collision data collected using the CMS detector at the LHC at a centre-of-mass energy of 8\TeV and correspond to an integrated luminosity of 19.4 or 19.7\fbinv depending on the study~\cite{LUMIPAS}.
Each search is separately optimized for different kinematic regimes of top or bottom squark masses, as well as for mass differences between the squark and LSP, where the LSP is taken to be the \PSGczDo.
They are:
(1) a search for top-squark pair production in multijet events with at least one tagged hadronically decaying top quark (hereafter referred to as the ``multijet t-tagged'' search), which is sensitive to scenarios with a large mass difference between the top squark and the LSP;
(2) a search for dijet events with exactly one or two tagged bottom-quark jets (b jets) possibly accompanied by additional jets radiated in the initial state (hereafter referred to as the ``dijet b-tagged'' search), which is sensitive to scenarios with large or intermediate mass differences between the bottom squark and the LSP;
and (3) a search for events with a single jet (hereafter referred to as the ``monojet'' search), which is sensitive to scenarios with highly compressed spectra, \ie to scenarios in which the mass difference between the top or bottom squark and the LSP is small.
The results from the three searches are combined and interpreted in the context of simplified model spectra (SMS)~\cite{bib:SMS}.
Previous searches for top- and bottom-squark pair production at the LHC are presented in Refs.~\cite{ATLAS1,ATLAS2,ATLAS5,ATLAS5a,ATLAS6,ATLAS7,ATLAS8,CMS-STOP-lepton,CMS-alphaT,CMS-sbottom-ssdilep,CMS-tophiggsino,CMS-SUS-13-024,CMS-SUS-14-010}.

This paper is organised in the following way.
Section~\ref{sec-detector} describes the CMS detector, and Section~\ref{sec-eventReco} discusses event reconstruction algorithms.
The simulations of signal and background events are outlined in Section~\ref{sec-simulation}.
A summary of the strategies shared by all three searches, including common event selections and backgrounds, are discussed in Section~\ref{sec-strategy}.
The multijet t-tagged search is presented in Section~\ref{sec-SUS-13-015},
the dijet b-tagged search in Section~\ref{sec-SUS-13-018}, and the monojet search in
Section~\ref{sec-SUS-13-009}.
Finally, the results are shown in Section~\ref{sec-results} and interpreted using SMS in Section~\ref{sec-interpretations}, with a summary in Section~\ref{sec-conclusions}.
Additional information for model testing can be found in Appendix~\ref{infoAdd}.

\section{The CMS detector\label{sec-detector}}

The central feature of the CMS apparatus is a superconducting solenoid of 6\unit{m} internal diameter, providing a magnetic field of 3.8\unit{T}. Within the superconducting solenoid volume are a silicon pixel and strip tracker, a lead tungstate crystal electromagnetic calorimeter (ECAL), and a brass and scintillator hadron calorimeter (HCAL), each composed of a barrel and two endcap sections. Muons are measured in gas-ionization detectors embedded in the steel flux-return yoke outside the solenoid. Extensive forward calorimetry complements the coverage provided by the barrel and endcap detectors.

The polar angle $\theta$, defined with respect to the anticlockwise-beam direction,
the pseudorapidity $\eta$, defined as $\eta = -\ln[ \tan (\theta/2)]$, and the
azimuthal angle $\phi$ in the plane perpendicular to the beam axis, define the
coordinates used to describe position within the detector.
The transverse momentum vector \ptvec of a particle is defined as the projection of its four-momentum on to the plane perpendicular to the beams.
Its magnitude is referred to as \pt.

The silicon tracker measures charged particles within the pseudorapidity range $\abs{\eta}< 2.5$.
Isolated particles of $\pt = 100$\GeV emitted with $\abs{\eta} < 1.4$ have track resolutions of 2.8\% in \pt and 10 (30)\mum in the transverse (longitudinal) impact parameter \cite{TRK-11-001}.
The ECAL and HCAL measure energy deposits in the pseudorapidity range $\abs{\eta}< 3$.
Quartz-steel forward calorimeters extend the coverage to $\abs{\eta}=5$.
The HCAL, when combined with the ECAL, measures jets with a resolution $\Delta E/E \approx 100\% / \sqrt{E\,[\GeVns]} \oplus 5\%$~\cite{Chatrchyan:2013dga}.
Muons are measured in the pseudorapidity range $\abs{\eta}< 2.4$.
Matching muons to tracks measured in the silicon tracker results in a relative transverse momentum resolution for muons with $20 <\pt < 100\GeV$ of 1.3--2.0\% in the barrel and better than 6\% in the endcaps.
The \pt resolution in the barrel is better than 10\% for muons with \pt up to 1\TeV~\cite{CMSPAPER:MUO_10_004}.

The events used in the searches presented here were collected using a two-tier trigger system: a hardware-based level-1 trigger and a software-based high-level trigger.
A full description of the CMS detector and its trigger system can be found in Ref.~\cite{bib:CMS_TDR}.

\section{Event reconstruction\label{sec-eventReco}}
Events are reconstructed with the CMS particle-flow algorithm~\cite{bib:Particle_Flow,bib:PAS_PFT_10_001}.
Using an optimized combination of information from the tracker, the calorimeters, and the muon systems,
each particle is identified as a
charged hadron, neutral hadron, photon, muon, or electron.
Charged hadrons that do not originate from the primary vertex,
defined by the pp interaction vertex with the largest sum of charged-track $\pt^{2}$ values, are not considered.
The remaining particles are clustered into jets using the anti-\kt algorithm with distance parameter 0.5~\cite{bib:akjets}.
Calorimeter energy deposits corresponding to neutral particles originating from overlapping pp interactions, ``pileup", is subtracted on an event-by-event basis using the jet-area method~\cite{Cacciari:2011ma}.
Jets are corrected to take into account the detector response as a function of jet \pt and $\eta$,
using factors derived from simulation.
The jets must satisfy loose identification criteria that remove calorimeter noise.
An additional residual energy correction, derived from dijet and $\gamma$+jets events, is applied to
account for differences in the jet energy scales~\cite{JETJINST} between simulation and data.

Both the multijet t-tagged and dijet b-tagged analyses employ tagging of b quark jets (b tagging).
Utilising the precise inner tracking system of the CMS detector,
the combined secondary vertex (CSV) algorithm~\cite{BJet} uses secondary vertices, track-based lifetime information, and jet kinematics
to distinguish between jets from b quarks and those from light quarks or gluons.

In the multijet t-tagged analysis, a jet is tagged as a b quark jet if it satisfies $\pt>30$\GeV, $\abs{\eta}<2.4$, and the medium working point requirements of the algorithm~\cite{bTagPAS}.
Averaged over \pt in \ttbar events, the b quark identification efficiency is 67\% for the medium working point,
and the probability for a jet originating from a light quark or gluon to be misidentified as a b quark jet is 1.4\%.
The dijet b-tagged analysis uses the loose and medium working point versions of the algorithm.
The b-tagging efficiency is 80--85\% (46--74\%) for the loose (medium) working point~\cite{BJet}, and the probability for a jet originating from a light quark or gluon to be misidentified
as a b quark is 8--12\% (1--2\%).
Values are quoted for jets with $\pt>70$\GeV and are dependent on the jet \pt.

Muons are reconstructed by finding compatible track segments in the silicon tracker
and the muon detectors~\cite{CMSPAPER:MUO_10_004}.
Both the dijet b-tagged and multijet t-tagged analyses require muons to lie within $\abs{\eta}<2.1$, whereas the monojet analysis uses muons up
to $\abs{\eta}<2.4$.
Electron candidates are reconstructed from a cluster of energy deposits in the ECAL
that is matched to a track in the silicon tracker~\cite{CMSPaper:EGM_13_001}.
Electron candidates are required to satisfy $\abs{\eta}<1.44$ or
$1.56<\abs{\eta}<2.5$, where the intermediate range of $\abs{\eta}$ is excluded to avoid the transition region between the ECAL barrel and endcap, in which the reconstruction efficiency is difficult to model.
Muon and electron candidates are required to originate within 2\unit{mm} of the beam axis in the
transverse plane.
In the monojet analysis, hadronically decaying $\tau$ leptons are reconstructed using the ``hadron-plus-strips''
algorithm~\cite{bib:HPStaus}, which reconstructs candidates with one or three charged pions and up to two neutral pions.

A relative lepton isolation parameter is defined as the sum of the \pt of all photons and all charged and neutral hadrons, computed in a
cone of radius $\Delta R = \sqrt{\smash[b]{( \Delta \eta ) ^{2} + ( \Delta \phi) ^{2}}} = 0.4$ around the lepton direction, divided by the lepton \pt.
Values are corrected for the effect of pileup.
Lepton candidates with relative isolation values below 0.2 are considered isolated in the monojet and dijet b-tagged analyses.

In the multijet t-tagged analysis, a key ingredient for providing good background rejection and simultaneously preserving good signal selection involves vetoing prompt leptons from $\PW$ or $\Z$ boson decays, while accepting possible secondary leptons from b quark decays.
Hence events containing a muon or electron with $\pt> 5$\GeV are vetoed based on the spatial distribution of particles around the lepton.
A directional isolation parameter $\text{Iso}^{\text{dir}}$ is defined by considering particles in a region of radius $\Delta R$ centred on the lepton direction, where $\Delta R$ is 0.2 for muons and 0.2 (0.3) for electrons with $\abs{\eta} \le 1.44\,({>}1.56)$.
A sum is performed over the particle transverse momenta multiplied by the square of the angle in the $\eta$--$\phi$ plane between the particle and the \pt-weighted centroid of all particles contributing to the sum~\cite{bib:dirIsoPAS}.
Leptons from heavy-quark decays usually are closer to hadronic activity in $\eta$--$\phi$ space than leptons from on-shell $\PW$ or $\Z$ boson decays.
The requirements on $\text{Iso}^\text{dir}$ have been chosen to retain high rejection efficiency, especially for high-\pt leptons, and a small misidentification rate for leptons from b quark decays.
This is the first CMS publication to make use of this variable.

The hermetic nature of the CMS detector allows event reconstruction over nearly the full solid angle.
Conservation of momentum in the transverse plane can therefore be used to detect a momentum imbalance,
which can be associated with particles that exit the detector without interaction.
The missing transverse momentum vector \ptvecmiss is defined as the projection on the plane perpendicular to the
beam axis of the negative vector sum of the momenta of all reconstructed particles in an event.
Its magnitude is referred to as \ptmiss.
For the monojet analysis, an alternative definition of \ptvecmiss is used, \ptvecmissmu, which differs from the nominal definition in that the contribution of muons is excluded.
This alternative definition allows the same trigger, for which missing transverse momentum is defined without muons, to be used for both signal and control samples, reducing systematic uncertainties.
The alternative definition \ptvecmissmu is also used to evaluate some electroweak backgrounds for the multijet t-tagged and dijet b-tagged analyses, as described below.

\section{Simulation of signal and background event samples\label{sec-simulation}}

Monte Carlo (MC) simulations of signal and background events are used to optimize selection criteria,
determine signal efficiencies, and develop background estimation techniques.

\begin{figure*}[tbh]
  \centering
  \includegraphics[width=0.4\textwidth]{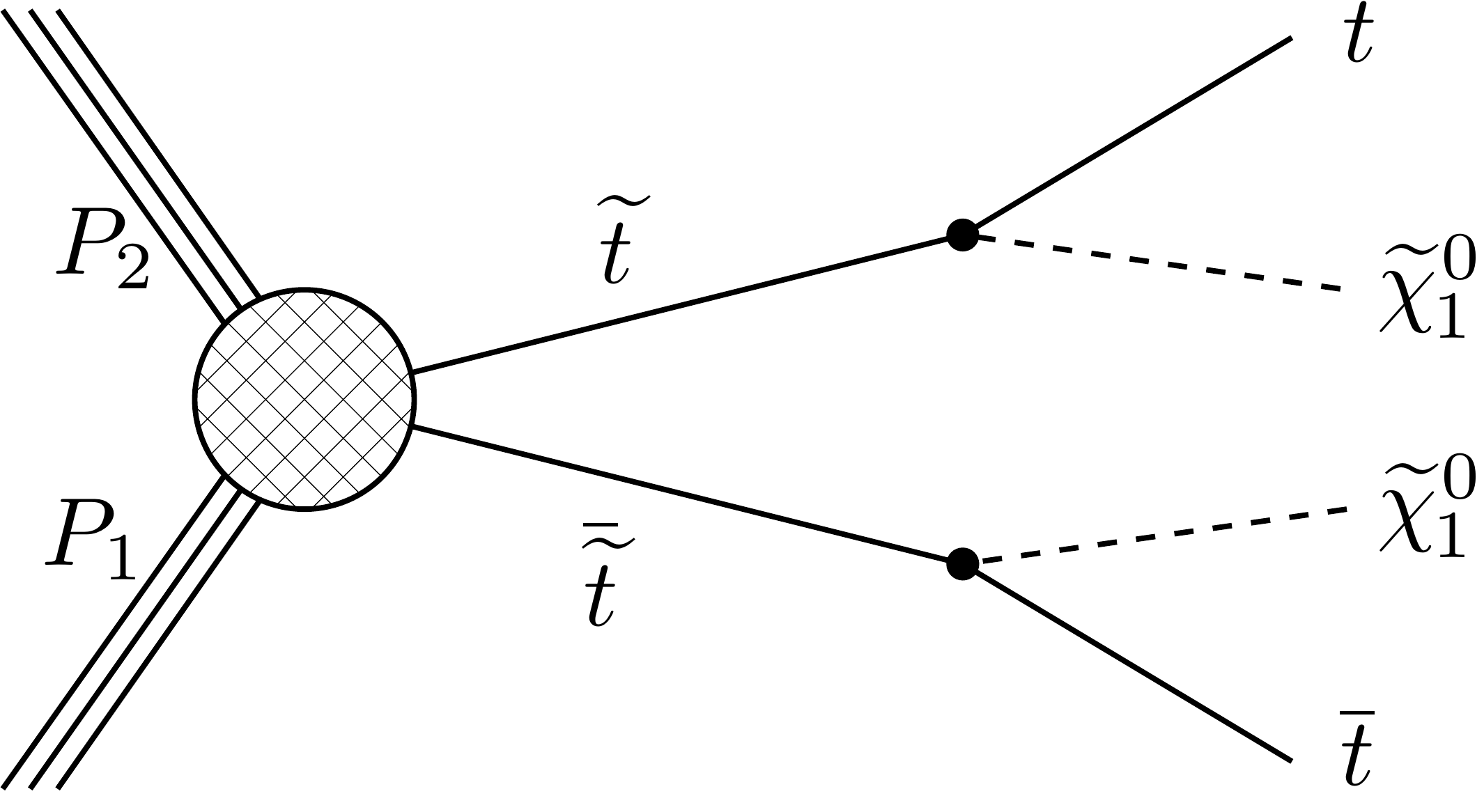}
  \includegraphics[width=0.4\textwidth]{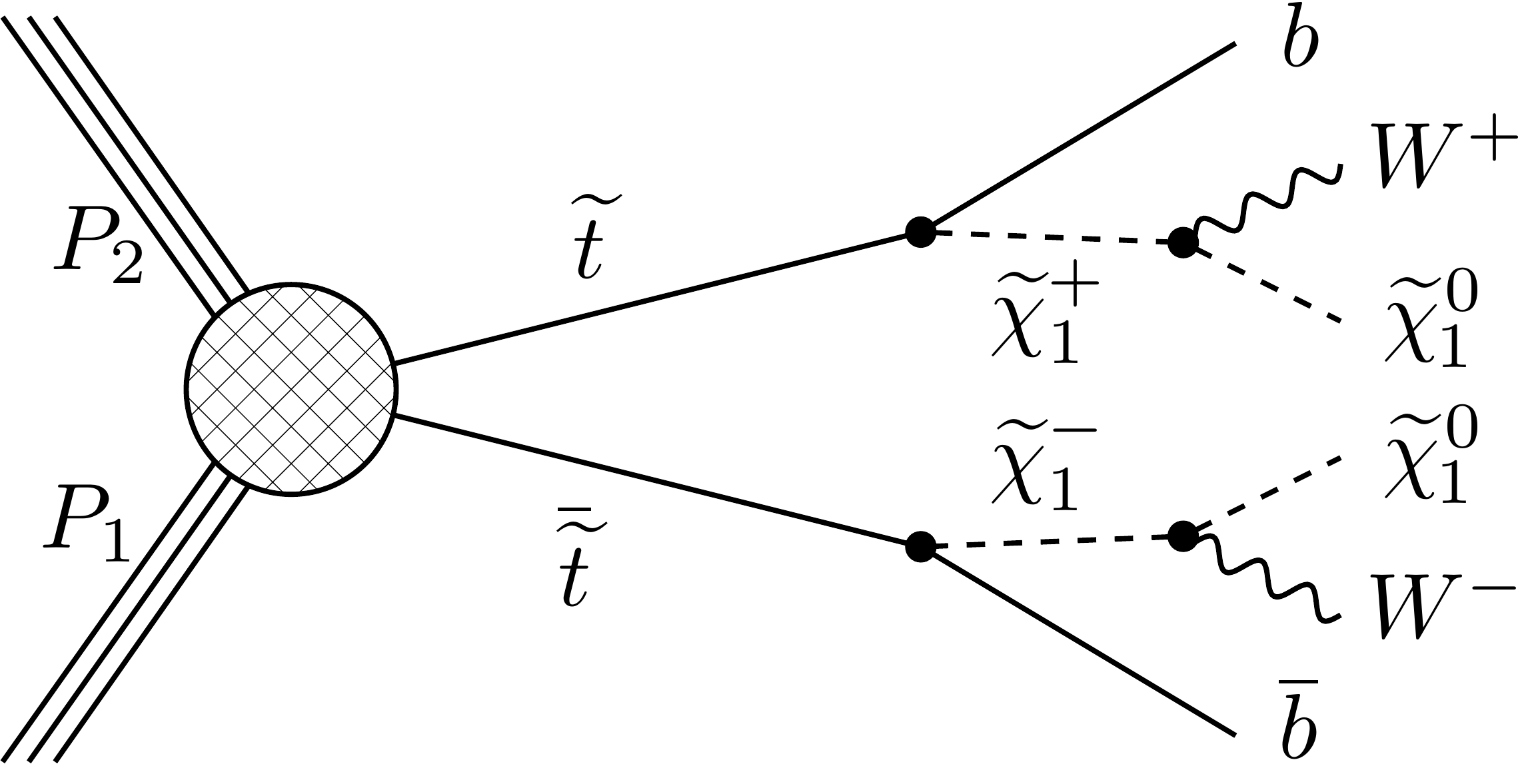}
  \includegraphics[width=0.4\textwidth]{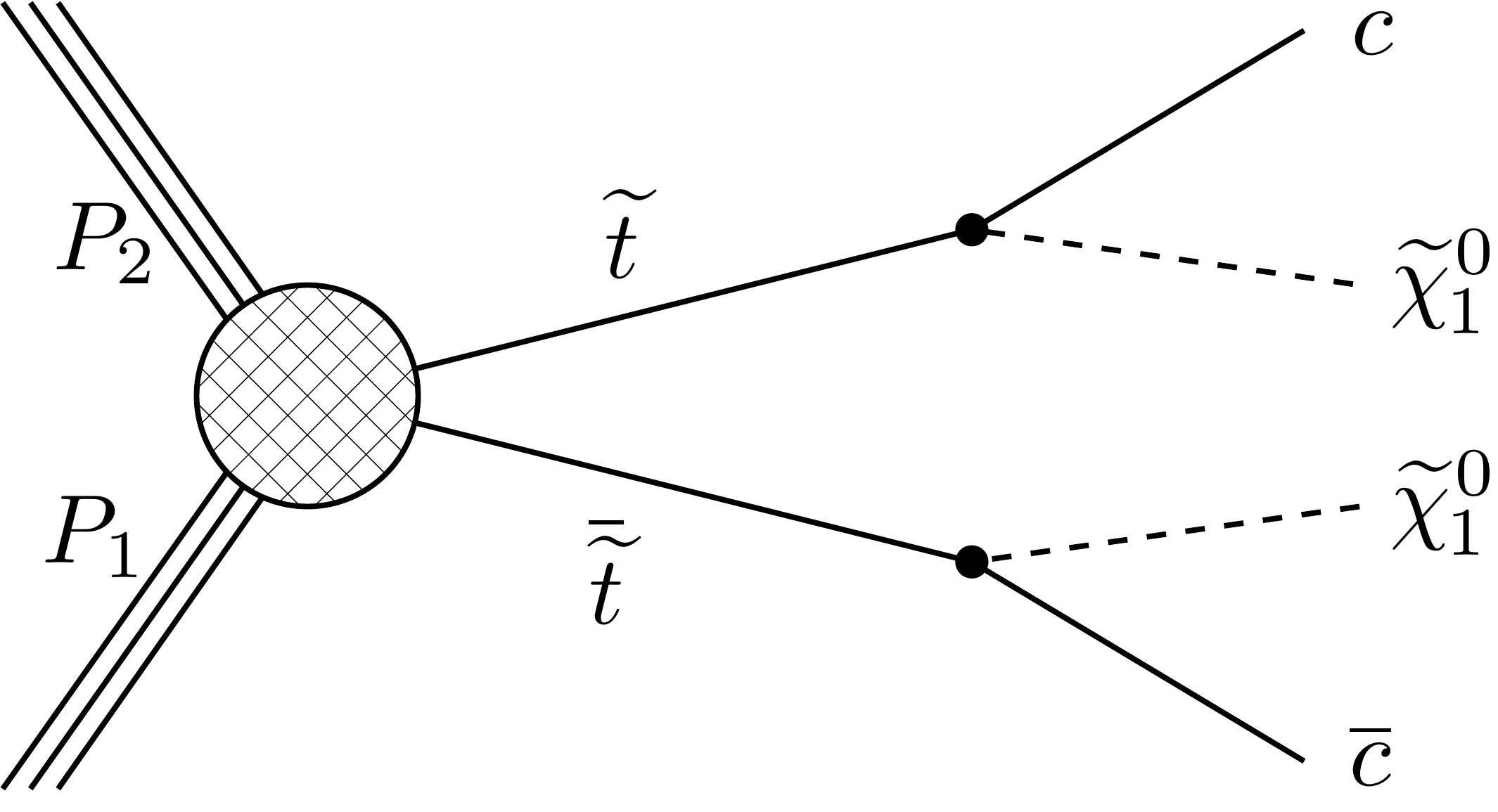}
  \includegraphics[width=0.4\textwidth]{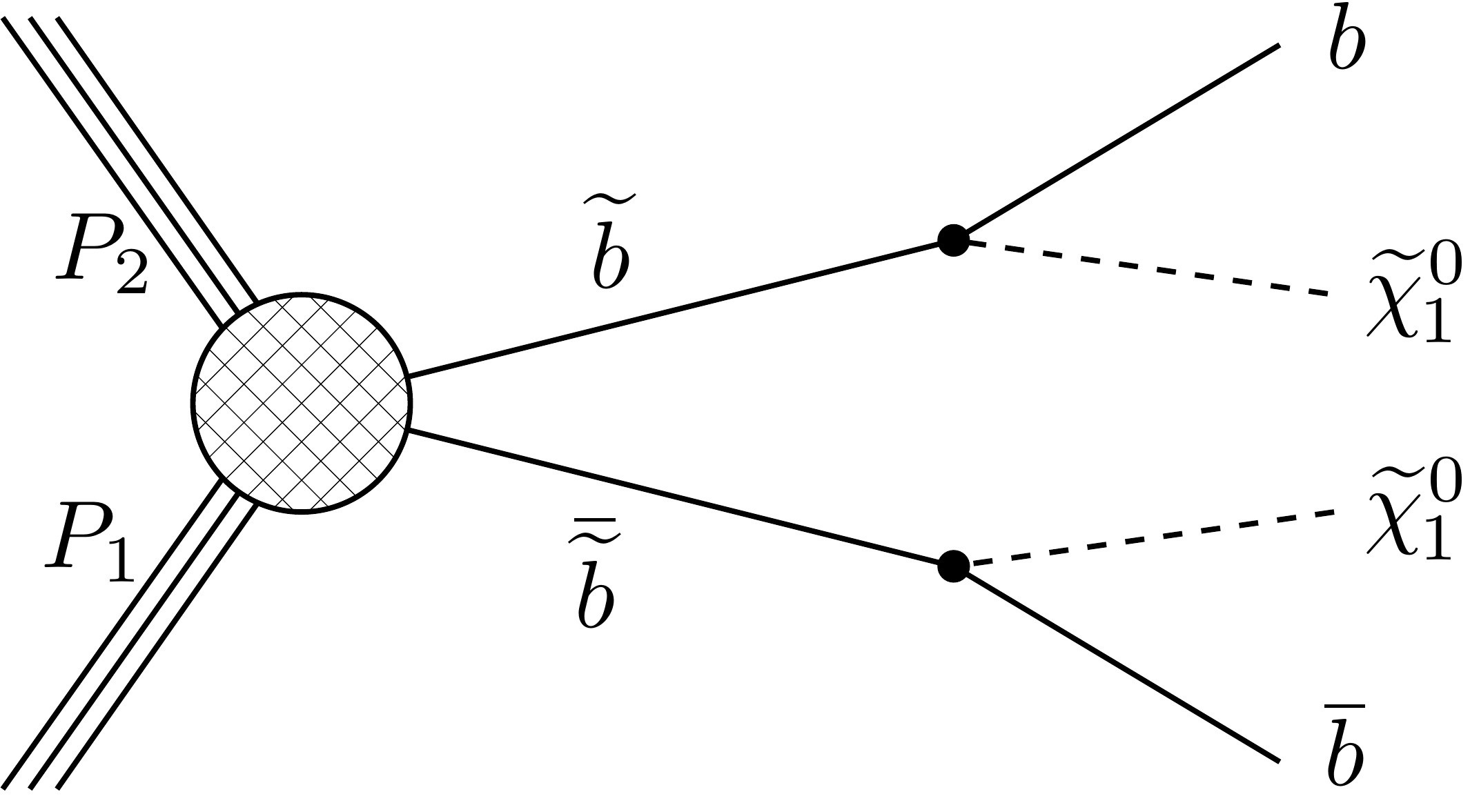}
  \caption{Feynman diagrams showing the pair production of top or bottom squarks followed by their decays according to
$\PSQt\to\PQt\PSGczDo$ (top, left),
$\PSQt\to\PQb\PSGcpmDo$
with $\PSGcpmDo\to\PQb\PW^{\pm}$ (top, right),
$\PSQt\to\PQc\PSGczDo$ (bottom, left), a flavour changing neutral current loop-induced process, and
$\PSQb\to\PQb\PSGczDo$ (bottom, right).
\label{feynman_diagrams}}

\end{figure*}

Within the context of natural SUSY, several SMS scenarios are examined. They are based on the pair production of top or bottom squarks followed by the decay of the top or bottom squarks according to
$\PSQt\to\PQt\PSGczDo$,
$\PSQt\to\PQb\PSGcpmDo$ with $\PSGcpmDo\to\PQb\PW^{\pm}$,
$\PSQt\to\PQc\PSGczDo$, and
$\PSQb\to\PQb\PSGczDo$, where \PSGcpmDo is the lightest chargino.
The Feynman diagrams for these processes are shown in Fig.~\ref{feynman_diagrams}.
Simulated samples of signal events are generated with the \MADGRAPH 5.1.3.30~\cite{madgraph2} event generator, with up to two additional partons incorporated at the matrix element level.
All SUSY particles other than those included in the SMS scenario under consideration are assumed to be too heavy to participate in the interaction.

SM events are simulated using a number of MC event generators.
Top-antitop quark pair production (\ttbar), \PW/\Z+jets, $\Z\gamma$, $\PW\gamma$, $\ttbar\Z$, and $\ttbar\PW$ samples are produced using the \MADGRAPH{}5
event generator with CTEQ6L~\cite{CTEQ6} parton distribution functions (PDFs).
Single top quark events are generated with the \POWHEG~\cite{POWHEG} program using
the CT10~\cite{CT10} and CTEQ66~\cite{CTEQ66} PDFs.
Multijet events from QCD processes and events with WW, WZ and ZZ (diboson) production are
simulated with the \PYTHIA 6.4.24~\cite{pythia} program using the CTEQ6L PDFs.

For both the signal and SM simulated samples, the parton shower, hadronization, and multiple-parton interactions are described using \PYTHIA.
Decays of $\tau$ leptons are handled by the \TAUOLA 27.121.5 package~\cite{TAUOLA}.
The generated events are interfaced to the CMS fast detector simulation~\cite{FASTSIM}
for the signal samples and to a \GEANTfour-based~\cite{Agostinelli2003250} detector simulation for the SM background samples.

\section{Search strategy\label{sec-strategy}}

The analyses presented here are designed to be efficient for possible signals,
while maintaining manageable background levels.
All three searches require at least one high-\pt jet and a large value of \ptmiss.
Background from QCD multijet events is reduced by a minimum angle between the directions of the \ptvecmiss vector and highest \pt jet(s).
Electroweak backgrounds are reduced by vetoing events with leptons.
Use of b tagging and kinematic variables further distinguishes signal from background.

The sources of SM background, and the background evaluation procedures, are also similar in the three searches.
Events with a $\Z$ boson that decays to neutrinos, denoted \znunubr+jets,
contain genuine \ptmiss and constitute a significant background.
This background is estimated
using dimuon control samples, exploiting the similar kinematics of \znunu
and \zmumu events as well as the known branching fractions.
In regions where \ttbar contamination is small, \wpj events with $\PW \to \mu \nu$ can similarly be used to estimate the \znunubr+jets background.
Another significant source of background is from \wpj events when the $\PW$ boson decays leptonically, denoted \wlnubr+jets events.
Here, the lepton (electrons and muons, including those from leptonically decaying $\tau$ leptons) fails the lepton veto
and hence is ``lost'', i.e. it is not isolated, not identified, or outside of the acceptance of
the analysis.
Hadronically decaying $\tau$ leptons (\tauh) from $\PW$ boson decay in \ttbar and \wpj events form another significant background source.
Both the lost-lepton and \tauh backgrounds are evaluated using single-muon control samples.
Dijet and multijet backgrounds are reduced using topological selections,
with the remaining
contributions estimated using data control regions enhanced in QCD events.
Very small backgrounds from processes such as diboson, \ttbarZ, \ttbarW, and single top quark are estimated from simulation.
The data control regions used in the estimates of the SM backgrounds are defined in such a manner to minimize the contributions of signal events, and thus possible signal event contributions to control regions are ignored.

\section{Search for top-squark pair production using top-quark tagging \label{sec-SUS-13-015}}
This search for pairs of hadronically decaying top quarks with large \ptmiss in the final state is motivated by the scenario of top-squark pair production, assuming that the mass difference between the top squark and the stable LSP is larger than the mass of the top quark, $m_{\PSQt} - m_{\PSGczDo} \ge m_{\PQt}$.
The decay channel $\PSQt\to\PQt\PSGczDo$ is therefore kinematically available, allowing a search for top squarks through top quark tagging, which provides an important discriminant against the multijet background.
If \PSGcpmDo states exist with a mass between the top squark and the LSP masses, the top squark can also decay via $\PSQt\to\PQb\PSGcpDo\to\PQb\PWp\PSGczDo$ (plus its charge conjugate), yielding a different event signature since no top quark is produced.
By requiring just one fully reconstructed top quark, the search maintains sensitivity to
$\PSQt\to\PQt\PSGczDo$ as well as $\PSQt\to\PQb\PSGcpmDo$ decays.

\subsection{Event selection \label{sec:t2tt_eventSel}}

The event sample used for this analysis is collected by triggering on events with $\ptmiss{}>80$\GeV, where \ptmiss is reconstructed using the particle-flow algorithm, and at least two central ($\abs{\eta}< 2.6$) jets with $\pt>50$\GeV.
This trigger is $(98\pm1)\%$ efficient as measured in data once the analysis requirements described below have been applied.
The selected events are required to have:
(i)~no identified electrons or muons with $\pt>5$\GeV{} that are isolated according to the directional isolation parameter described in Section~\ref{sec-eventReco};
(ii) at least five jets with $\pt> 30$\GeV and $\abs{\eta}<2.4$, of which the two highest \pt jets must have $\pt>70$\GeV and the next two highest \pt jets $\pt>50$\GeV{};
(iii) at least one b-tagged jet, $\nbjets\ge1$; and
(iv) azimuthal angle $\Delta\phi(\ptvec^{j},\ptvecmiss)$ between the directions of the three highest \pt jets and the \ptvecmiss vector larger than 0.5, 0.5, and 0.3, respectively, with $\pt^1>\pt^2>\pt^3$.
The electron and muon vetoes minimize backgrounds from SM \ttbar and \PW{}+jets production, where the \PW{} boson decays into a neutrino and a lepton.
Events containing a hadronically decaying $\tau$ lepton are not explicitly rejected.
The jet multiplicity and b-tagging requirements help to select signal events, since the \susy signatures of interest tend to include multiple jets in the central $\eta$ range, high-\pt leading jets and b jets.
The $\Delta\phi$ requirement strongly suppresses the background from QCD multijet events,
which mostly arises from the mismeasurement of jet \pt, leading to large \ptvecmiss aligned along a jet axis.
Events that satisfy the above requirements are denoted the ``preselection'' sample.

Reconstruction of hadronically decaying top quarks is performed as suggested in Refs.~\cite{Kaplan:2008ie,Plehn:2010st,Kaplan:2012gd}.
To maximize signal acceptance,
one ``fully reconstructed'' and one ``partially reconstructed'' top quark are required.
The collection of five or more jets in the preselection sample is divided into all possible sets of three jets and a remnant, where the remnant must contain at least one b-tagged jet. The fully reconstructed top quark is one of the three-jet (trijet) combinations.
The partially reconstructed top quark is then built from the remnant using the b-tagged jet as a seed.
If the remnant contains multiple b-tagged jets, the one with highest \pt is used as the seed.
Once events with two candidate top quarks are identified,
they are used to form additional kinematical variables that distinguish between signal and the remaining SM background, which arises primarily from \ttbar production.

\subsubsection{Top quark reconstruction}
To be considered as a fully reconstructed top quark, the trijet system must satisfy the following requirements.
(i)~Each jet must lie within a cone in ($\eta,\phi$) space of radius 1.5 centred on the momentum direction formed by the trijet combination.
The radius requirement implies a moderate Lorentz boost of the top quark as expected for the large $\Delta m = m_{\PSQt} - m_{\PSGczDo}$ region targeted in this search.
(ii) The trijet system mass ($m_{\text{3-jet}}$) must be within the range 80-270\GeV.
(iii) The trijet system must satisfy one of the three following criteria:
\begin{equation*} \label{eq:taggerEQ}
\begin{split}
\text{(a)}\quad&0.2< \arctan\left(\frac{m_{13}}{m_{12}}\right) < 1.3\quad\text{and}\quad R_\text{min} < \frac{m_{23}}{m_{\text{3-jet}}} < R_\text{max},\\
\text{(b)}\quad& R_\text{min}^{2}\left[1+\left(\frac{m_{13}}{m_{12}}\right)^2\right]
     < 1-\left(\frac{m_{23}}{m_{\text{3-jet}}}\right)^2
     < R_\text{max}^{2}\left[1+\left(\frac{m_{13}}{m_{12}}\right)^2\right]
     \quad\text{and}\quad\frac{m_{23}}{m_{\text{3-jet}}} > 0.35, \\
\text{(c)}\quad&R_\text{min}^{2}\left[1+\left(\frac{m_{12}}{m_{13}}\right)^2\right]
     < 1-\left(\frac{m_{23}}{m_{\text{3-jet}}}\right)^2
     < R_\text{max}^{2}\left[1+\left(\frac{m_{12}}{m_{13}}\right)^2\right]
     \quad\text{and}\quad
     \frac{m_{23}}{m_{\text{3-jet}}} > 0.35. \\
\end{split}
\end{equation*}
Here, $m_{12}$, $m_{13}$, and $m_{23}$ are the dijet masses, where the jet indices 1, 2, and 3 are \pt ordered.
The numerical constants have values $R_\text{min} = 0.85(m_{\PW}/m_{\PQt})$,
$R_\text{max} = 1.25(m_{\PW}/m_{\PQt})$,
$m_{\PW} = 80.4$\GeV, and $m_{\PQt} = 173.4$\GeV~\cite{PDG}.

\begin{figure}[htbp!]
\centering
   \includegraphics[width=0.4\textwidth]{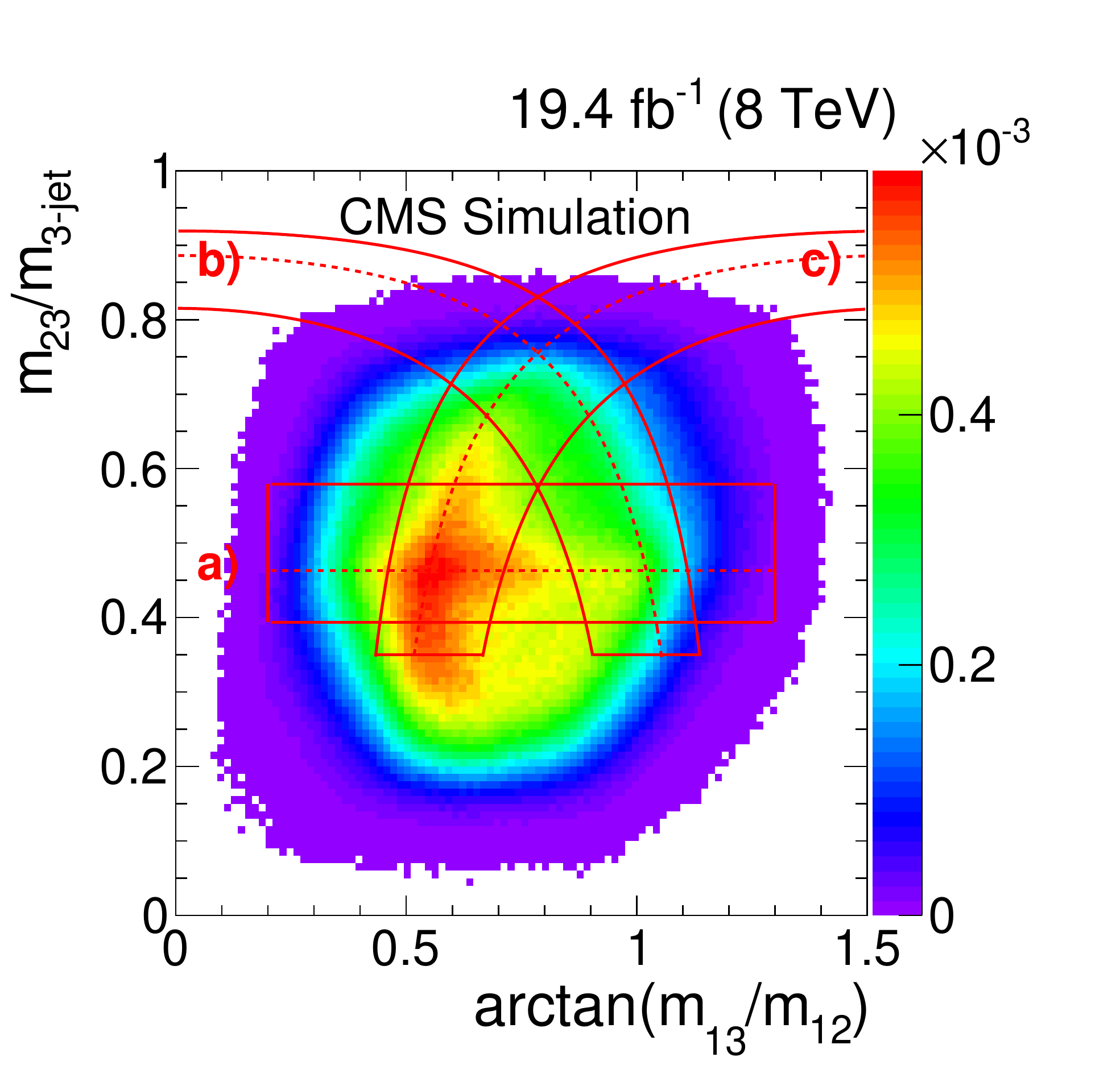}
   \includegraphics[width=0.4\textwidth]{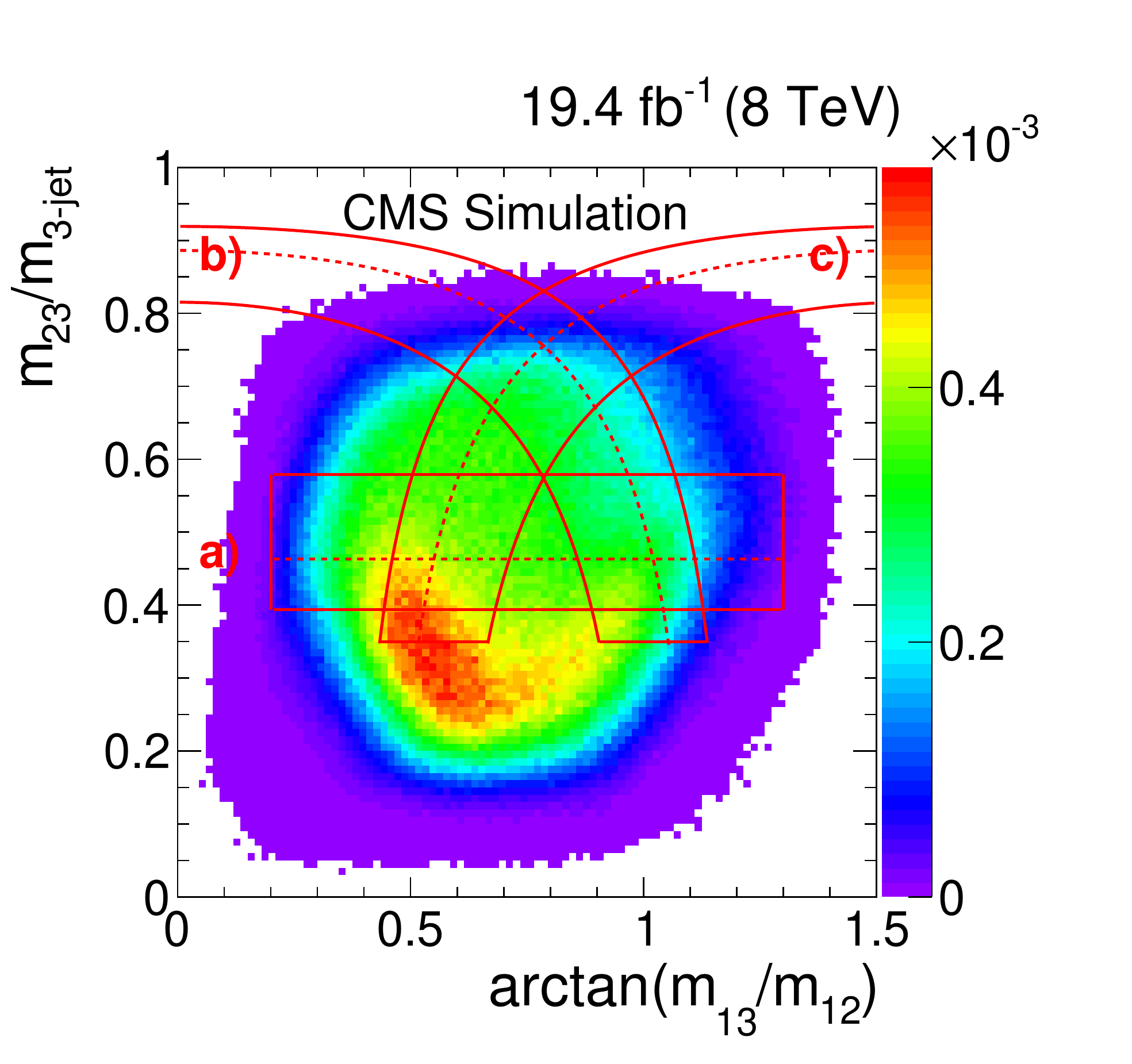} \\
   \caption{
   Distributions of $m_{23}/m_{\text{3-jet}}$ versus $\arctan(m_{13}/m_{12})$ for simulated SM \ttbar (left),
   and multijet (right) events in the multijet t-tagged search.
   The red contours (a), (b), and (c) limit the regions in which conditions (a), (b) and (c) are satisfied, respectively.
   The central dashed lines represent where the ratios involved in conditions (a), (b) and (c) are equal to $m_{\PW}/m_{\PQt}$, as described in the text.
    }
\label{fig:kinVars1}

\end{figure}

The top quark tagging (t tagging) conditions of (a), (b), or (c)
can be reduced (under certain approximations detailed in Ref.~\cite{Plehn:2010st} ) to the requirement that $m_{23}/m_{\text{3-jet}}$, $m_{12}/m_{\text{3-jet}}$, or $m_{13}/m_{\text{3-jet}}$,
respectively, be consistent with the $m_{\PW}/m_{\PQt}$ ratio.
The other conditions are motivated by the Lorentz structure of the tW coupling and suppress contributions from light-quark and gluon jets~\cite{Plehn:2010st}.
These t tagging conditions are illustrated in Fig.~\ref{fig:kinVars1} for simulated SM \ttbar (left) and QCD (right)
events.
   The lower box defines the region dictated by the criterion (a),
   with the central dashed horizontal line representing the ratio $m_{\PW}/m_{\PQt}$.
   Similarly, the curved regions defined by criteria (b) and (c) are also shown, where the central dashed line indicates where $m_{12}/m_{\text{3-jet}}$ is equal to $m_{\PW}/m_{\PQt}$ for region (b), and where $m_{13}/m_{\text{3-jet}}$ is equal to $m_{\PW}/m_{\PQt}$ for region (c).
The requirement that events lie within the boundaries defined by (a), (b), or (c) is seen to be effective at selecting the SM \ttbar events, which are very similar to signal events due to similar $m_{23}/m_{\text{3-jet}}$ and $m_{13}/m_{12}$ ratios,
while rejecting the bulk of the multijet background.
If multiple trijet combinations satisfy these criteria, the triplet with mass closest to the top quark mass
is selected.
The four-momentum of the selected trijet system, $\pTopFourVec = (E^{\text{3-jet}}, \vec{p}^{\text{3-jet}})$, is used in the subsequent calculation of kinematical variables that refine the event selection, described below.

The partial reconstruction of a second top quark is attempted in the remnant system, denoted $\text{R-sys}$.
The four-momentum of the collective decay products in $\text{R-sys}$ is denoted $\pRsystemFourVec=(E^{\text{R-sys}}, \vec{p}^{\text{R-sys}})$
and is constructed from either 3, 2, or 1 jet(s) in $\text{R-sys}$.
If $\text{R-sys}$ has $\ge$3 jets, all possible trijet combinations containing the b-tagged jet are considered.
To retain maximum signal acceptance, the full reconstruction criteria of requirements (a), (b) and (c) are not used.
Instead we merely select the trijet system with mass closest to that of the top quark.
In addition, to reduce the misconstruction of top quark candidates, requirements are placed on the hadronic decay of the \PW{} boson candidate in the trijet system:
excluding the b-tagged jet, the remaining pair of jets is required to have a dijet mass between 50 and 120\GeV.
If this condition is satisfied, the four-momentum of the trijet system defines \pRsystemFourVec.
Otherwise the trijet system is rejected and we examine 2-jet combinations involving the b-tagged jet.
In the latter case, the separation between the b-tagged jet and the other jet is required to satisfy $\Delta R \equiv \sqrt{\smash[b]{(\Delta\eta(\PQb,\jet))^2+(\Delta\phi(\PQb,\jet))^2}}\le 2.0$ and
the dijet mass must be less than the top quark mass.
If multiple jet pairs satisfy these requirements, the pair with smallest $\Delta R$ is selected
and the four-momentum of the pair defines \pRsystemFourVec.
If no jet pair satisfies the requirements, the b-tagged jet is selected as the complete remnant system, and its four-momentum defines \pRsystemFourVec.

\subsubsection{Kinematic requirements}

\begin{figure*}[tbh]
\centering
  \includegraphics[width=0.4\textwidth]{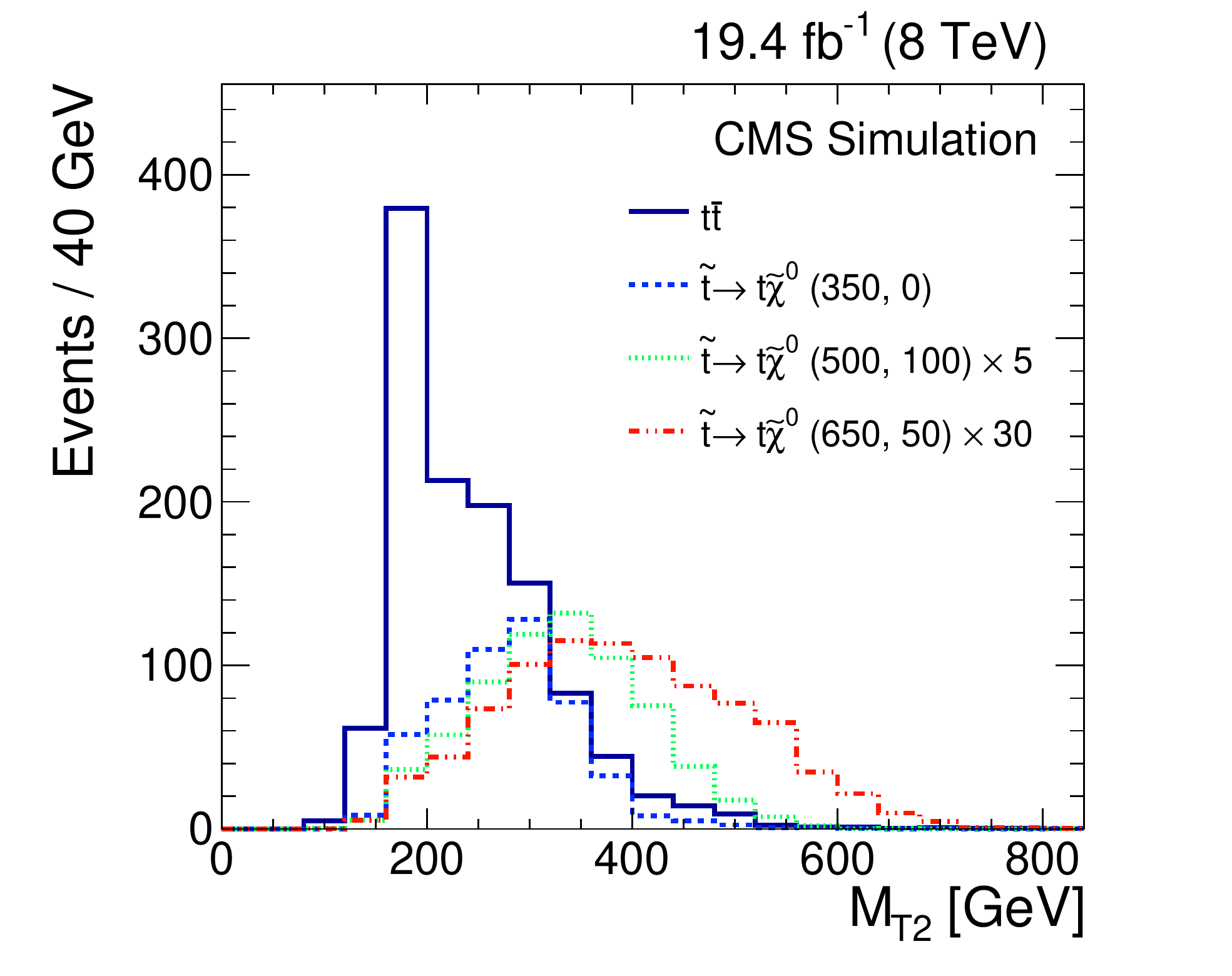}
   \includegraphics[width=0.4\textwidth]{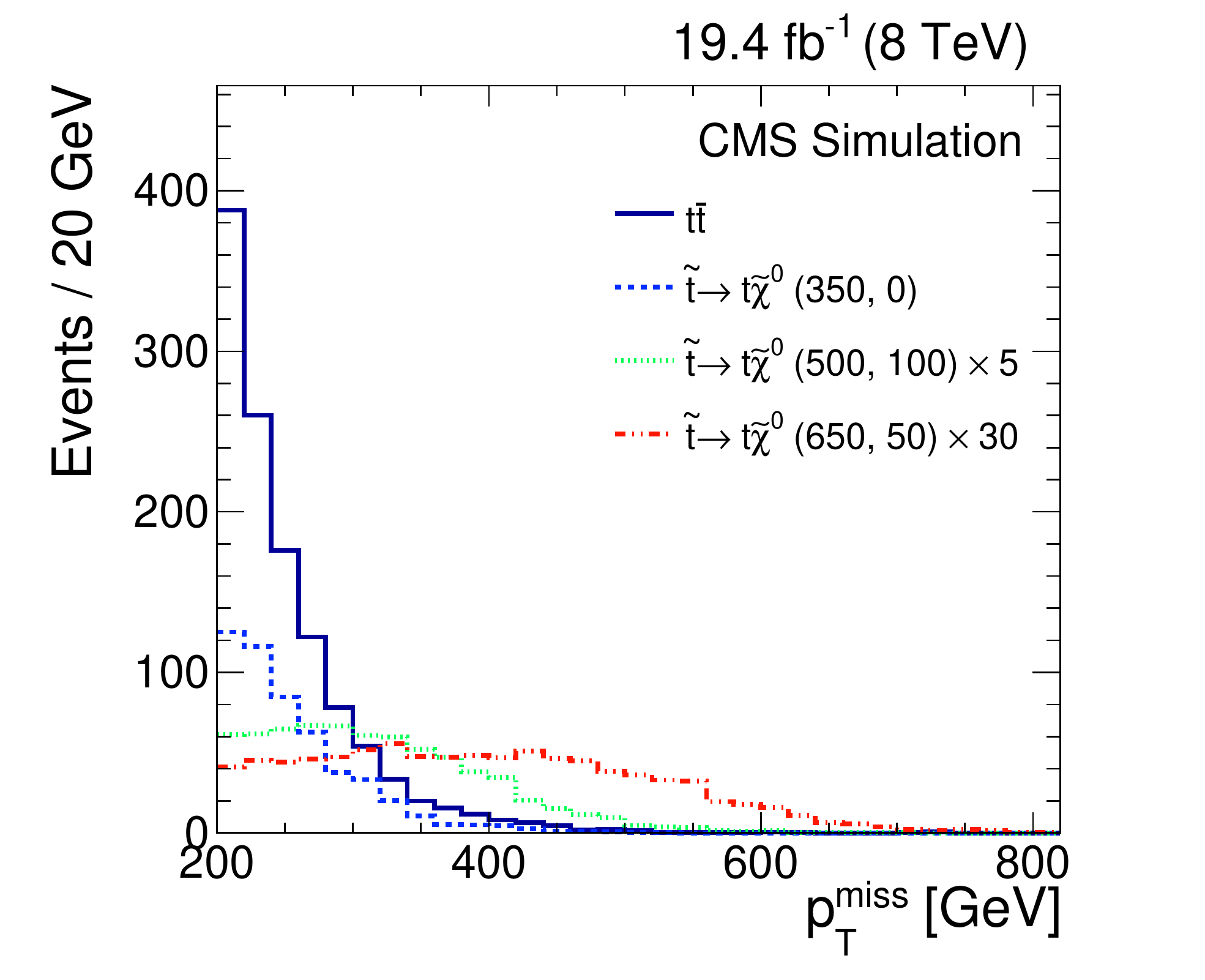}
\\
   \includegraphics[width=0.4\textwidth]{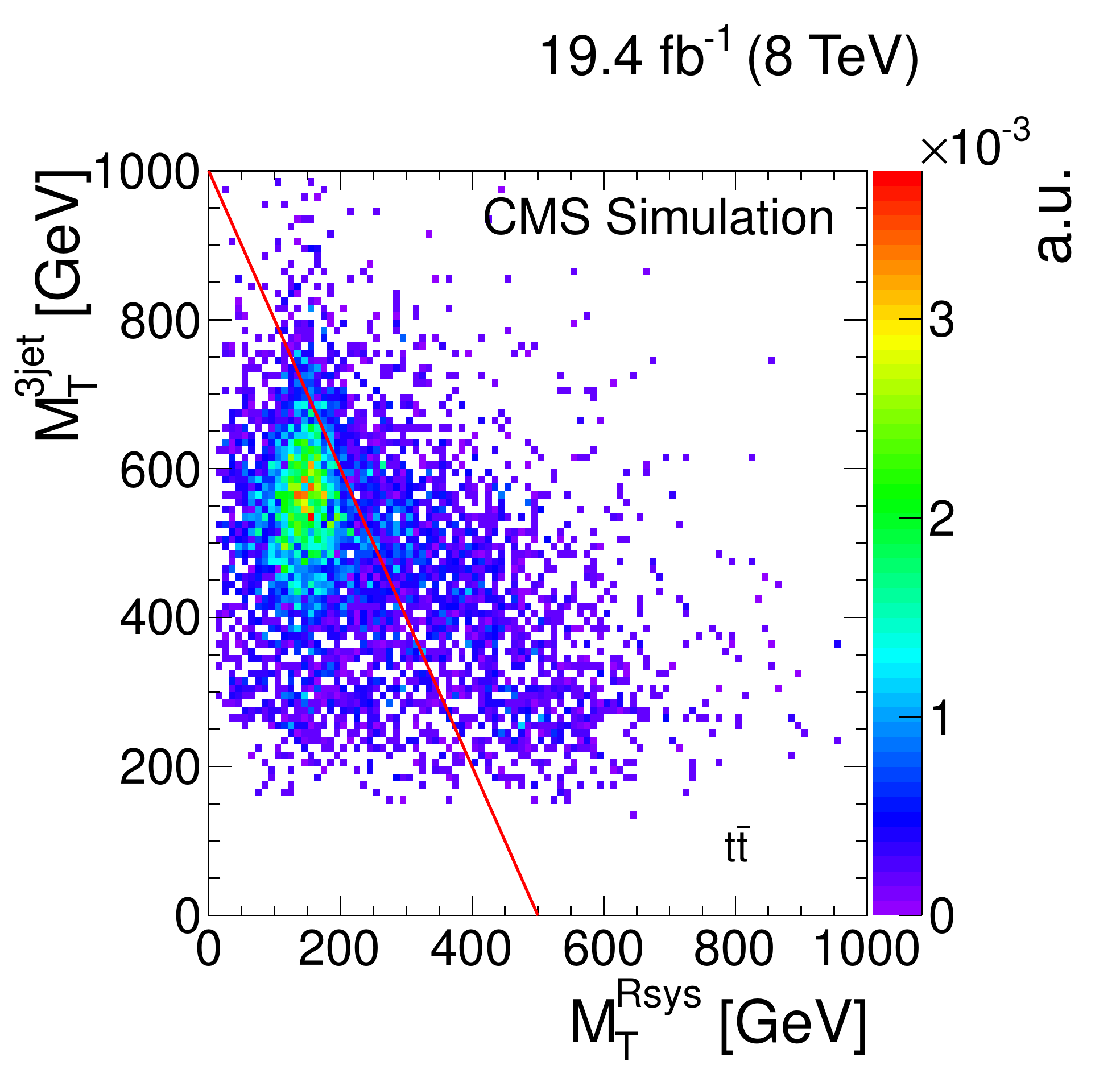}
   \includegraphics[width=0.4\textwidth]{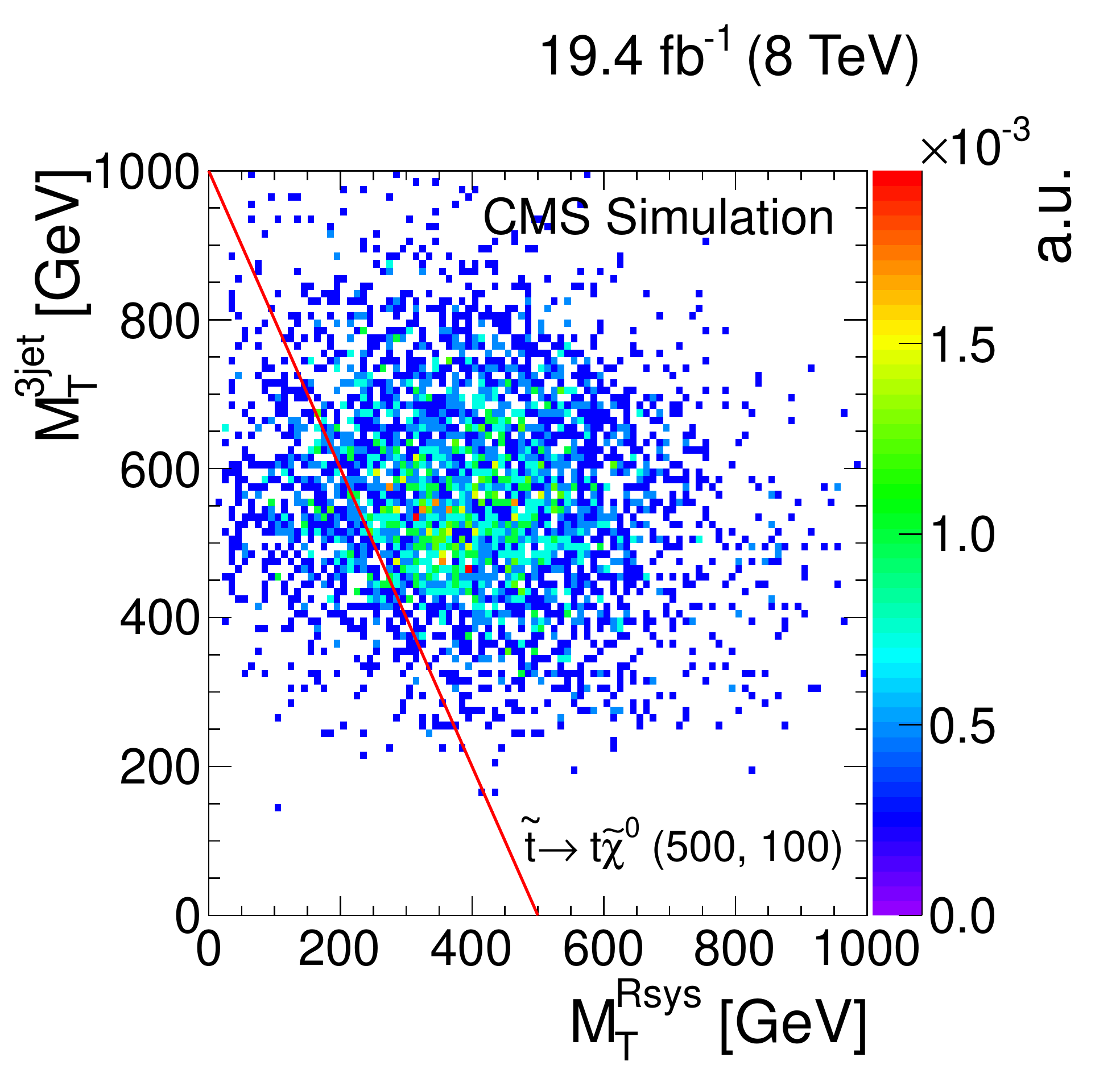}
   \caption{
   The top row shows one-dimensional distributions for \MTTwo (left) and \ptmiss (right) for the simulated processes of \ttbar and three signal models in the multijet t-tagged search.
   The bottom row shows two-dimensional distributions of \MTt versus \MTb for \ttbar (left) and a signal model with $(m_{\PSQt},m_{\PSGczDo}) = (500,100)\GeV$ (right). Events below the lines are rejected.
   The distributions are shown after applying the preselection requirements together with a cut $\ptmiss>200\GeV$, and are normalized to equal area; the axis label ``a.u." means arbitrary units.
    }
\label{fig:kinVars2}

\end{figure*}

After requiring one fully reconstructed and one partially reconstructed top quark, kinematic information is used to distinguish between
signal and SM contributions.
The \MTTwo~\cite{Lester:1999tx,Barr:2003rg} variable, an extension of the transverse mass used for the W boson mass determination~\cite{WbosonMT}, is sensitive to the pair production of heavy particles with decay products that include undetected particles like neutrinos or the \PSGczDo.
The \MTTwo variable is constructed using \pTopFourVec, \pRsystemFourVec, and the \ptvecmiss vectors in an event, assuming the undetected particles to be massless.
The top-left plot in Fig.~\ref{fig:kinVars2} shows a comparison of the shapes of the two \MTTwo distributions of simulated signal and SM \ttbar events after applying the preselection criteria and requiring $\ptmiss>200$\GeV.
The results for signal events are shown for various mass hypotheses for the top squark and LSP.
For the \ttbar background, the \MTTwo distribution peaks around the top quark mass and decreases relatively quickly for larger \MTTwo values.
For the signal, the distribution peaks at higher values.
As one of the top quarks is only partially reconstructed, the kinematic endpoint of \MTTwo is only approximately reconstructed.
To reduce the SM \ttbar background while maintaining good signal efficiency for a range of sparticle mass hypotheses, we require $\MTTwo \ge 300$\GeV.
The top-right plot in Fig.~\ref{fig:kinVars2} shows the \ptmiss distribution in the same conditions.

The variable \MTt, defined using the \ptvecmiss and the fully reconstructed trijet system of the identified top quark,
\begin{equation}  \label{eq:forMTt}
 (\MTt)^2 = (m^{\text{3-jet}})^2 + 2(\ET^{\text{3-jet}}\ptmiss - \pt^\text{3-jet}\ptmiss\cos\Delta\phi),
\end{equation}
is also used to distinguish between signal and SM \ttbar events, where
$(\ET^{\text{3-jet}})^2 \equiv (m^{\text{3-jet}})^2 + (\pTopT)^2$.
Here, $\pTopT$ is the magnitude of \pTopv in the transverse plane and $\Delta\phi$ is the azimuthal angle between \ptvecmiss and \pTopv.
The variable \MTb is similarly defined using Eq.~(\ref{eq:forMTt}), by replacing the ``3-jet'' variables with those of the partial top quark decay products in $\text{R-sys}$.
The bottom row in Fig.~\ref{fig:kinVars2} shows distributions of \MTt versus \MTb for SM \ttbar simulated events (left) and for simulated events from a typical signal (right).
All events are required to satisfy the preselection requirements and to have $\ptmiss>200\GeV$.
For signal events, the \ptmiss requirement typically forces the two top quarks to lie in the hemisphere opposite to \ptvecmiss.
This leads to larger values of \MTt and \MTb due to the large azimuthal angle differences involved.
In contrast, for SM \ttbar events, \ptvecmiss typically lies close to one of the two top quarks, and thus either \MTt or \MTb tends to have a smaller value.
The resulting correlations can be used to further reduce the SM \ttbar background.
Based on simulation, a simple linear requirement $(0.5\MTt+\MTb)\ge500$\GeV is imposed [see the diagonal lines in Fig.~\ref{fig:kinVars2} (bottom)]. This requirement is found to be more effective than simple restrictions on \MTt and \MTb separately.

Four exclusive search regions are selected, defined by $200\le\ptmiss\le350$\GeV and $\ptmiss>350\GeV$ with exactly one or at least two b-tagged jets.
The $\nbjets\ge2$ requirement increases the sensitivity for high-mass top squark production.
We further define a ``baseline'' selection $\ptmiss>200$\GeV and $\nbjets\ge1$ that encompasses all exclusive regions.
Yields for different processes in each of the search regions are shown in Table~\ref{tab:selection:cutflows2}.

\begin{table}[htb]
  \topcaption{For illustrative purposes, event yields from different MC simulated samples for each of the four exclusive search regions, defined by the multijet t-tagged analysis in the text, are shown.  All numbers are scaled to an integrated luminosity of \fulllumi\fbinv, and only statistical uncertainties are shown. The signal points correspond to $\ttwott$ and are labelled as $(m_{\PSQt}, m_{\PSGczDo})$ in units of\GeV. }
  \label{tab:selection:cutflows2}
  \centering
     \resizebox{\textwidth}{!}{
     \begin{tabular}{c,,,,}
      \hline
 & \multicolumn{1}{c}{$200<\ptmiss<350$\GeV}    & \multicolumn{1}{c}{$\ptmiss>350$\GeV}    & \multicolumn{1}{c}{$200<\ptmiss<350$\GeV}    & \multicolumn{1}{c}{$\ptmiss>350$\GeV} \\
       & \multicolumn{1}{c}{$\nbjets=1$}      & \multicolumn{1}{c}{$\nbjets=1$}      & \multicolumn{1}{c}{$\nbjets\geq2$}      & \multicolumn{1}{c}{$\nbjets\geq2$} \\
      \hline
      \ttbar                & 77.8,4.0 & 12.6,1.6 & 57.1,3.5 & 6.3,1.2 \\
      \wlnubr+jets    & 14.3,2.3 &  4.6,1.3 &  2.9,1.0 & 1.1,0.6 \\
      \znunubr+jets   & 13.4,0.9 &  7.1,0.5 &  3.2,0.4 & 1.3,0.2 \\
      Multijet              &  1.1,0.6 &  \multicolumn{1}{c}{$0.0^{+0.5}_{-0.0}$} & \multicolumn{1}{c}{$0.0^{+0.5}_{-0.0}$} & \multicolumn{1}{c}{$0.0^{+0.5}_{-0.0}$} \\
      Single top quark      &  7.0,2.5 &  3.5,1.7 &  5.2,2.1 & 1.8,1.2 \\
      \ttbarZ               &  2.7,0.2 &  0.9,0.1 &  2.8,0.2 & 1.4,0.2 \\
      \ttbarW               &  1.1,0.2 &  0.2,0.1 &  1.0,0.2 & 0.1,0.1 \\
      ZZ                    &  0.5,0.1 &  0.2,0.1 &  0.1,0.1 & \multicolumn{1}{c}{$0.0^{+0.1}_{-0.0}$} \\
      WZ                    &  0.4,0.2 &  0.1,0.1 &  0.1,0.1 & \multicolumn{1}{c}{$0.0^{+0.1}_{-0.0}$} \\
      WW                    &  0.3,0.2 &  0.1,0.1 &  0.3,0.2 & \multicolumn{1}{c}{$0.0^{+0.2}_{-0.0}$} \\
      \hline
      Total                 &119.0,5.4 & 29.3,2.8 & 72.7,4.2 & 12.0,1.8 \\
      \hline
      Signal (350, 0)       & 74.6,4.8 &  3.8,1.1 & 76.9,4.9 &  7.5,1.5 \\
      Signal (500, 100)     & 21.1,0.8 & 13.9,0.7 & 28.3,1.0 & 19.8,0.8 \\
      Signal (650, 50)      &  2.8,0.1 &  6.5,0.2 &  3.8,0.1 &  9.3,0.2 \\
      \hline
    \end{tabular}}

\end{table}

\subsection{Background predictions~\label{t2tt_backgrounds}}

The background is evaluated using a combination of control samples in data and results from MC simulation, following procedures established in Refs.~\cite{RA2,RA2_2011}.
The SM backgrounds from \ttbar, \wlnubr+jets, and QCD multijet production are estimated using data control regions.
The background from \znunubr+jets production is estimated using simulated events that are scaled to match the data in control regions.
The SM backgrounds from rare processes, such as \ttbarZ, $\PW\Z$ and $\Z\Z$ production with at least one $\znunu$ or $\PW \to \ell\nu$ decay, are small and estimated directly from simulation.

The background from SM events with a \tauh lepton is estimated from a data control sample
selected using a trigger requiring a muon with $\pt>17\GeV$, $\abs{\eta}<2.1$ and at least three jets, each with $\pt>30\GeV$.
To define the control sample, we require the muon to be isolated (as defined in Section~\ref{sec-eventReco}) and to have $\pt^{\mu}>20\GeV$ and $\abs{\eta}<2.1$.
To select events with a $\PW\to\mu\nu$ candidate, the transverse mass
$\MT=\sqrt{\smash[b]{2\pt^{\mu}\ptmiss(1-\cos\Delta\phi)}}$ is required to be less than 100\GeV,
where $\Delta\phi$ is the azimuthal angle between the $\vec{\pt}^\mu$ and the \ptvecmiss directions.
Since the $\mu$+jets and \tauh+jets events arise from the same physics processes, the hadronic component of the two samples is the same except for the response of the detector to the muon or $\tauh$ lepton.
To account for this difference, the muon in the data control sample is replaced by a simulated \tauh lepton (a ``\tauh jet'').
The resulting $\pt^{\jet}$ is simulated using a $\pt^{\jet}/\pt^{\tauh}$ response function obtained from MC simulated events.
The $\tauh$ jet in the MC simulated event is reconstructed and matched to the generated $\tau$ lepton, in bins of the generated $\tau$ lepton \pt.
Corrections are applied to account for the trigger efficiency, acceptance and efficiency of the $\mu$ selection, $\MT$ requirement efficiency,
contamination from $\tau\to\mu\nu\nu$ decays, and the ratio of branching fractions $\mathcal{B}(\PW\to\tauh\nu)/\mathcal{B}(\PW\to\mu\nu)
= 0.65$~\cite{PDG}.
The \njets, \ptvecmiss, \MTTwo, \MTt, and \MTb results for each event in the $\mu$+jets data control sample are then recalculated with this simulated $\tauh$ jet, and the search region selection criteria are applied to predict the \tauh background.
The \tauh background estimation method is validated by applying it to simulated \ttbar and \PW{}+jets samples.
For the \ptmiss and \MTTwo variables, the predicted distributions reproduce the expected distributions within statistical uncertainties.

Due to the multiple sampling of the response template, the uncertainty in the prediction is evaluated with a set of pseudo-experiments using a bootstrap technique~\cite{tEFR82a}.
The main systematic uncertainties in the $\tauh$ background estimation arise from the statistical precision of the validation method (6--21\%), the $\mu$ acceptance (3--4\%), and the $\tauh$-jet response function (2--3\%)~\cite{bib:HPStaus}.
An additional uncertainty of 3--14\% is assigned to the \tauh background prediction to account for differences between the simulation and data for the efficiency of the $\MT$ requirement,
which arise as a consequence of finite resolution in \ptmiss and because of uncertainties in the fraction of dileptonic \ttbar events.
\newsavebox{\closureBox}

The lost-lepton background arises from SM \ttbar and \PW{}+jets events.
It is
estimated using a $\mu$+jets control sample selected with the same
trigger and selection criteria as those used for the search,
except requiring (rather than vetoing) exactly one well reconstructed, isolated muon with $\pt^{\mu}>5$\GeV.
As in the estimation of the $\tauh$ background, only events with $\MT<100$\GeV are considered.
Leptons lost due to non-identification and non-isolation are treated separately.
The reconstruction and isolation efficiencies of the electrons and muons
respectively, $\varepsilon_\text{reco}^{\Pe,\Pgm}$ and $\varepsilon_\text{iso}^{\Pe,\mu}$,
are taken from \ttbar simulation in the lepton \pt bins after the baseline selection.
To estimate the number of events with unidentified leptons in the search regions, the ratio
$\left(1/\varepsilon_\text{iso}^{\mu}\right) [(1-\varepsilon_\text{reco}^{\Pe,\mu})/\varepsilon_\text{reco}^{\mu}]$
is applied to the number of events in the control sample;
similarly,
the number of events with non-isolated leptons is estimated using
$\left(\varepsilon_\text{reco}^{\Pe,\mu}/\varepsilon_\text{reco}^{\mu}\right) [(1-\varepsilon_\text{iso}^{\Pe,\Pgm})/\varepsilon_\text{iso}^{\mu}]$.
The acceptance and efficiencies are validated with ``tag-and-probe''
studies of \zll ($\ell =\Pe,\mu$) events in data and simulation~\cite{EWK-10-005}.
The method is validated by predicting the lost-lepton background using a single-muon sample from simulated \ttbar and \wpj events.
The predicted distributions and the true distributions (taken directly from the simulation) agree within the uncertainties.

The dominant uncertainties in the lost-lepton background prediction arise from the differences in lepton reconstruction and isolation efficiencies between data and MC simulation.
The uncertainties due to lepton reconstruction efficiency are determined by comparing tag-and-probe efficiencies in \zll events at the \Z boson mass peak in data and simulation.
For isolation uncertainties, the isolation variables in the simulation are scaled to match the distribution from the data, and the resulting differences in predictions are taken as a systematic uncertainty.
Variations of the PDFs following the recommendation of Refs.~\cite{bib:PDF4LHCa,bib:PDF4LHC} change the muon acceptance, but lead to less than 3\% uncertainty in the final prediction.
An additional uncertainty of 3\% is assigned to account for possible differences between data and simulation for the $\MT$ requirement, evaluated in the same manner as for the $\tauh$ background.

The \znunubr{}+jets background is estimated from \zmumubr{}+jets simulation, with a normalization that is adjusted to account for differences with respect to data using a scale factor \RmumuDataMC determined from a dimuon control sample.
The dimuon control sample is selected using the preselection criteria of Section~\ref{sec:t2tt_eventSel}, except that the lepton veto is removed and instead, a $\mu^{+}\mu^{-}$ pair is required to be present.
The $\mu^{+}$ and $\mu^{-}$ must satisfy $\pt>20$\GeV, $\abs{\eta}<2.1$, a relative isolation parameter $<$0.2 (as defined in Section~\ref{sec-eventReco}),
and the dimuon mass must lie in the $\Z$ boson mass range 71--111\GeV.
To mimic the effect of neutrinos, \ptvecmissmu is used.
The dimuon control sample includes events from \ttbar and \ttbarZ production, which must be subtracted.
The \ttbar contribution is evaluated using simulation, with a normalization that is validated using a single-lepton (electron or muon) control sample with lepton $\pt>20$\GeV.
In the single-lepton control sample, we also validate the normalization of the simulation after requiring either $\nbjets=1$ or $\nbjets\geq2$.
The normalization in the single-muon control sample is found in all cases to be consistent with unity.
A statistical uncertainty in this unit normalization (6--8\%) is propagated as a systematic uncertainty in the normalization of the \ttbar contribution to the dimuon control sample.
The \ttbarZ contribution to the dimuon control sample is estimated directly from simulation.
The \RmumuDataMC scale factor is defined by the ratio of data to MC events in the dimuon control sample, after subtraction of the \ttbar and \ttbarZ components.
The scale factor is found to be statistically consistent with unity for events with exactly zero b-tagged jets.
Events with one b-tagged jet are found to have a scaling factor of $1.33 \pm 0.17$ (stat).
In events with two or more b-tagged jets, the scaling factor is found to be  $1.47 \pm 0.49$ (stat).

Systematic uncertainties in \RmumuDataMC include uncertainties in the normalization and subsequent removal of the \ttbar and \ttbarZ processes (1--5\%),
uncertainties in the simulation to account for muon acceptance (10\%), trigger efficiency uncertainties (1\%), and data-versus-simulation shape disagreements.
The shape disagreements are divided into an overall normalization uncertainty (26--33\%) to account for discrepancies in the normalization due to the remaining event selection requirements,
and a residual shape uncertainty (up to 80\%)
which accounts for potential normalization or shape discrepancies in the
tails of the analysis variables.
The residual shape uncertainty is taken from the envelope of a first-order polynomial fit to the data/MC ratio of the analysis variables.
An asymmetric systematic uncertainty is assigned to account for the difference between this fit envelope and the overall normalization uncertainty.

The QCD multijet background is expected to be small due to the \ptmiss and \DeltaPhi requirements.
This background is estimated by measuring the number of QCD multijet events in a data control region and scaling the yield by a factor \RQCD, which translates the yield to the search region.
The control region is identical to the search region except that one of the three highest
\pt jets must fail the respective \DeltaPhi requirement specified in Section~\ref{sec:t2tt_eventSel}.
The \RQCD factor is defined as $\RQCD = \RQCD^\mathrm{SB} F_\mathrm{SR}$, where $\RQCD^\mathrm{SB}$ is the ratio of the number of measured QCD multijet events found with the standard and inverted \DeltaPhi requirements in a sideband $175<\ptmiss<200$\GeV,
and $F_\mathrm{SR}$ is a MC-derived extrapolation factor that translates $\RQCD^\mathrm{SB}$ to the search region $\ptmiss>200\GeV$.
The analysis requires a reconstructed top quark, at least one b-tagged jet, and large \ptmiss, so the
sideband and inverted \DeltaPhi control regions are dominated by \ttbar, \znunubr{}+jets, and \PW{}+jets events.
To determine the number of QCD multijet events in the sideband and control regions,
the number of events observed in data is corrected for non-QCD contributions using the method described above for the \ttbar contribution to the dimuon control sample in the \znunubr{}+jets background estimate.
Using simulation, the ratio of events in the standard and inverted \DeltaPhi regions is determined
as a function of \ptmiss.
The results are fit with a first-order polynomial.
The $F_\mathrm{SR}$ factor, whose value is defined by the slope of this polynomial, is consistent with zero.

The statistical uncertainty from simulation, the jet energy scale uncertainty, and jet energy resolution uncertainty are combined to define a systematic uncertainty in $\RQCD$.

The individual contributions to the background, evaluated as described above, are listed in Table~\ref{tab:TAUpred} for each of the four search regions.
Both statistical and systematic uncertainties are indicated.
For the QCD multijet background, the predicted event yields for $\nbjets\geq 2$ are small, around 0.10 events.
The corresponding total uncertainties of around 0.45 events are much larger, with about equal contributions from statistical and systematic terms, and so we merely quote these latter results as one standard deviation upper limits on the background estimates.
\begin{table}[htbp]
\topcaption{Predicted SM backgrounds corresponding to an integrated luminosity of \fulllumi\fbinv for all four of the multijet t-tagged search regions defined in the text.
Both statistical and systematic uncertainties are quoted.
\label{tab:TAUpred}}
\centering
{
\begin{tabular}{lccc}
\hline
 Background source            & $\nbjets$ & $200\le\ptmiss\le350\GeV$     & $\ptmiss>350\GeV$ \\
\hline
   $\tau\to\text{hadrons}$ & ${=}1$ & $62.2 \pm 5.6 \pm 5.6$          & $12.3 \pm 1.7 \pm 2.6$            \\ [0.5ex]
Lost lepton            & ${=}1$ & $48 \pm 6{} \pm 11$    &  $7.0 \pm 2.4{}^{+3.2}_{-3.1}$    \\ [0.5ex]
\znunubr+jets    & ${=}1$ & $17.9 \pm 1.4{}^{+5.1}_{-8.4}$  & $11.3 \pm 1.0{}^{+3.8}_{-5.5}$    \\ [0.5ex]
Multijets              & ${=}1$ & $17 \pm 3 \pm 24$           &  $2.0 \pm 1.1 \pm 2.7$            \\ [0.5ex]
Rare processes         & ${=}1$ &  $1.9 \pm 0.9$                  & $0.8 \pm 0.4$                     \\
\hline
\rule[1.5\baselineskip]{0pt}{2.5ex}Total                  & ${=}1$ & $148^{+29}_{-24}$         & $33.4{}^{+7.0}_{-7.8}$            \\[0.5ex]
\hline
$\tau\to\text{hadrons}$ & ${\ge}2$ & $41.5 \pm 4.3 \pm 5.3$         & $4.3 \pm 1.4{}^{+1.0}_{-1.1}$   \\ [0.5ex]
Lost lepton            & ${\ge}2$ & $32.6 \pm 5.1{}^{+8.6}_{-8.2}$ & $1.2 \pm 0.8 \pm0.5$            \\ [0.5ex]
\znunubr+jets    & ${\ge}2$ &  $4.6 \pm 0.6{}^{+2.8}_{-2.4}$ & $1.8 \pm 0.4{}^{+1.6}_{-1.0}$   \\ [0.5ex]
Multijets              & ${\ge}2$ & $<0.5$        & $<0.5$       \\ [0.5ex]
Rare processes         & ${\ge}2$ &  $1.9 \pm 0.9$                 & $1.2 \pm 0.6$                   \\
\hline
\rule[1.5\baselineskip]{0pt}{2.5ex}Total                  & ${\ge}2$ & $81^{+13}_{-12}$             & $8.6^{+2.6}_{-2.4}$              \\[0.5ex]
\hline
\end{tabular}
}

\label{tab:stuff}
\end{table}

\section{Search for bottom-squark pair production using bottom-quark tagging \label{sec-SUS-13-018}}
We next describe the dijet b-tagged analysis. This analysis requires large \ptmiss and one or two jets identified as originating from bottom quarks.
The possible presence of a hard light-flavour third jet, arising from initial-state radiation (ISR), is incorporated.
The search is motivated by the possibility of bottom-squark pair production, where each bottom squark decays directly to the \PSGczDo LSP with the emission of a bottom quark, $\PSQb \to \PQb\PSGczDo$.
The signal production rate depends on the bottom squark mass, while the transverse momenta and hence the signal acceptance of the search depend on the mass difference $\Delta m = m_{\PSQb} - m_{\PSGczDo}$.

\subsection{Event selection \label{sec:t2bb_eventSel}}

The data used in the dijet b-tagged search are collected using the same trigger described in Section~\ref{sec:t2tt_eventSel} for the multijet t-tagged search.
The trigger efficiency is measured to be larger than 95\% after application of the selection criteria described below, as measured in data.
A set of loose selection criteria are applied to define a baseline data set that is used in addition as a validation sample to compare data and simulation for various kinematic quantities.
Exactly two central jets are required with $\pt>70$\GeV and $\abs{\eta}<2.4$,
and events are vetoed if they have an additional jet with $\pt>$ 50\GeV and $\abs{\eta} <5.0$.
One or both of the leading jets are required to be tagged as originating from a b quark, using the medium CSV algorithm working point.
Events containing an isolated electron, muon, or track (representing single-prong $\tau$-lepton decays or unidentified electrons or muons) with $\pt>10$\GeV are rejected to suppress background processes such as \ttbar{} and \wlnubr{}+jets production.
In addition, the scalar sum \HT{} of the \pt values of the two highest-\pt jets ($\jet_{1}$ and $\jet_{2}$, with $\pt^{\jet_{1}}>\pt^{\jet_{2}}$) is required to be more than 250\GeV,
and
\ptmiss{} is required to be larger than 175\GeV.
To reject QCD dijet events, we require $\Delta\phi(\jet_{1},\jet_{2})< 2.5$ radians.
To further suppress the SM background from \ttbar{} and \wlnubr{}+jets events, the transverse mass defined by
$\MT(\jet_2, \ptmiss{}) = \sqrt{{{[\ET^{\jet_2}+\ptmiss]^2-[\ptvec^{\jet_2}+\ptvecmiss]^{2}}}}$
is required to be larger than 200\GeV.

Events are characterized using the boost-corrected contransverse mass \MCT~\cite{MCT1, MCT}, which for processes involving two identical decays of heavy particles such as $\PSQb\PASQb\to \jet_1 \jet_2 \PSGczDo \PSGczDo$, is defined as
$(\MCT)^2 = [\ET^{\jet_1}+\ET^{\jet_2}]^2-[\ptvec^{\jet_1}-\ptvec^{\jet_2}]^{2}
= 2\pt^{\jet_1} \pt^{\jet_2}\left[1+\cos\phi(\jet_1,\jet_2)\right]$.
For signal events, the \MCT distribution is characterized by an endpoint at $(m_{\PSQb}^{2}-m_{\PSGczDo}^{2})/m_{\PSQb}$.

To obtain sensitivity to different mass hypotheses, the search is conducted in four regions of \MCT:
 $\MCT <250$, $250<\MCT <350$, $350<\MCT <450$, or $\MCT>450$\GeV.
For each \MCT region, we require either $\nbjets = 1$ or $\nbjets = 2$, for a total of eight exclusive search regions.

\begin{figure}[hbtp]
 \centering
 \includegraphics[width=0.45\textwidth]{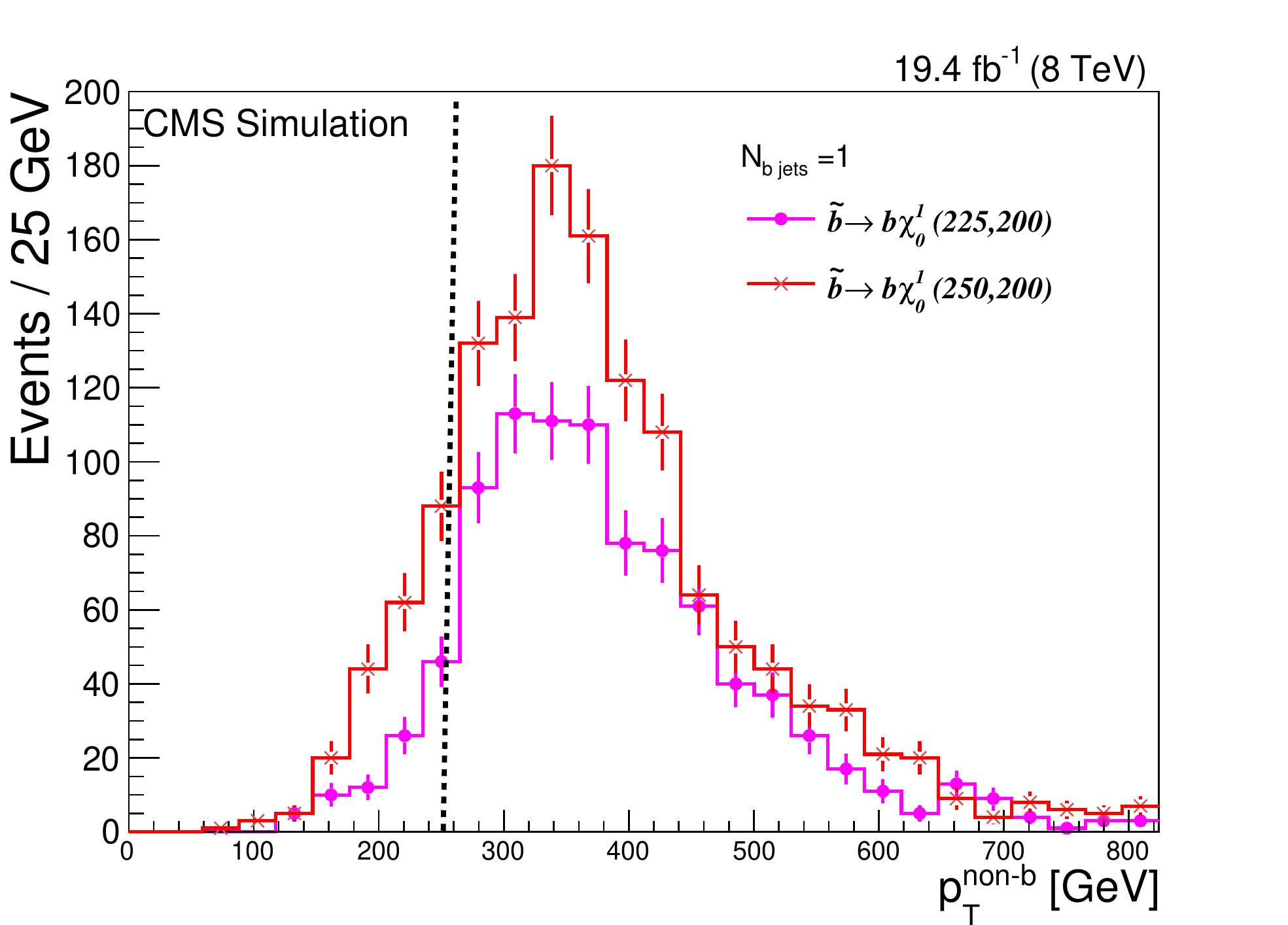}
 \includegraphics[width=0.45\textwidth]{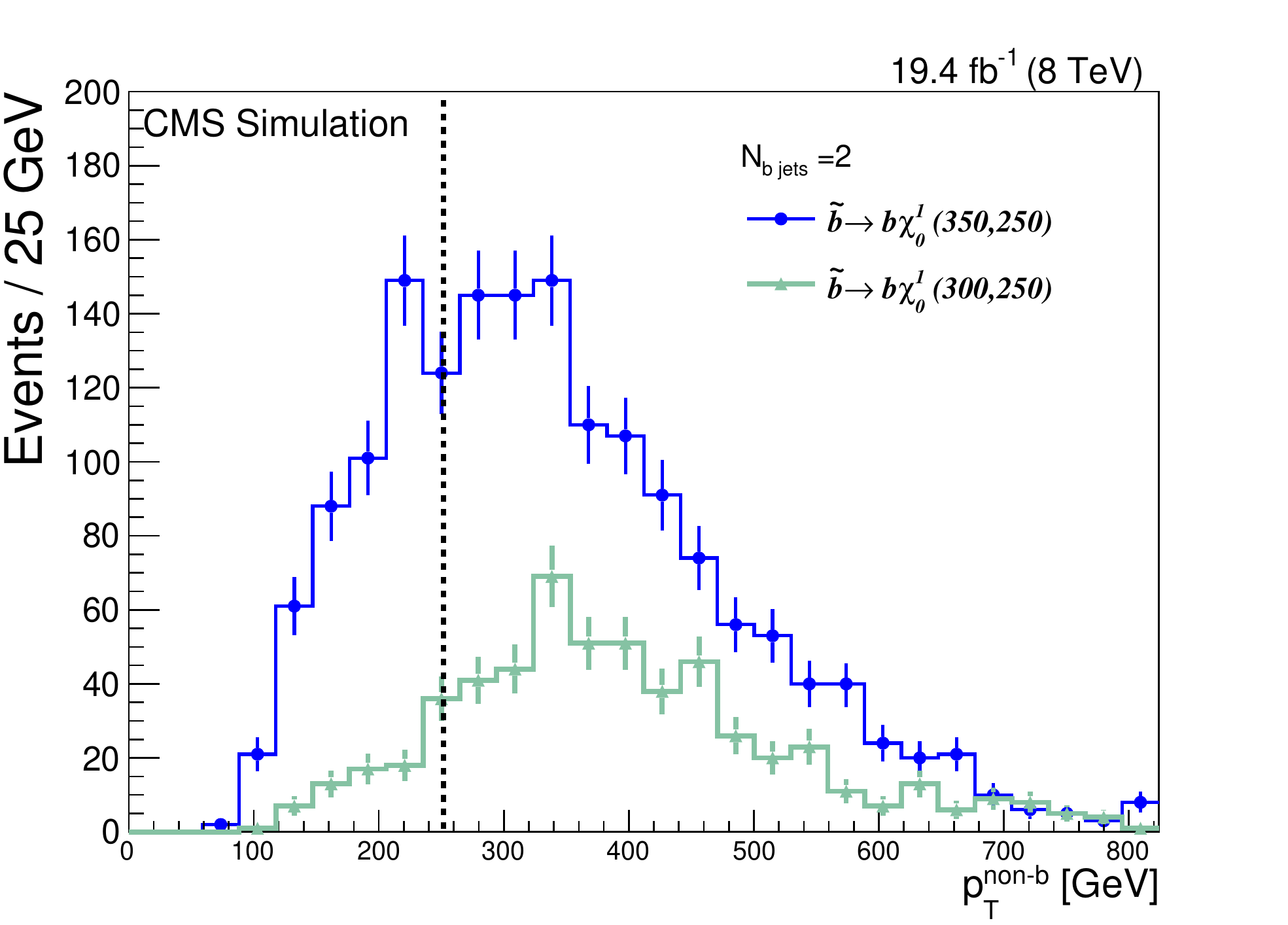}
   \caption{
The distribution of $\pt^{\text{non-b}}$ (see text) for the ISR search regions with $\nbjets=1$ (left) and $\nbjets=2$ (right) in the dijet b-tagged analysis.
The selection requirement $\pt^{\text{non-b}}>250\GeV$ is indicated by the vertical dashed lines.}
   \label{fig:t2bbISR}

\end{figure}

For $m_{\PSQb} - m_{\PSGczDo} \lesssim 100\GeV$, the \pt values of jets from the squark decay become too small to efficiently satisfy the selection requirements.
However, events containing a high-$\pt$ jet from ISR can provide a transverse boost to the recoiling $\PSQb\PASQb$ system, enabling such events to satisfy the trigger and selection conditions.
Additional search regions, hereafter denoted ``ISR'' search regions, are therefore considered by
modifying the baseline selection requirements to allow an additional third jet from ISR:
exactly three jets with \pt$>30$\GeV and $\abs{\eta}<2.4$ are then required, where the two highest \pt
jets must have $\pt > 70\GeV$ and the highest \pt jet is required not to be b-tagged using the CSV loose definition.
At least one of the two other jets must be b-tagged according to the medium CSV working point, and the events are classified according to whether one or both of these jets are so tagged, defining two ISR search regions.
As in the nominal dijet case, events are rejected if they contain isolated leptons or tracks, or if $\HT{}<250\GeV$.
An additional requirement is $\pt^{\text{non-b}}>250\GeV$, where $\pt^{\text{non-b}}$ is the modulus of the vector sum over the transverse momenta of all jets that are not b-tagged.
This requirement increases the probability of selecting events with hard ISR jets and is expected to be reasonably efficient for signal processes, as shown for two representative $\ttwobb$ mass hypotheses in Fig.~\ref{fig:t2bbISR}.
In addition, events must satisfy $\ptmiss>250\GeV$.
To reduce the multijet background, we require
$\Delta\phi(\ptvec^{\jet_i},\ptvecmiss)>0.5$ radians, where $i=1,2,3$.
Finally, no requirement is placed on \MCT for the two ISR search regions.

For purposes of illustration, the background estimates predicted by simulation for the 10 search regions are listed in Table~\ref{tab:2tc}.
The contribution from QCD multijet production to the $\nbjets=2$ search regions is expected to be negligible,
so only the upper limits on this background contribution are quoted.

\begin{table}[htb!]
\centering
\topcaption{Predicted background yields from simulation for the dijet b-tagged analysis.
The results are scaled to an integrated luminosity of \fulllumi\fbinv.
The uncertainties are statistical.
The results for the \ttwobb signal events are labelled as $(m_{\PSQb}, m_{\PSGczDo})$, in\GeV, and the units of the \MCT variable are also\GeV.}\label{tab:2tc}
\begin{tabular}{lc,,,,,}
\hline
& \multirow{2}{*}{\nbjets{}} & \multicolumn{1}{c}{\MCT}   & \multicolumn{1}{c}{\MCT} & \multicolumn{1}{c}{\MCT}  & \multicolumn{1}{c}{\MCT} & \multicolumn{1}{c}{\multirow{2}{*}{ISR}} \\
&   & \multicolumn{1}{c}{$<$250} & \multicolumn{1}{c}{$\in[250,350]$} & \multicolumn{1}{c}{$\in[350,450]$}  & \multicolumn{1}{c}{$>$450} &  \\ \hline
\znunubr{}+jets   & 1 &   818,12    &  367.0,7.8      & 59.0,2.8    & 16.0,1.5   & 161.0,2.6     \\
\wlnubr{}+jets      & 1 &   398.0,8.4   &  149.0,4.9      & 17.0,1.5    &  6.0,0.9 &  90.0,3.4     \\
\ttbar{}              & 1 &   221.0,2.5   &  176.0,2.2      & 17.0,0.7    &  2.2,0.2 &  71.0,1.4     \\
Single top quark                    & 1 &   33.0,3.7    &   13.0,2.3      &  0.3,0.3  &  \multicolumn{1}{c}{$<$0.5}      &  24.0,4.4     \\
VV                            & 1 &   18.0,0.7    &   17.0, 0.6     &  0.9,0.1  &  0.3,0.1 &   4.8,0.4   \\
ttZ                         & 1 &    0.8,0.1  &    0.5,0.1    &  0.2,0.1  &  \multicolumn{1}{c}{$<$0.04}     &   0.6,0.1   \\
Multijets                   & 1 &   12.0,8.2    &    6.0,6.0    &   \multicolumn{1}{c}{$<$0.5}     &   \multicolumn{1}{c}{$<$0.5} &   0.3,0.4   \\ \hline
Total                         & 1 & 1500,17     &  729,11       & 94.0,3.2    & 25.0,1.6   & 352.0,6.2  \\ \hline
Signal (275,250)          & 1 &   11.0,0.8    &   10.0,0.7      &  1.3,0.2  & 0.0,0.0  &  54.0,1.9     \\
Signal (750,50)           & 1 &    0.6,0.1  &    1.1,0.1    &  1.7,0.2  & 2.7,0.2  &   3.1,0.3   \\
\hline
\znunubr{}+jets     & 2 &   58.0,3.2   & 28.1,2.1    &  4.8,0.8  & 1.1,0.3 &  7.7,0.5 \\
\wlnubr{}+jets      & 2 &   13.0,1.4   &  4.7,1.0  &  1.0,0.3  & \multicolumn{1}{c}{$<$0.2} &  2.7,0.6 \\
\ttbar{}                      & 2 &   12.1,0.6   & 11.0,0.5    &  1.8, 0.2 & 0.3,0.1 & 15,0.6   \\
Single top quark                    & 2 &    1.3,0.7 &  2.2,1.1  &  \multicolumn{1}{c}{$<$0.5}       & \multicolumn{1}{c}{$<$0.5}  &  0.7,0.5 \\
VV                            & 2 &    1.5,0.1 &  3.2, 0.2 &  0.1,0.0  & \multicolumn{1}{c}{$<$0.1}      &  0.2,0.1 \\
ttZ                           & 2 &    0.3,0.1 &  0.2, 0.1 &  \multicolumn{1}{c}{$<$0.1}     & \multicolumn{1}{c}{$<$0.1}     &  0.2,0.1 \\
Multijets                   & 2 &    \multicolumn{1}{c}{$<$0.5}      &  \multicolumn{1}{c}{$<$0.5}       &  \multicolumn{1}{c}{$<$0.5}       & \multicolumn{1}{c}{$<$0.5}      &   \multicolumn{1}{c}{$<$0.5}     \\ \hline
Total                         & 2 &   86.0,3.6   & 49.0,2.5    &  7.7,0.8  & 1.4,0.4 & 27.0,1.1   \\ \hline
Signal (275,250)          & 2 &    1.5,0.2 &  1.4,0.2  &  0.0,0.0  & 0.0,0.0 &  4.6,0.6 \\
Signal (750,50)           & 2 &    0.7,0.1 &  1.1,0.2  &  1.5,0.2  & 3.6,0.3 &  0.5,0.1 \\
\hline
\end{tabular}

 \end{table}

\subsection{Background predictions \label{sec:t2bbBgd}}
As compared to the multijet t-tagged search, due to jet multiplicity and lepton veto requirements including an isolated track veto, backgrounds involving top quarks are significantly reduced.
Instead, in all 10 search regions the dominant background is from \znunubr{}+jets events, followed in importance by contributions from W{}+jets and \ttbar{} processes.
The SM background due to these processes, as well as the contribution from single-top quark production, are determined using data with assistance from simulation.
From studies with simulation and data control samples, the contribution of QCD multijet events is expected to be negligible.
The contribution of diboson and \ttbarZ events in the search regions is less than $3\%$ and is estimated from simulation assuming a 50\% systematic uncertainty.

For nine of the search regions, the eight \MCT search regions and the ISR search region with $\nbjets=2$,
the \znunubr{}+jets background is evaluated using a control sample enriched in \wmunubr+jets events as they have similar kinematic properties.
For this control sample, which is selected using an isolated muon trigger, the muon is required to have $\pt>30$\GeV and $\abs{\eta}<2.1$ to ensure a trigger efficiency near unity.
To exclude Drell--Yan processes, an event is vetoed if it contains an additional muon candidate that in combination with the required muon forms a system having invariant mass within 25\GeV of the mass of the \Z{} boson.
To reject muons from decays-in-flight and from semileptonic decays within heavy-flavour jets, the selected muon must be separated by $\Delta R >0.3$ from all jets.
The remaining events are accepted and classified using the same criteria that define each of the nine search regions, except that a b-tag veto (using the loose CSV working point) is applied, to minimize the contribution of \ttbar{} or single top quark processes.
The muon \ptvec is removed from the event to mimic the signature of neutrinos from decays of the \Z{} boson.
All kinematic variables are modified accordingly, where \METmu{} is used.
Selection thresholds for the resulting \METmu{}, $\HT$, and $\MT(\jet_2,\METmu{})$ variables are the same as those used to define the search regions.
In the case of the doubly b-tagged ISR search region, the $\pt^{\text{non-b}}$ requirement
(which, in this case, is effectively a requirement that the leading jet \pt be larger than 250\GeV)
is common to both the search region and the control sample.
The muon selection, in conjunction with the restrictions on \METmu and $\MT$, ensures that the contributions of QCD multijet events are negligible.
The $\Delta\phi$ requirement thus has minimal impact and is not implemented for the control sample selection.
The estimated number of \znunubr{}+jets background events is:
\begin{eqnarray}\label{eqn:zinvFromwlnu}
N_{\text{SR}}^{\text{pred}}(\znunu; \MCT, \pt^{\text{non-b}}, \nbjets{}) = N_{\text{CR}}^\text{obs}(\MCT,\pt^{\text{non-b}})~R^\mathrm{MC}_\mathrm{SR/CR}(\MCT ,\pt^{\text{non-b}},\nbjets{}),
\end{eqnarray}

where $R^\mathrm{MC}_\mathrm{SR/CR}$ is the ratio of the number of \znunubr{}\,+\,b jets events in the search region to the total number of events in the control sample, taken from simulation and determined separately for each search region defined by either \MCT and \nbjets{} (in the case of the eight \MCT search regions) or by $\pt^{\text{non-b}}$ and \nbjets{} (in the case of the doubly b-tagged ISR search region).
The term $N_{\text{CR}}^\text{obs}(\MCT,\pt^{\text{non-b}},\nbjets{})$ represents the number of events observed in data, in each control region. The number of simulated events in the control sample is corrected for differences between simulation and data in the muon isolation and identification efficiencies as a function of muon \pt, muon $\eta$, and trigger efficiency.

The \wmunubr{}+jets control sample described above, when used to evaluate
the \znunubr{}+jets background in the $\nbjets=1$ ISR search region,
overlaps with the \wmunubr{}+jets control sample used to evaluate the \znunubr{}+jets background in the $\nbjets=2$ ISR search region.
Therefore, an alternative data control sample of \zmumubr{}+jets events is used to evaluate this background in the $\nbjets=1$ ISR region to provide sufficient discrimination between control regions.
Using the same single-muon triggered control sample,
we require the identical selection requirements as for the singly b-tagged ISR search region,
except that we demand two opposite-sign, well-identified, isolated central ($\abs{\eta}<2.1$) muons with $\pt>30$\GeV and $\pt>20$\GeV, respectively, that have an invariant dimuon mass between 76 and 106\GeV.
One b-tagged jet is required using the medium CSV definition.
In an analogous way to Eq.~(\ref{eqn:zinvFromwlnu}),
the number of \znunubr{}+jets events is estimated by applying muon and trigger efficiencies,
and by scaling the observed number of events in the control region by the factor $R^\mathrm{MC}_\mathrm{SR/CR}$,
which is the ratio from simulation of the number of \znunu events in the search region to the total number of events in the control region.

Tests of the method are performed with simulation, treating MC events as data and comparing the predicted number of background events with the true number.
Systematic uncertainties are assigned based on the level of agreement:
2--13\% for the $\nbjets{}=1$ search regions and 8--30\% for the $\nbjets{}=2$ search regions,
where the uncertainties are dominated by the statistical precision available.
To determine a systematic uncertainty in the number of non-\wmunubr{}+jets events in the single-muon control sample,
the production cross sections of Drell--Yan, diboson, \ttbar{}, and single-top simulation samples
are varied up and down by 50\%; less than 10\% variation is observed for one or two b jets, across all search regions.
The sensitivity of $R^\mathrm{MC}_\mathrm{SR/CR}$ in both the \wmunubr{}+jets and \zmumubr{}+jets enriched control samples to muon isolation and identification is also studied.
Varying these muon criteria within their uncertainties, and taking the deviations from the central values in each search bin, systematic uncertainties of 3--10\% for $\nbjets{}=1$ and 5--10\% for $\nbjets{}=2$ are assigned for both the
\MCT and ISR search regions.
Another source of systematic uncertainty in the ratio $R^\mathrm{MC}_\mathrm{SR/CR}$ can arise from differences between data and simulation in the production of Z bosons in association with one or two b jets.
The data are observed to agree with the simulation to better than about 5\% for \zmumu{} events having at least one b jet and covering \MCT values up to 250\GeV; we thus apply a 5\% systematic uncertainty for all \MCT and ISR search regions.
Other theoretical systematic uncertainties largely cancel in the ratio of cross sections but are nevertheless considered.
Higher-order corrections from QCD are expected to be less than 5\%, and the uncertainty from the choice of the PDFs is negligible
as higher-order electroweak corrections are similar for W and Z boson production and largely cancel in the cross section ratios \cite{WZxsPDF}.

\PW{}+jets, \ttbar{}, and single-top processes make up the lost-lepton background, as defined in Section~\ref{sec-strategy}.
This lost-lepton background is evaluated together with the background due to \tauh events
via control samples defined by the same dijet-with-\ptmiss trigger used to define the 10 search regions.
The event selection criteria for each control region are identical to those used to define the respective search region, except for the following three conditions.
First, a single muon is required (rather than vetoed) using tight muon identification criteria.
Second, in the cases of the eight \MCT search regions, the requirement on $\Delta\phi (\ptvec^{\jet_1}, \ptvec^{\jet_2})$ is removed.
Third, in all 10 control regions, exactly one or exactly two jets must be b-tagged using the loose CSV working point.
The prediction in each search region for the number of lost-lepton and \tauh background events due to \PW{}+jets, \ttbar{}, and single-top processes is given by:
\begin{equation}\label{eqn:wtoppred}
N_{\text{SR}}^{\text{pred}}(\text{lost-lep \& \tauh}; \MCT, \pt^{\text{non-b}}, \nbjets{}) = N_{\text{CR}}^\text{obs}(\MCT,\pt^{\text{non-b}},\nbjets{})~R^\mathrm{MC}_\mathrm{SR/CR}(\MCT,\pt^{\text{non-b}},\nbjets{}),
\end{equation}
where the factor $R^\mathrm{MC}_\mathrm{SR/CR}$ (determined from simulation) is the ratio of the number of \PW{}+jets, \ttbar{}, and single-top events in a particular search region to the number of \PW{}+jets, \ttbar{}, single-top, diboson, and Drell--Yan events in the corresponding control region;
finally, $N_{\text{CR}}^\text{obs}(\MCT,\nbjets{})$ represents the number of events observed in data for each control region.

The data and simulation samples as well as the control and search regions are all defined to be kinematically similar, so most of the uncertainties due to mismodelling of event kinematics or instrumental effects are expected to largely cancel.
However, the relative \ttbar{} and \PW{}+jets contribution depends on the b jet multiplicity, which can be different between a search region and its corresponding control region.
The accuracies of the factors $R^\mathrm{MC}_\mathrm{SR/CR}$ are tested in data using two independent single-muon triggered samples containing exactly one b jet (expected to contain roughly equal \ttbar{} and \PW{}+jets contribution) and exactly two b jets (expected to have a dominant \ttbar{} contribution).
A related source of uncertainty arises from possible differences in the modelling of lepton isolation and the isolated track veto between data and simulation.
To probe this effect, the numbers of events with exactly one muon are predicted starting from a control sample
with an isolated track and no isolated muon or electron using a transfer factor derived from MC.
The average weighted uncertainty of the two studies results in 4--20\% differences in the predicted background in various search regions.
Statistical uncertainties in the transfer factors, due to the finite size of simulation samples, result in 2--16\% and 10--80\% uncertainties in the predicted backgrounds, for search regions with one and two b jets, respectively.
Uncertainties related to the efficiency of the CSV algorithm to identify b jets result in 2--20\% uncertainties in the final background predictions.
And finally, uncertainties in the background prediction due to the contributions of dibosons and other rare processes, taken from simulation with 50\% uncertainty, are less than 2\% across all search regions.
The predicted numbers of \ttbar{}, single-top, and \wlnubr{}+jets events in the various search regions are listed in Table \ref{tab:1/z}, along with the statistical and total systematic uncertainties.

Background yields from QCD multijet processes are expected to be less than a percent of the total across all search bins.
An estimate of the contribution from the QCD background is made by measuring the number of multijet events in a QCD enriched control region, and scaling this number by a transfer factor.
The control regions are identical to the search regions except that the $\Delta\phi (\ptvec^{\jet_1}, \ptvec^{\jet_2})$ requirement is inverted (for the dijet search regions),
and $\Delta\phi(\ptvec^{\jet_{1,2,3}},\ptvecmiss)$ is inverted (for the ISR search regions).
In the case of the dijet searches, the transfer factor is taken from a zero b-jet sideband.
In the case of the ISR searches, the transfer factor is taken from a sideband defined by $175<\ptmiss<200\GeV$.

From studies from simulation, QCD events in the region with the standard $\Delta \phi$ requirement survive only because of mismeasurement, where under-measurement of one of the two leading jets results in it being reconstructed as the third jet,
where the third leading jet of the event must have $\pt<50$\GeV.
This behaviour is observed to have no correlation with the b quark content of events.
A dijet sideband region with zero b jets is therefore used to estimate the number of QCD events in the search regions.
This dijet sideband is divided into two regions: a QCD subdominant sideband region for which $\Delta\phi (\ptvec^{\jet_1}, \ptvec^{\jet_2}) <2.5$ together with $\Delta\phi(\ptvec^{\jet_3},\ptvecmiss) < 0.3$ to enrich the QCD content, and a QCD dominant sideband region defined by $\Delta\phi (\ptvec^{\jet_1}, \ptvec^{\jet_2})>2.5$.
In the QCD subdominant sideband region, the contribution from non-QCD processes (\Z{}+jets, \ttbar{}, and \PW{}+jets events) is significant and is subtracted (via simulation normalized to data) from the observed numbers of events.
Contributions from non-QCD processes in the QCD dominant sideband region are negligible.
The QCD transfer factors, characterized in bins of \MCT and \nbjets for the eight dijet searches, are then defined as the ratio of the number of multijet events between these two sideband regions.

Using a method similar to the QCD background determination in the multijet t-tagged search, described in Section~\ref{t2tt_backgrounds}, the ISR sideband of $175<\ptmiss<200\GeV$ is divided into two regions: a regular sideband region for which $\Delta\phi(\ptvec^{\jet_{1,2,3}},\ptvecmiss) >0.5$, and an inverted sideband region for which $\Delta\phi(\ptvec^{\jet_{1,2,3}},\ptvecmiss) <0.5$.
While QCD processes dominate the inverted sideband region (due to the $\Delta\phi(\ptvec^{\jet_{1,2,3}},\ptvecmiss)>0.5$ requirement), non-QCD processes dominate the regular sideband region (due to the large \ptmiss conditions).
Using simulation, \Z{}+jets, \ttbar{}, and \PW{}+jets processes are subtracted from the data yields for both sideband regions.
The QCD transfer factors are then defined by the ratio of the remaining data yield in the regular sideband region to the remaining data yield in the inverted sideband region.
Due to possible correlations between \ptmiss and $\Delta\phi(\ptvec^{\jet_{1,2,3}},\ptvecmiss)$, the transfer factors are parametrized as a linear function of \ptmiss using simulation.
The transfer factor is then extrapolated from the value obtained in the sideband to the value at the average expected \ptmiss from QCD processes in the ISR search regions.

The systematic uncertainty in the QCD background prediction comes from (i) the limited number of observed events in the data control samples, as well as (ii) the limited number of simulated non-QCD events that are subtracted from the sideband regions used to determine the transfer factors.
For the ISR search regions, uncertainties associated with the determination of the linear parametrization of the transfer factor are propagated as an additional source of systematic uncertainty in the QCD background prediction.

The background yields using the methods outlined above are summarized in Table~\ref{tab:1/z}.
\begin{table}[htb]
         \centering
        \topcaption{\label{tab:1/z}
        Predicted SM backgrounds corresponding to an integrated luminosity of \fulllumi\fbinv for the
        10 dijet b-tagged search regions defined in the text, with \MCT given in units of \GeVns.
        The first uncertainty is statistical and the second systematic.
        }
\resizebox{\textwidth}{!}{
        \begin{tabular}{lcccccc } \hline
        \multirow{2}{*}{}   & \multirow{2}{*}{\nbjets{}} & \MCT & \MCT  & \MCT  & \MCT  & \multirow{2}{*}{ISR} \\
           &   & $<$250 & $\in[250,350]$ & $\in[350,450]$  & $>$450 &  \\ \hline

 \znunubr+jets          & 1 & 848$\pm$12$\pm$79     &  339$\pm$8$\pm$52     & 48.0$\pm$3.0$\pm$6.0      & 8.1$\pm$1.6$\pm$1.7   & 176$\pm$24$\pm$21     \\
\ttbar, \wlnubr+jets    & 1 & 645$\pm$24$\pm$57     &  381$\pm$17$\pm$38    & 36.0$\pm$4.9$\pm$5.7      & 7.8$\pm$2.6$\pm$2.0   & 171$\pm$5$\pm$25  \\
QCD multijets                 & 1 &  25.0$\pm$9.4$\pm$5.2   &   16.0$\pm$7.4$\pm$2.8  &  0.0$\pm$1.0$\pm$1.2  & negligible            &   3.2$\pm$0.2$\pm$4.6 \\
Rare processes            & 1 &  18.0$\pm$9.2             &   18.0$\pm$8.9            &  1.1$\pm$0.5          & 0.3$\pm$0.1           &   5.4$\pm$2.7         \\ \hline
Total                         & 1 &1540$\pm$100           &  754$\pm$68           & 85$\pm$10             &  16.0$\pm$4.1             & 356$\pm$41            \\ \hline

\znunubr+jets           & 2 & 60.0$\pm$3.4$\pm$7.1      & 28.0$\pm$2.4$\pm$3.8  & 3.9$\pm$0.9$\pm$1.0   & 0.7$\pm$0.6$\pm$0.6   &  6.6$\pm$0.4$\pm$1.2  \\
\ttbar, \wlnubr+jets  & 2 & 29.0$\pm$2.9$\pm$5.5    & 17.0$\pm$2.5$\pm$3.3      & 2.4$\pm$0.9$\pm$0.6   & 0.0$\pm$0.2$\pm$0.2   & 19.0$\pm$1.8$\pm$3.4  \\
QCD multijets                 & 2 &  1.9$\pm$0.7$\pm$0.4  &  1.2$\pm$0.8$\pm$0.2  & 0.0$\pm$0.1$\pm$0.1   & negligible            &  0.4$\pm$0.1$\pm$0.7  \\
Rare processes                & 2 &  1.8$\pm$0.9          &  3.4$\pm$1.7          & 0.1$\pm$0.1           & 0.1$\pm$0.1           &  0.4$\pm$0.4          \\ \hline
Total                         & 2 & 93$\pm$10             & 50.0$\pm$6.4              & 6.5$\pm$1.7           & 1.0$\pm$0.9           & 26.0$\pm$4.1          \\ \hline

        \end{tabular}
}
\end{table}

\graphicspath{{SUS-13-009/plots/}}

\section{Search for top- and bottom-squark pair production in compressed spectrum scenarios \label{sec-SUS-13-009}}
We next describe the monojet search.
Given the lack of observation of a SUSY signature in more conventional searches,
it is important to search for SUSY with compressed mass spectra, i.e., SUSY scenarios in which the parent sparticles are close in mass to the daughter sparticles.
Small mass splittings $\Delta m = m_{\PSQt} - m_{\PSGczDo}$ or $\Delta m = m_{\PSQb} - m_{\PSGczDo}$ between the top or bottom squark and the LSP leave little visible energy in the detector, making signal events difficult to distinguish from SM background.
However, events with an energetic ISR jet recoiling against the \ptvecmiss vector from the LSP can provide a clear signal for compressed events.
We thus perform a search for events with a single jet and significant \ptmiss.

For $m_{\PSQt} - m_{\PSGczDo}<m_{\PW}$,
the dominant \PSQt decay mode is the flavour changing neutral-current process
$\PSQt \to \PQc \PSGczDo$.
In the case of the $\PSQb$, the kinematically similar decay
$\PSQb \to \PQb \PSGczDo$
dominates for compressed scenarios,
so the monojet topology is used to search for both top and bottom squarks.
The search represents an optimization of the studies presented in Refs.~\cite{EXO-12-048,bib:CMS_EXO11003,bib:CMSEXO11059}.
Relative to these previous studies, we increase the threshold on \njets, and define search regions using
the \pt of the highest \pt jet rather than \ptmiss.

\subsection{Event selection \label{sec:t2cc_eventSel}}
Data used in the analysis are selected by a combination of two triggers.
The first trigger requires $\METmu > 120$\GeV, where $\METmu$ is calculated using calorimetric information only.
The second trigger requires a jet to satisfy $\pt > 80$\GeV, $\abs{\eta} < 2.6$, and to have less than 95$\%$ of the jet momentum carried by neutral hadrons.
In addition, the second trigger requires $\METmu > 105$\GeV, where $\METmu$ is calculated using the particle-flow algorithm.
Selection criteria of $\METmu > 250$\GeV, and a leading jet (which has the highest momentum of all jets in the event and is denoted $\jet_{1}$) with $\pt^{\jet_{1}}>110$\GeV and $\abs{\eta}<2.4$, ensure a fully efficient trigger.
To suppress the instrumental and beam-related backgrounds, and to remove noisy events and misidentified high-\pt electrons and photons, events are rejected based on the properties of $\jet_{1}$: if less than 20\% of its energy is carried by charged hadrons, or if more than 70\% of its energy is carried by either neutral hadrons or photons, the event is rejected.

\begin{figure*}[tbh]
  \centering
  \includegraphics[width=0.65\textwidth]{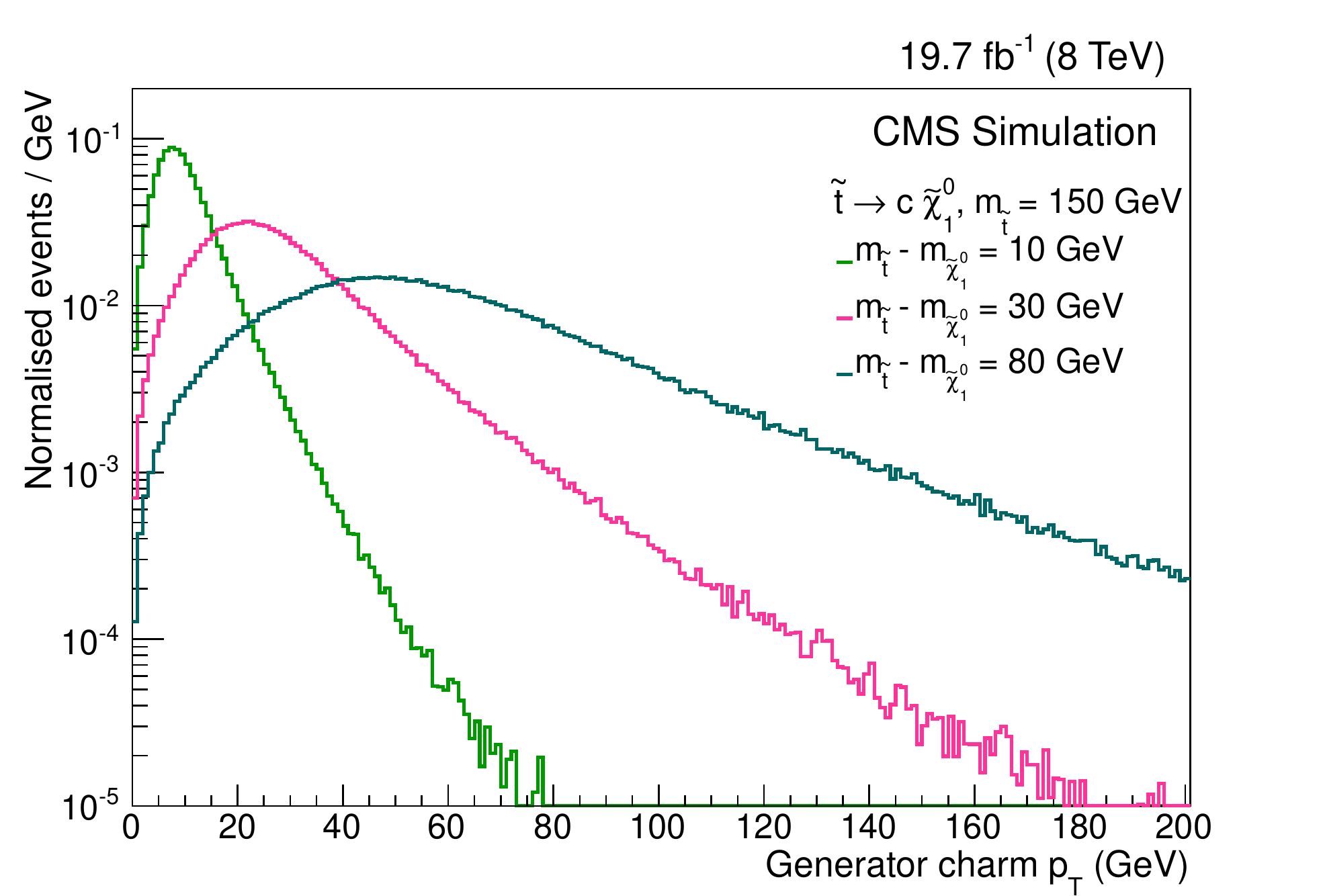}
  \caption{Charm quark \pt distribution for charm quarks emitted in the decay of top squarks of mass 150\GeV, for mass differences, $m_{\PSQt}-m_{\PSGczDo}=10, 30, 80\GeV$ in the monojet analysis.
         \label{stopj2pT}}

\end{figure*}

Although event selection is based upon a single high-momentum jet,
signal acceptance is increased by accepting events in which there is a
second jet $\jet_{2}$ originating from ISR.
In addition, the signal also has
soft final-state jets produced by the charm or bottom quarks
originating from the sparticle decays.
Ideally, these soft jets should
not be taken into account in the jet counting.
To suppress them a \pt threshold is introduced for the jet counting.
Figure~\ref{stopj2pT} shows the \pt distribution of charm quarks, taken from simulation, for a few representative mass hypotheses
in the process
$\ttwocc$.
Placing the jet counting
threshold at 60\GeV for jets with $\abs{\eta} < 4.5$ provides a compromise
between a high threshold to reject soft jets and a low threshold to
reject QCD multijet events.
Using this threshold condition, events with
up to two jets are accepted.
To suppress the QCD dijet background, $\Delta\phi(\ptvec^{\jet_1}, \ptvec^{\jet_2})$ is required to be less than 2.5.
To reduce electroweak and top backgrounds, events with electrons satisfying
$\pt>10$\GeV and $\abs{\eta} < 2.5$, or muons reconstructed with $\pt >10$\GeV and $\abs{\eta}<2.4$, are rejected.
Events with a well-identified $\tauh$ lepton with $\pt >20$\GeV and $\abs{\eta} < 2.3$ are removed.
The analysis is performed in search regions that reflect the hardness of the radiated jet in an event, in seven inclusive regions of leading jet \pt: $\pt^{\jet_1} > 250,$ 300, 350, 400, 450, 500, and 550\GeV.

Following the above selection criteria, expected event yields from various SM processes, as predicted by simulation in each of the search regions, are shown in Table~\ref{tab:t2ccBgds}.

\begin{table*}[htb]
        \centering
        \topcaption{Predicted background yields from simulation for the monojet analysis.
        The results are scaled to an integrated luminosity of 19.7\fbinv.
        The uncertainties are statistical.
        The results for the \ttwocc
 		signal events are labelled as $(m_{\PSQt},m_{\PSGczDo})$, in \GeVns.
        }
\label{tab:t2ccBgds}
\resizebox{\textwidth}{!}{
\begin{tabular}{l,,,,,,,} \hline
 \multicolumn{1}{c}{$\pt^{\jet_1}$ (\GeVns{})}   &  \multicolumn{1}{c}{$>$250}  &  \multicolumn{1}{c}{$>$300} &  \multicolumn{1}{c}{$>$350} &  \multicolumn{1}{c}{$>$400} &  \multicolumn{1}{c}{$>$450} &  \multicolumn{1}{c}{$>$500} &  \multicolumn{1}{c}{$>$550}  \\ \hline
 \znunubr{}+jets   & 22600, 56   & 11100, 37  & 5230, 24  & 2620, 17  & 1340, 12  & 727, 8.7   & 406, 6.5  \\
\wlnubr{}+jets    & 13600, 70   &  6870, 50  & 3180, 34  & 1500, 23  &  751, 17  & 376, 12    & 204, 8.7  \\
WW,WZ,ZZ              &   819, 27   &   546, 18  &  332, 12  &  181.0, 6.5 &   92.0, 3.4 &  61.0, 2.3   &  34.0, 1.2  \\
\ttbar              &   639.0, 5.7  &   369.0, 4.3 &  206.0, 3.2 &  113.0, 2.4 &   64.0, 1.8 &  36.0, 1.3   &  21.0, 1.0  \\
Multijets             &   602, 19   &   344, 15  &  178, 10  &   91.0, 7.3 &   48.0, 5.2 &  27.0, 4.0   &  18.0, 3.5  \\
Single top quark            &   172.0, 7.6  &    97.0, 5.7 &   49.0, 4.1 &   21.0, 2.7 &   11.0, 2.2 &   5.2, 1.4 &   3.2, 1.2 \\
\zellellbr{}+jets &   127.0, 6.1  &    75.0, 4.7 &   40.0, 3.5 &   25.0, 2.8 &   17.0, 2.4 &   11.0, 2.0  &   7.4, 1.6 \\ \hline

Total &38600,96 & 19400,67 & 9220,45 & 4550,31 & 2320,22 & 1240,16 & 693,12 \\ \hline
Signal (200, 120)  & 1130,22 & 663,17 & 352 , 12 & 193.0 , 9.2 & 111.0,7.0 & 62.9,5.1 & 35.5 , 3.9\\
Signal (250, 240) & 1640,15 & 1070,12 & 657.0,9.6 & 403.0,7.5 & 256.0,6.0 & 156.0,4.6 & 98.0,3.7 \\
\hline
\end{tabular}
}
\end{table*}

\subsection{Background predictions~\label{t2cc_backgrounds}}

The dominant SM backgrounds are due to \znunubr{}+jets
and \wlnubr{}+jets processes.
These backgrounds are estimated from data,
utilizing a control sample of $\mu$+jets events in which
\zmumubr{} and \wmunubr{} events are used to estimate the \znunubr{}+jets
and \wlnubr{}+jets backgrounds, respectively.
Small contributions from diboson, QCD multijet, and \ttbar events are estimated
using simulation corrected for any differences between simulation and data.
Very small backgrounds arising from single top quark and \zll processes are
taken from simulation directly.

The \znunubr{}+jets background is estimated using a data control sample of dimuon events, selected using the same trigger as the search regions.
The redefinition of the \ptmiss to exclude muons and mimic neutrinos at both the trigger level and in analysis variables allows the use of the same trigger, not possible in the multijet t-tagged or dijet b-tagged analyses, and reduces systematic uncertainties.
The \zmumubr{}+jets enriched control sample is selected by applying the full signal selection, except for the muon veto, instead demanding two oppositely charged muons with $\pt > 20$\GeV and $\abs{\eta} < 2.4$.
At least one of the muons must be isolated,
and the dimuon reconstructed invariant mass must lie within a window of 60--120\GeV, to be consistent with the Z boson mass.
The number of observed dimuon events in the data control sample ($N^\text{obs}$) is corrected for non-\zmumubr{} processes ($N^{\text{bgd}}$), estimated using simulation.
The event yield is corrected for the acceptance ($A$) and efficiency ($\epsilon$) of the muon selection criteria, taken from \zmumubr{} simulation and corrected for differences in muon identification between data and simulation.
The number of \znunubr{}+jets events is estimated using:

\begin{equation}
N (\znunubr{}\text{+jets}) = \frac{N^{\text{obs}} - N^{\text{bgd}}}{A\epsilon}\, R,
\end{equation}

where $R$ is the ratio of branching fractions of \znunu{} to \zmumu{} decays~\cite{PDG}, corrected for the contributions of virtual photon exchange in the \Z+jets sample and for the Z mass window requirement.

The uncertainty in the prediction includes both statistical and systematic contributions:
(i) the statistical uncertainty in the number of \zmumu{}+jets events in the data and simulation,
(ii) a 50\% uncertainty from each of the non-\Z backgrounds estimated using simulation,
(iii) uncertainties
associated with PDF choice (2\%)~\cite{bib:pdfatLHC,bib:pdfBall,CTEQ6} as recommended in Refs.~\cite{bib:PDF4LHCa,bib:PDF4LHC},
(iv) a 2\% uncertainty due to hadronization,
and (v) a 2\% uncertainty in $R$.
The statistical uncertainty in the number of \zmumubr{}+jets events, 2--17\%, dominates the total uncertainty, which ranges from 5\% to 19\%.

The background due to lost leptons from \wpj events is estimated using a single-muon control sample enriched in \wmunubr{}+jets events selected with the same trigger as the search regions.
The full signal selection is applied, except that the muon veto
is replaced by the requirement of a well-identified muon with $\pt > 20$\GeV and $\abs{\eta} < 2.4$.
The transverse mass of the muon-\ptvecmissmu system, as defined in Section~\ref{t2tt_backgrounds}, is required to satisfy $50<\MT<100$\GeV.
Analogously to the \znunubr{}+jets background estimation, the observed single-muon event yield in data ($N^\text{obs}$) is corrected
for non-\wmunubr{} processes using simulation ($N^{\text{bgd}}$),
and for the acceptance ($A'$) and efficiency ($\epsilon'$) of the single-muon selection criteria using \wpj simulation, where differences between muon identification in data and simulation are taken into account.
The total \wmunubr{}+jets event yield is:
\begin{eqnarray}
N(\wmunubr{}\text{+jets}) = \frac{N^\text{obs} - N^\text{bgd}}{A'\epsilon'}.
\end{eqnarray}
The total lost-lepton and \tauh background is estimated by extrapolating the \wmunubr{} event yield to the total \wlnubr{} event yield using $\pt^{\jet_1}$-dependent generator level ratios of \wmunubr{} to $\PW(\Pe\PGn)$ and $\PW (\tauh \nu)$ events, correcting for the
inefficiencies of lepton vetoes used in the signal event selection (taken from \wpj simulation).

The uncertainty in the prediction includes both statistical and systematic contributions:
(i) the uncertainties in the numbers of single-muon events in the data and simulation samples,
(ii) a 50\% uncertainty in each simulated non-\wpj contribution to the control sample,
and (iii) statistical and systematic uncertainties (from PDFs) incorporated in the total uncertainties in acceptances and efficiencies.
Statistical uncertainties in the number of \wmunubr{}+jets events (1--8.6\%)
and
uncertainties in the acceptance and efficiency values (4.5--7.1\%)
dominate the total uncertainty, which ranges from 5.7\% to 12.0\%.

The background from QCD multijet production is expected to be small, contributing $\approx$2\% to the total background yield, and is predicted using the simulation normalized to data in control regions.
The normalization is determined from a QCD-enriched control sample,
defined using events that satisfy the signal event selection criteria except that the
$\Delta\phi(\ptvec^{\jet_1}, \ptvec^{\jet_2})<2.5$ and $\njets<3$ requirements are not applied in order to
maintain a sufficient number of events in the control sample,
which is defined by $\Delta\phi(\ptvec^{\jet_2},\ptvecmissmu)<0.3$, a region enriched with mismeasured jets.
The contribution from non-QCD dijet and multijet processes is subtracted from the data yield using simulation that has been normalized to data in QCD-free regions.
A set of $\pt^{\jet_1}$-dependent data-MC scale factors are extracted and applied to the simulated QCD yield in the search regions, using the ratio of QCD events found in data to the yield predicted using simulation in the control sample.

A systematic uncertainty of 50\% in the unnormalized QCD simulation is applied.
The uncertainty in the scale factors determined from data includes both statistical and systematic components, arising from a 50\% uncertainty assigned to each of the non-QCD contributions that are subtracted from the data yield in the control region.
The total uncertainty, including statistical uncertainties, in the QCD background prediction is $\approx 60\%$ in each search region.
A cross check of this prediction is performed in a QCD-rich sideband region defined by $\Delta\phi(\ptvec^{\jet_3},\ptvecmissmu)<0.3$ and found to agree within the uncertainties with the observed number of events.

The \ttbar contribution to total background is small ($\approx$2\%) and is estimated using simulation that has been validated using data.
A control sample of events with $\METmu >250$\GeV, $\pt^{\jet_1}>110$\GeV,
and $\Delta\phi(\ptvec^{\jet_1}, \ptvec^{\jet_2})<2.5$ is derived from the same trigger as used for the search regions.
A \ttbar-rich sample is created by then requiring an identified electron and an identified muon of opposite sign.
The invariant mass of the  $\Pe\mu$ system must be greater than 60\GeV.
The data and simulation in the control region are found to agree
within $3\pm20\%$, so no additional scale factor is applied to the next-to-next-to-leading-order cross section estimate~\cite{topxs:Kidonakis} used to normalize the yield to the integrated luminosities of the search samples.
To be consistent with the other small background estimations, a 50\% uncertainty is assigned that includes statistical and systematic uncertainties.
Diboson processes contribute $\approx $2\% to the total background.
The number of $\PW\PW$, $\PW\Z$, and $\Z\Z$ events are estimated using simulation, normalized to the luminosity with  next-to-leading-order (NLO) cross sections \cite{MCFM:diboson} and assigned a 50\% uncertainty, while
$\Z\gamma$ and $\PW\gamma$ events are estimated from data.
They are treated inclusively as part of the \znunubr{}+jets and \wlnubr{}+jets backgrounds,
which is found to agree with simulation within 15\%.
Single top quark and \zll{}+jets events account for $<1\%$ of total background
and are estimated directly from simulation.
A 50\% uncertainty is assigned to background predictions estimated from simulation.

The total background yields using the methods outlined above are shown in Table~\ref{tab:t2ccBgd} in each of the inclusive search regions.

\begin{table*}[t!bh]
\centering
\topcaption{SM background predictions for the monojet search regions defined in the text,
corresponding to an integrated luminosity of 19.7\fbinv.
For the \znunubr{}+jets and \wlnubr{}+jets terms, the first uncertainty is statistical and the second is systematic.
The uncertainties in the remaining backgrounds include both statistical and systematic terms.}
\label{tab:t2ccBgd}
\begin{tabular}{lcccc}
                \hline

   &  $\pt^{\jet_1}> 250$\GeV&   $\pt^{\jet_1}> 300$\GeV &  $\pt^{\jet_1}> 350$\GeV &  $\pt^{\jet_1}> 400$\GeV   \\ \hline

\znunubr{}+jets       & 21200$\pm$450$\pm$1000  &  10100$\pm$300$\pm$510  & 4600$\pm$210$\pm$250& 2250$\pm$150$\pm$130  \\
\wlnubr{}+jets        & 12300$\pm$120$\pm$690   &   5940$\pm$89$\pm$360   & 2690$\pm$62$\pm$170 & 1250$\pm$40$\pm$80  \\
\ttbar                    &   602$\pm$300           &   344$\pm$170           & 178$\pm$89          &   91$\pm$46     \\
\zellellbr{}+jets     &   127$\pm$64            &    75$\pm$38            & 40$\pm$20           &   25$\pm$13     \\
Single top quark                &   172$\pm$86            &    97$\pm$49            & 49$\pm$24           &   21$\pm$10     \\
Multijets                 &   786$\pm$470           &   508$\pm$310           & 304$\pm$180         &  162$\pm$99     \\
Diboson                   &   639$\pm$320           &   369$\pm$180           & 206$\pm$100         &  113$\pm$56     \\ \hline
Total                     & 35900$\pm$1500          & 17400$\pm$800           & 8060$\pm$440        & 3910$\pm$250    \\ \hline
\end{tabular}
\\[2ex]
\begin{tabular}{lccc} \hline
   &   $\pt^{\jet_1}> 450$\GeV &  $\pt^{\jet_1}> 500$\GeV &  $\pt^{\jet_1}> 550$\GeV  \\ \hline
\znunubr{}+jets       & 1250$\pm$110$\pm$84   & 663$\pm$80$\pm$48 & 334$\pm$57$\pm$28 \\
\wlnubr{}+jets        &  637$\pm$28$\pm$44    & 301$\pm$19$\pm$22 & 150$\pm$13$\pm$13 \\
\ttbar                    &   48$\pm$24           &   27$\pm$14       &  18.0$\pm$9.0       \\
\zellellbr{}+jets     &   17.0$\pm$8.3          &   11.0$\pm$5.6      &   7.4$\pm$3.7     \\
Single top quark          &   11.0$\pm$5.7          &    5.2$\pm$2.6    &   3.2$\pm$1.6     \\
Multijets                 &   80$\pm$49           &   52$\pm$32       &  28$\pm$18        \\
Diboson                   &   64$\pm$32           &   36$\pm$18       &  21$\pm$10        \\ \hline
Total                     & 2100$\pm$160          & 1100$\pm$100      & 563$\pm$71        \\ \hline
\end{tabular}
\end{table*}

\section{Results \label{sec-results} }
\begin{table*}[htb]
    \centering
        \topcaption{Event yields for the different search regions defined in Sections~\ref{sec:t2tt_eventSel}, \ref{sec:t2bb_eventSel}, and \ref{sec:t2cc_eventSel}.
        The multijet t-tagged search requires a combination of exclusive and inclusive bins in number of b-tagged jets ($\nbjets=1,~\nbjets\ge2$), whereas the dijet b-tagged searches require exclusive bins ($\nbjets=1,2$); the monojet search makes no requirements on b-tagged jets ($\nbjets\ge0$).  The SM background predictions and the yields observed in data correspond to integrated luminosities of 19.4, 19.4, and 19.7\fbinv for the multijet t-tagged, dijet b-tagged, and monojet searches, respectively.  The quoted uncertainties in the SM predictions reflect the total (statistical and systematic) uncertainties quadratically summed over all different background contributions.   }\label{tab:GrandResultTable}
        {
            \begin{tabular}{c|cc|cc|cc}
            \hline
   \multicolumn{1}{c}{\multirow{2}{*}{Search regions}} & \multicolumn{6}{c}{$\nbjets$}  \\
            \cline{2-7}
          & \multicolumn{2}{|c|}{$\ge0$} & \multicolumn{2}{c|}{$1$} & \multicolumn{2}{c}{$2$} \\
            \hline \hline
{Multijet t-tagged search} & & & {SM Pred.} & {Obs.} & {SM Pred.} & {Obs.} \\
            \hline
{$\ptmiss\in[200,350]$\GeV} & & & 148$^{+29}_{-24}$ & 141 & 81$^{+13}_{-12}$ & 68 \\
{$\ptmiss>350$\GeV}              & & &  33.4$^{+7.0}_{-7.8}$   &  30 &  8.6$^{+2.6}_{-2.4}$   & 15 \\
            \hline \hline
{Dijet b-tagged search}  & & & {SM Pred.} & {Obs.} & { SM Pred.} & {Obs.} \\
            \hline
{$\MCT <250$\GeV}              & & & 1540$\pm$100 & 1560 & 93$\pm$10   & 101 \\
{$\MCT \in[250,350]$\GeV} & & &  754$\pm$68  &  807 & 50.0$\pm$6.4  &  55 \\
{$\MCT \in(350,450]$\GeV} & & &   85$\pm$10  &  101 & 6.5$\pm$1.7 &   8 \\
{$\MCT >450$\GeV}              & & &   16.0$\pm$4.1 &   23 & 1.0$\pm$0.9 &   1 \\
            \hline
{ISR} & & & $356\pm41$& 359 & $26.0\pm4.1$ & 28 \\
            \hline \hline
{Monojet search} & {SM Pred.} & {Obs.} & & & &  \\
            \hline
{$\pt^{\jet_1}>250$\GeV}    & 35900$\pm$1500 & 36600 & & & & \\
{$\pt^{\jet_1}>300$\GeV}    & 17400$\pm$800  & 17600 & & & & \\
{$\pt^{\jet_1}>350$\GeV}    &  8060$\pm$440  &  8120 & & & & \\
{$\pt^{\jet_1}>400$\GeV}    &  3910$\pm$250  &  3900 & & & & \\
{$\pt^{\jet_1}>450$\GeV}    &  2100$\pm$160  &  1900 & & & & \\
{$\pt^{\jet_1}>500$\GeV}    &  1100$\pm$110  &  1000 & & & & \\
{$\pt^{\jet_1}>550$\GeV}    &   563$\pm$71   &   565 & & & & \\
            \hline
            \end{tabular}
        }

\end{table*}
Each search region definition was optimized and the SM backgrounds were evaluated before the data in the search regions were examined.
Table~\ref{tab:GrandResultTable} shows the observed yields compared with the SM background predictions in each of the 21 search regions defined by the three analyses.
All search regions are consistent with predictions of the SM, and no significant excesses are observed.

\begin{figure}[ht!]
  \centering
   \includegraphics[width=0.32\textwidth]{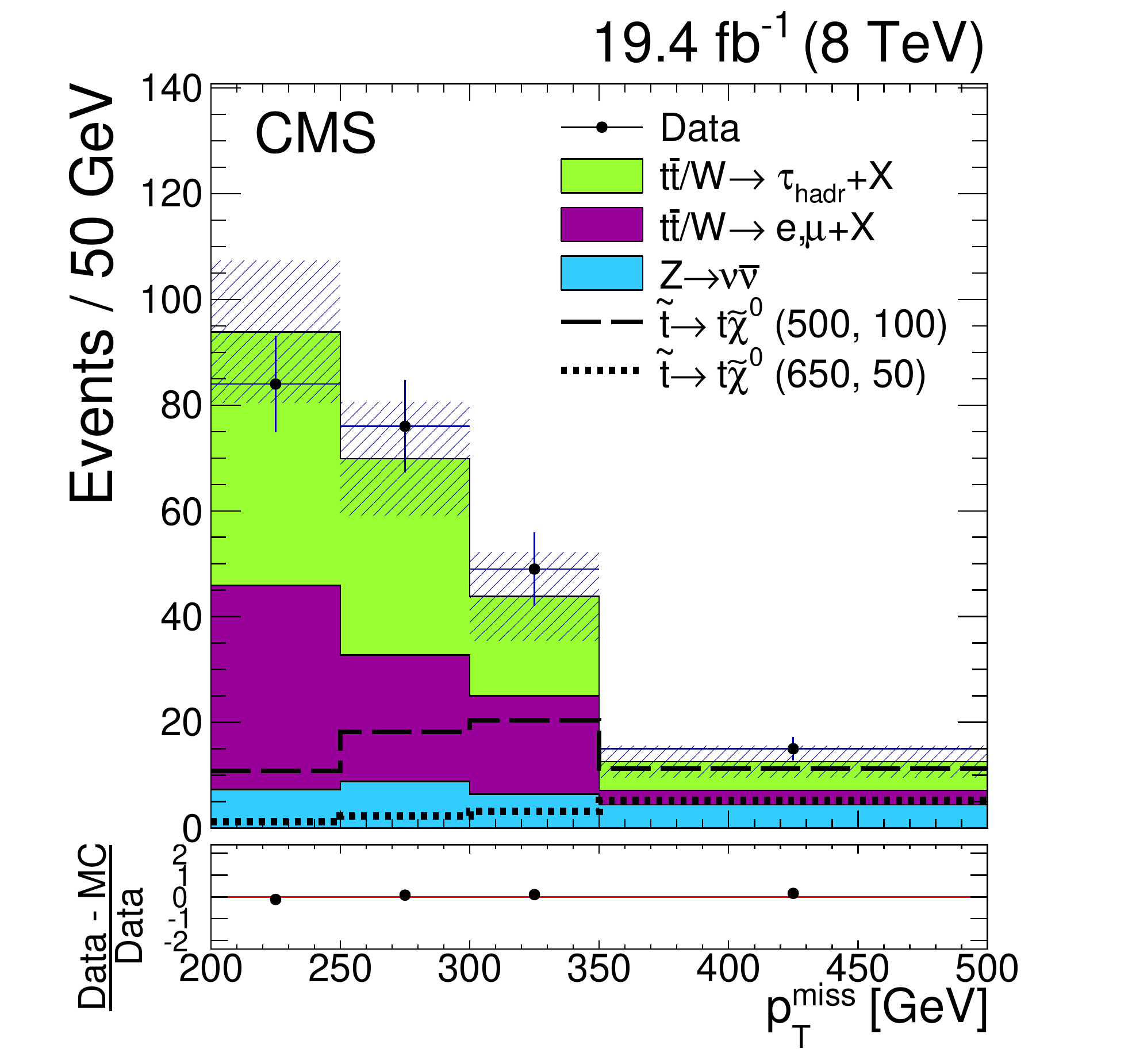}
   \includegraphics[width=0.32\textwidth]{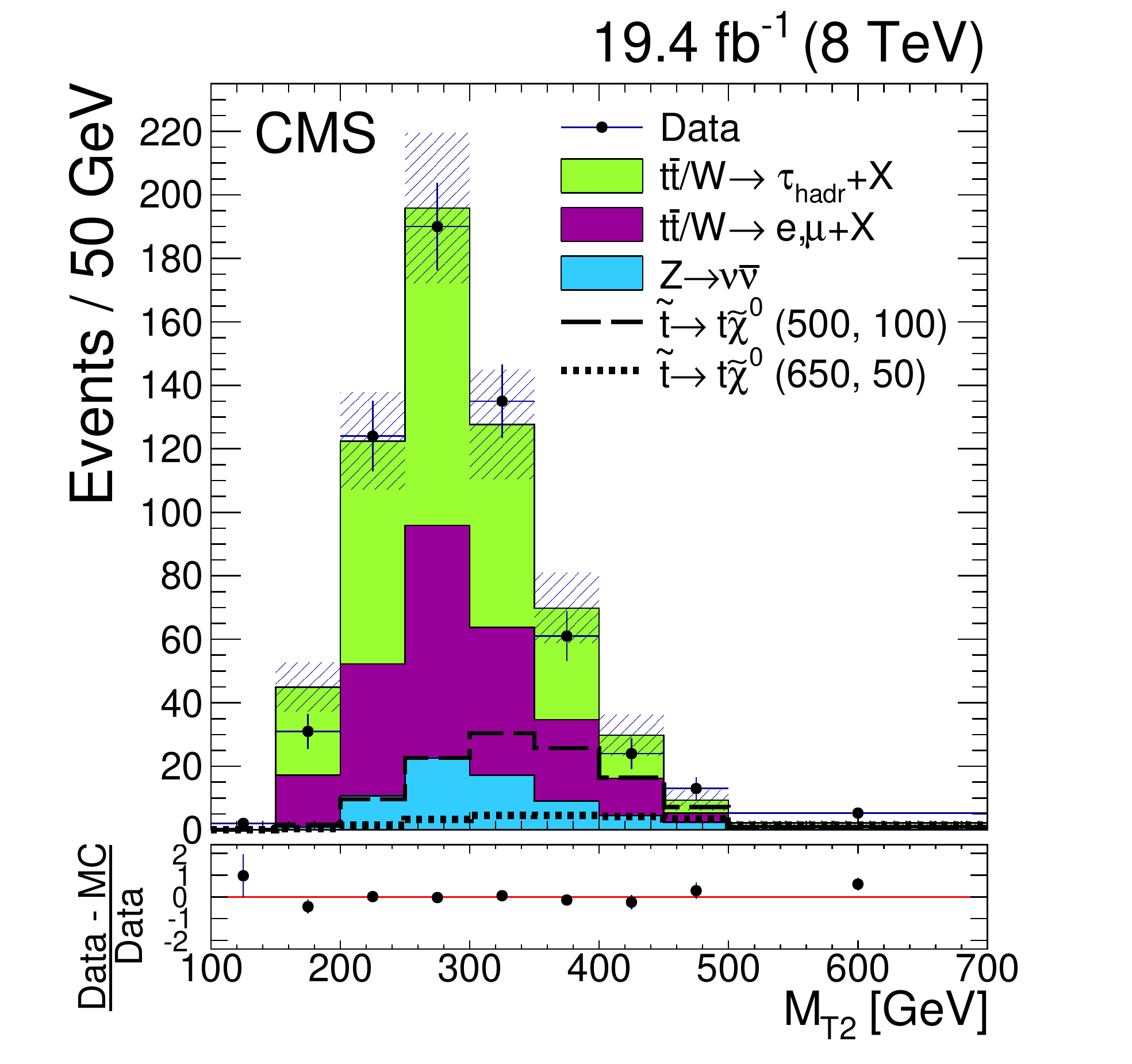}
  \includegraphics[width=0.32\textwidth]{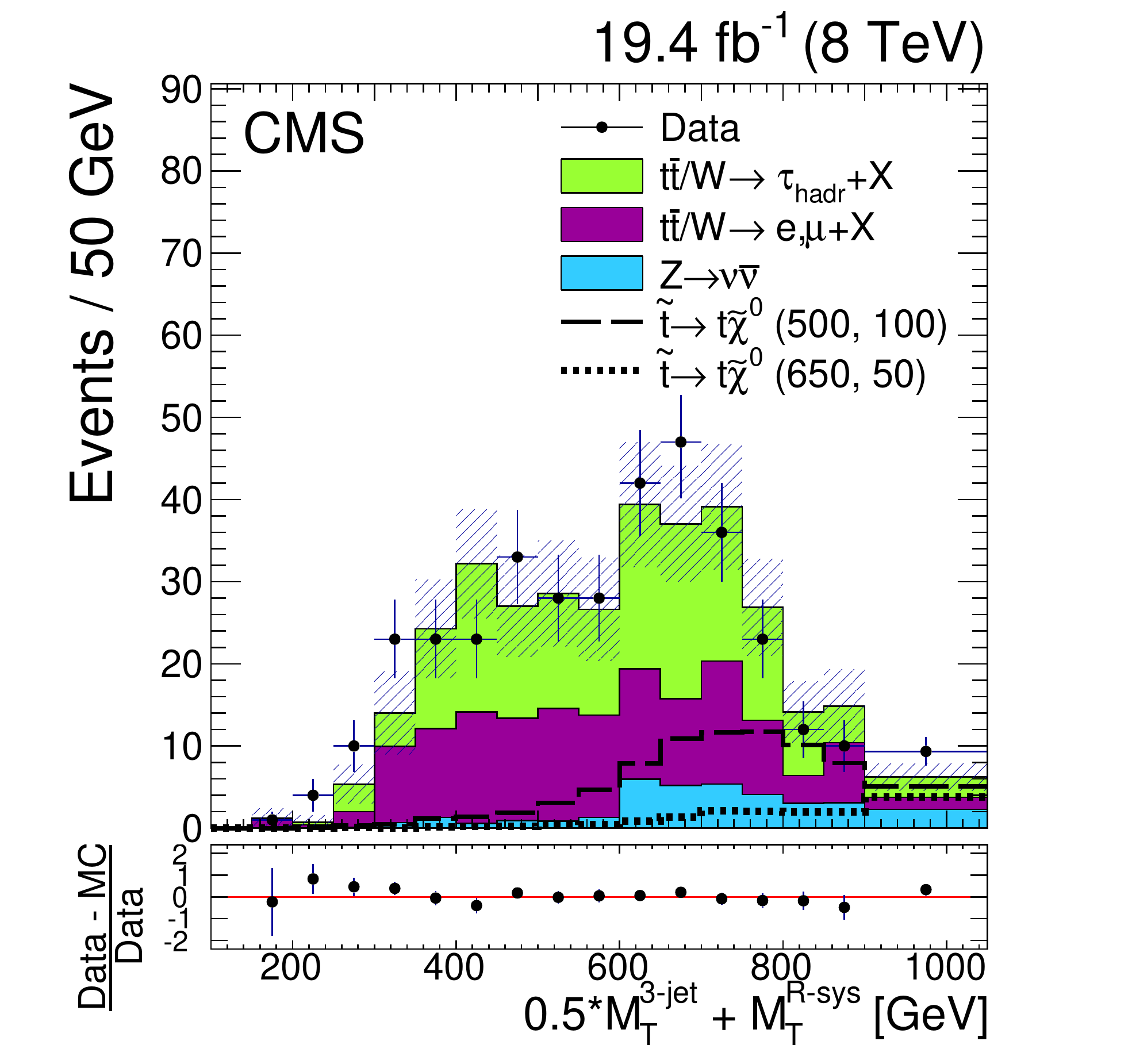}
  \caption{The \ptmiss (left), $\MTTwo$ (centre), and $0.5\MTt+\MTb$ (right) distributions from data (black dots), and predicted backgrounds (solid filled areas) in the multijet t-tagged search, where the total (statistical and systematic) uncertainty in the background prediction is shown by the hatched band.
  The distributions of two representative signals $(m_{\PSQt},m_{\PSGczDo}) = (500,100)$ and $(650,50)\GeV$ are overlaid (dashed and dotted lines respectively).
  The leftmost bin of each distribution contains the overflow.
}
  \label{fig:evselbaseline}
\end{figure}

Figure~\ref{fig:evselbaseline} shows distributions of some key variables in the multijet t-tagged search, for data and for the expected SM background estimated using the methods outlined in Section~\ref{t2tt_backgrounds}.
The hatched bands show both the statistical and systematic uncertainties from the predictions, taken from Table~\ref{tab:GrandResultTable}.
The \ptmiss distribution (Fig.~\ref{fig:evselbaseline}, left) is obtained after applying the baseline selection criteria described in Section~\ref{sec:t2tt_eventSel}.
The $\MTTwo$ distribution (Fig.~\ref{fig:evselbaseline}, centre) is obtained using the baseline selection criteria without the $0.5\MTt+\MTb \ge 500\GeV$ requirement, and
the $0.5\MTt+\MTb$ distribution (Fig.~\ref{fig:evselbaseline}, right) is obtained using the baseline selection criteria without the $\MTTwo \ge 300\GeV$ requirement.
The distributions simulated for two representative signal mass hypotheses for the case of $\ttwott$ production, scaled to an integrated luminosity of \fulllumi\fbinv, are superimposed for comparison.
The QCD prediction is not included in the plots shown in Fig.~\ref{fig:evselbaseline} since its contribution is negligible.

\begin{figure}[ht!b]
    \centering
        \includegraphics[width=0.35\textwidth]{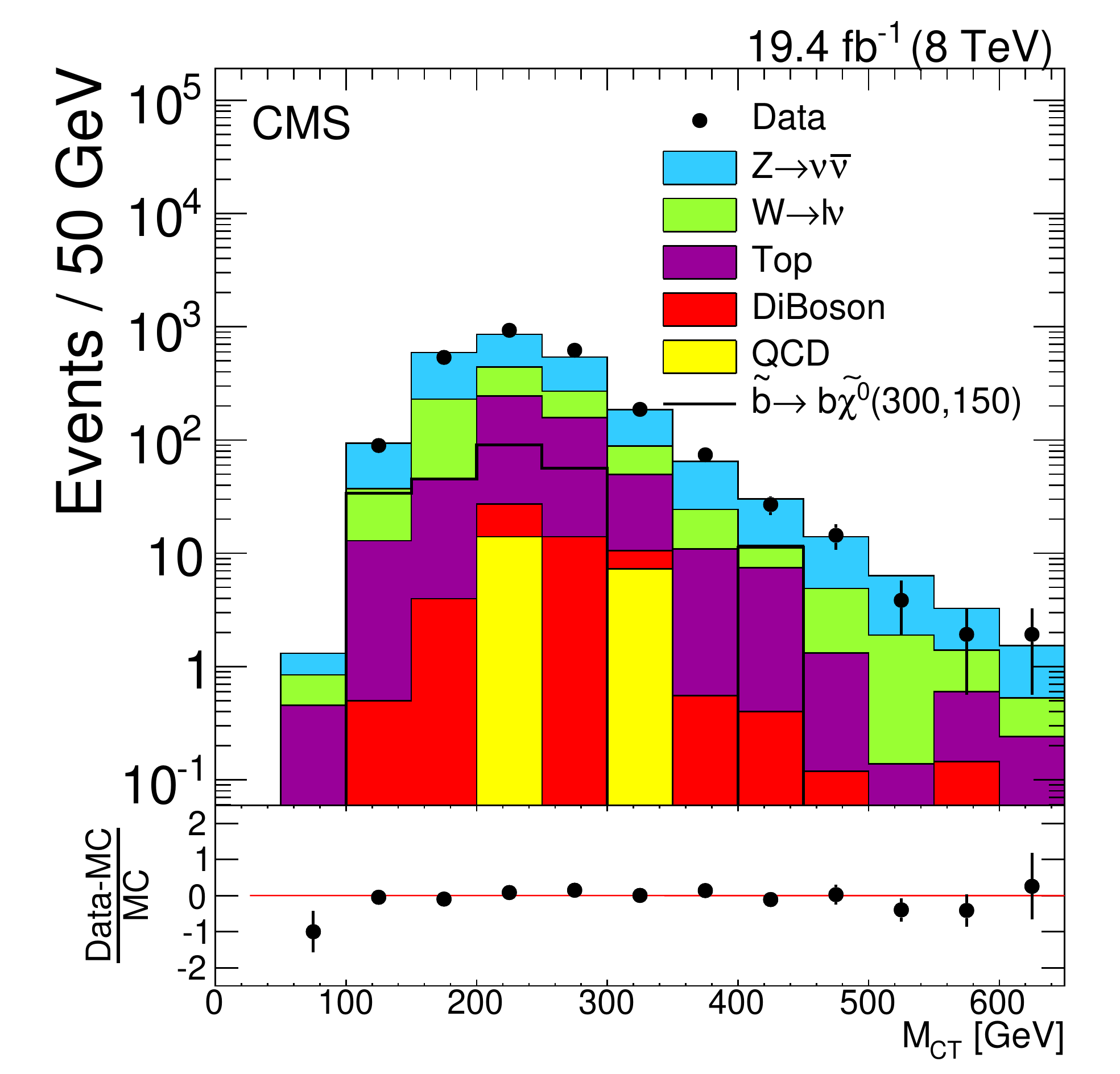}
        \includegraphics[width=0.35\textwidth]{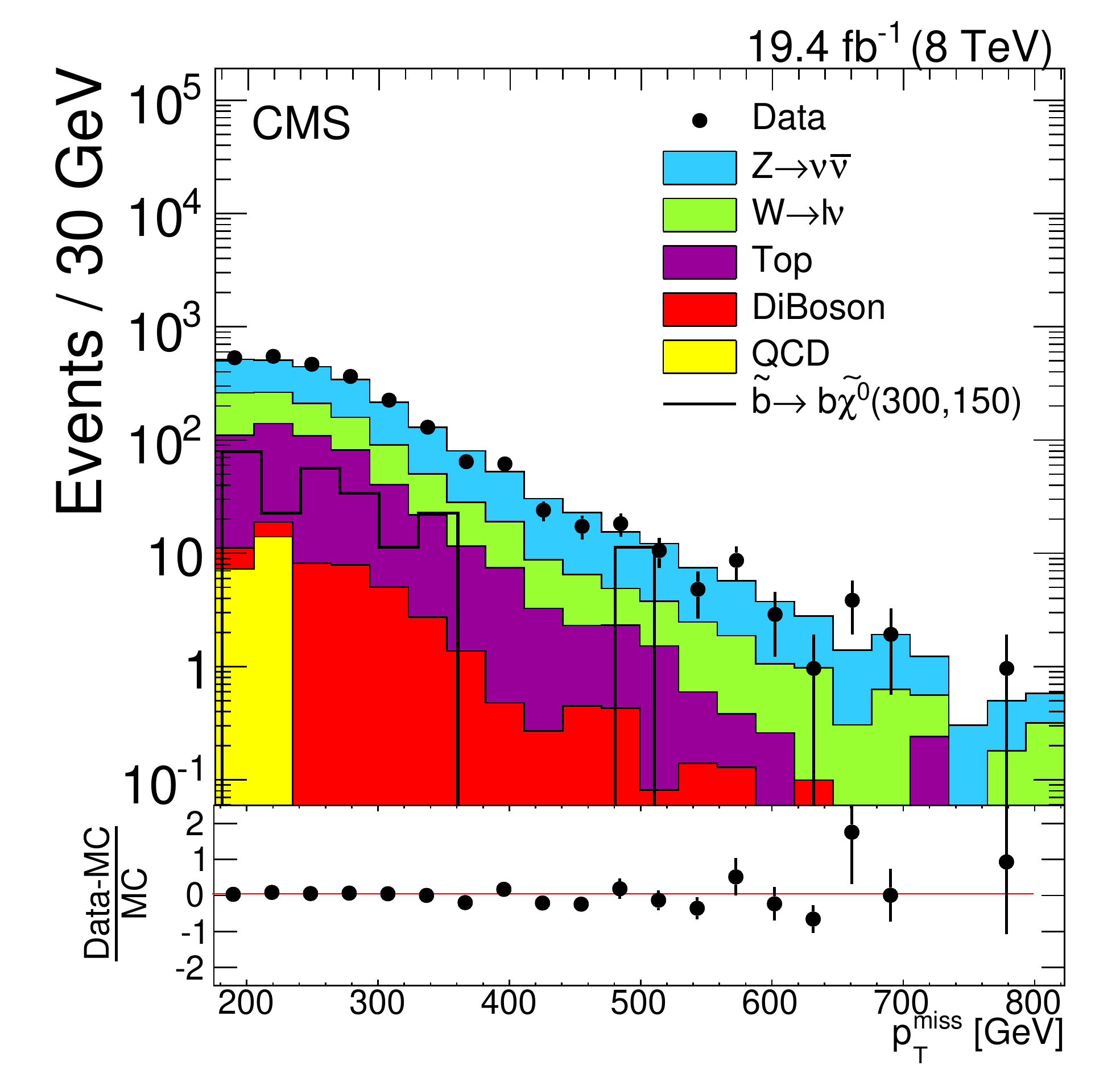}
        \includegraphics[width=0.35\textwidth]{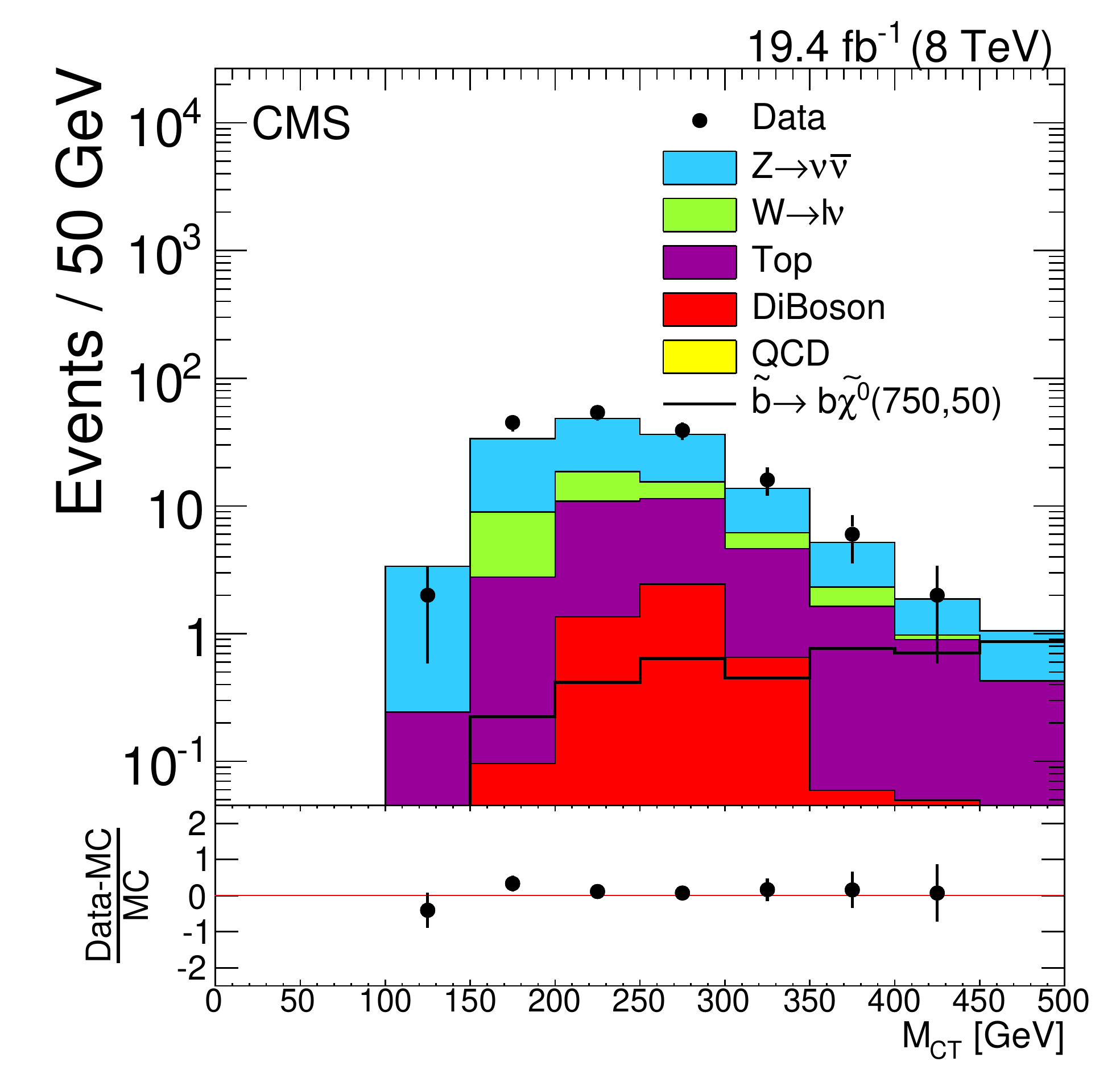}
        \includegraphics[width=0.35\textwidth]{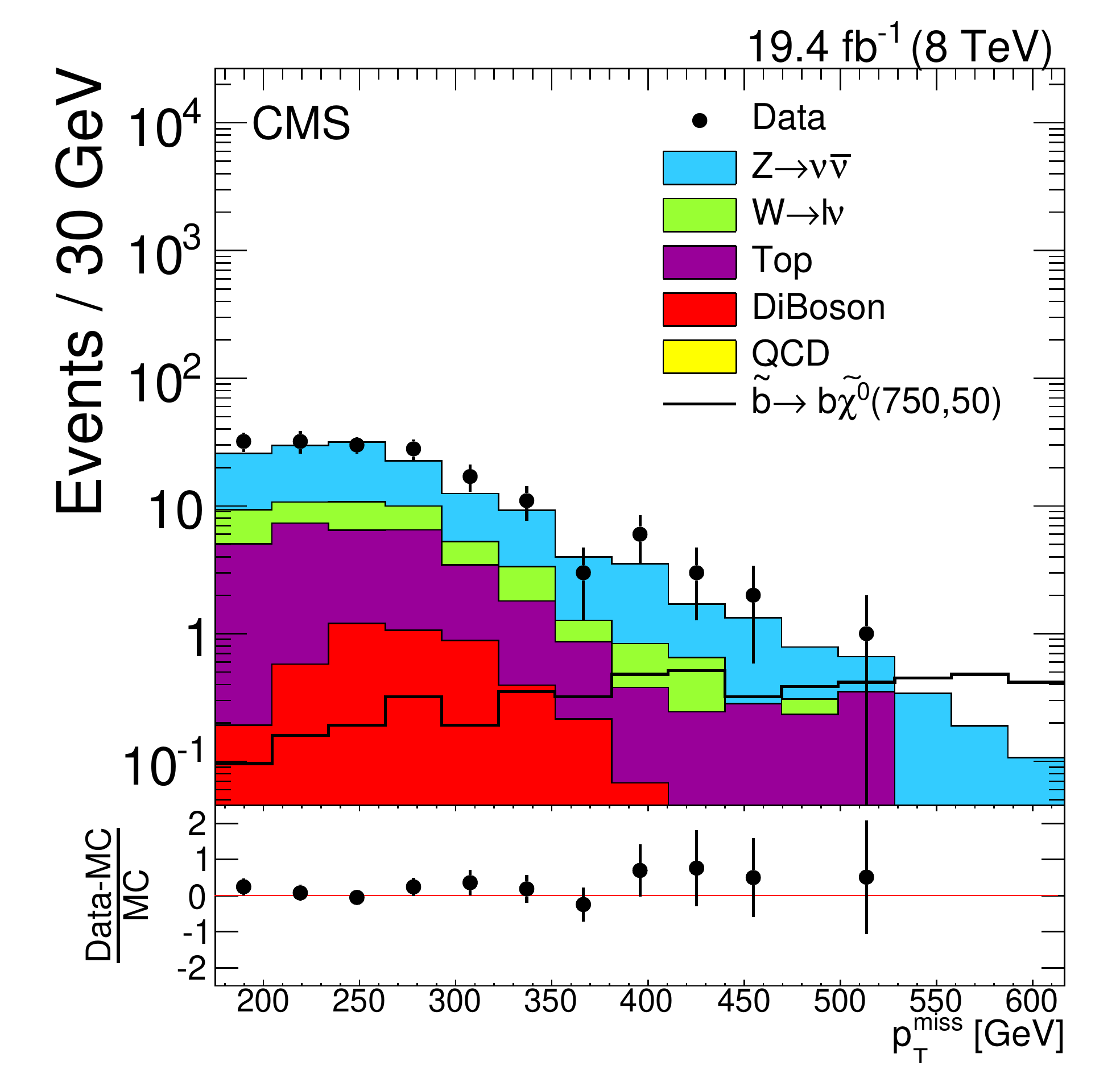}
        \caption{Distributions of (left column) \MCT and (right column) \ptmiss{} in data and MC simulation for baseline selected events in the dijet b-tagged search, with (top row) $\nbjets{}=1$ and (bottom row) $\nbjets{}=2$. Also shown (lines) are the corresponding distributions for two representative signals, ($m_{\PSQb}, m_{\PSGczDo}) = (750, 50)$ and $(300, 150)\GeV$. Statistical uncertainties are shown for the data.}
\label{fig:T41selection}
\end{figure}

Distributions of some representative variables sensitive to signals in the dijet b-tagged search
are shown in Fig.~\ref{fig:T41selection}, after the baseline selection criteria
(Section~\ref{sec:t2bb_eventSel}) have been applied.
The top (bottom) row shows results requiring $\nbjets=1$ ($\nbjets=2$).
The left-hand plots show the \MCT distributions, and the right hand plots the \ptmiss distributions.
The distributions of two representative signals for $\ttwobb$, scaled to an integrated luminosity of \fulllumi\fbinv, are superimposed for comparison.
While the total background prediction in Table~\ref{tab:GrandResultTable} is obtained using the methods outlined in Section~\ref{sec:t2bbBgd},
the background distributions in Fig.~\ref{fig:T41selection} are taken from simulation and normalized to an integrated luminosity of 19.4\fbinv.

\begin{figure}[htb]
  \centering
  \includegraphics[width=0.35\textwidth]{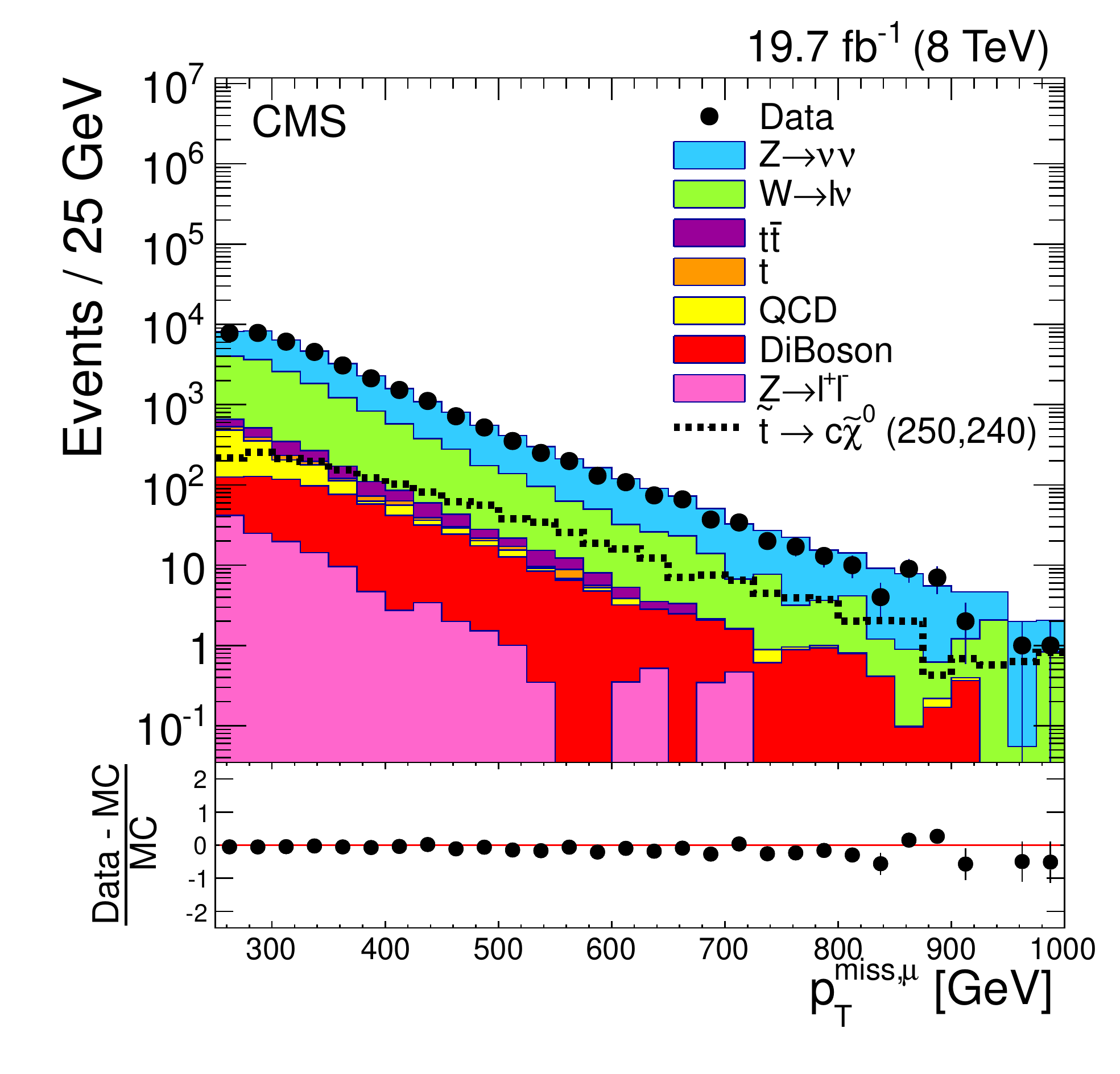}
   \includegraphics[width=0.35\textwidth]{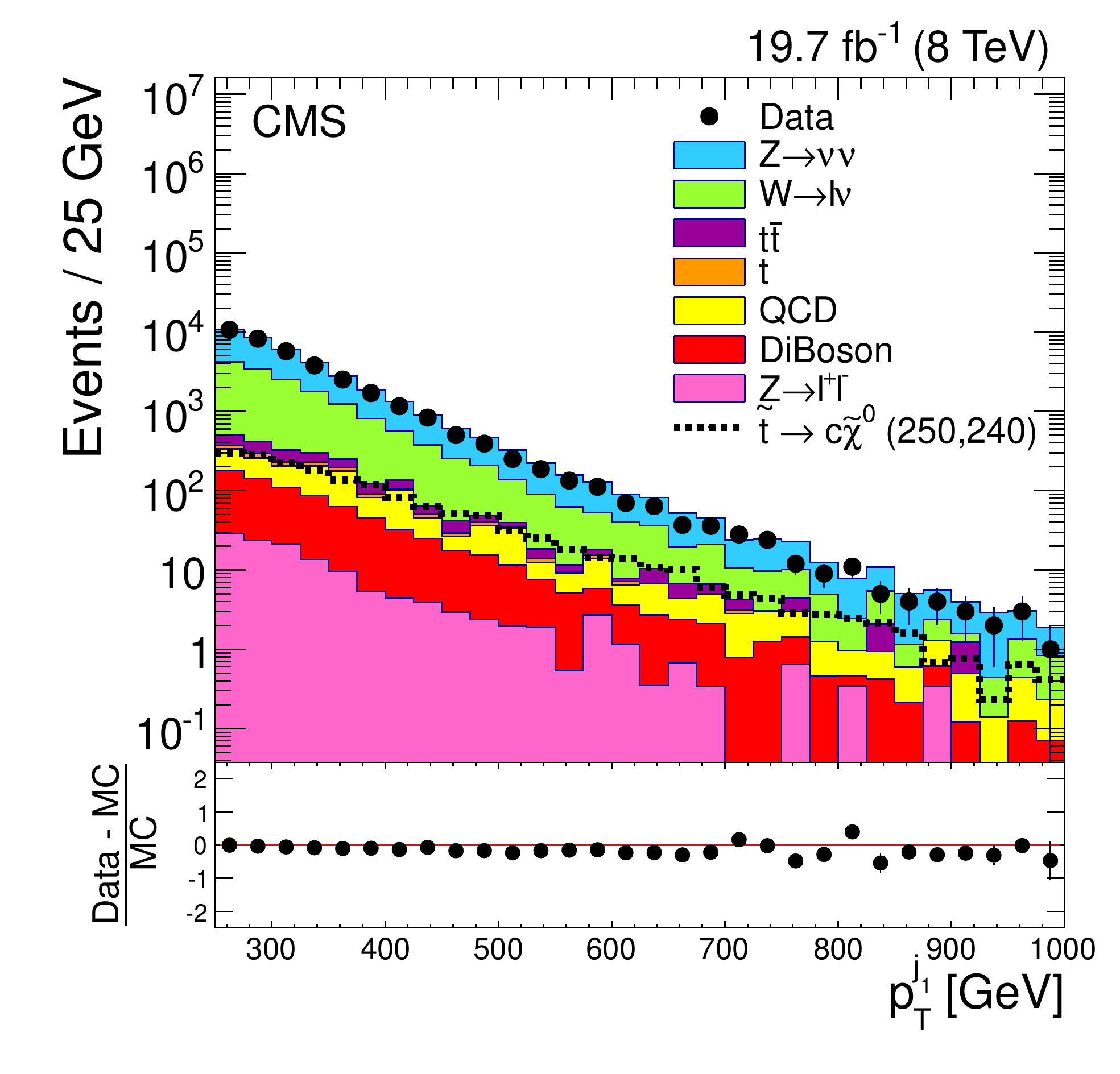}
  \caption{Distributions of (left) \METmu and (right) leading jet $\pt$ in the baseline monojet search region, $\pt^{\jet_1} >250$\GeV, for data and SM backgrounds. Background distributions are taken from simulation, and normalized to an integrated luminosity of 19.7\fbinv. A representative signal distribution for $\PSQt \to \PQc \PSGczDo$ is also shown (in the dotted line), where $m_{\PSQt} = 250\GeV$ and $m_{\PSGczDo} = 240\GeV$.  Statistical uncertainties are shown for the data.
\label{fig:ANA_MET_plots}}

\end{figure}

Figure~\ref{fig:ANA_MET_plots} shows the discriminating distributions in the monojet search, after the baseline selection criteria described in Section~\ref{sec:t2cc_eventSel} have been applied.
The left plot shows the \METmu distribution and the right plot the transverse momentum of the leading jet.
Analogously to Fig.~\ref{fig:T41selection}, the background distributions are taken directly from simulation and normalized to an integrated luminosity of 19.7\fbinv.

These three searches are individually designed to optimize the sensitivity to new physics for various signal topologies and third-generation sparticle mass hypotheses.
In Fig.~\ref{fig:evselbaseline}, the data are observed to agree with the SM background predictions, and in Figs.~\ref{fig:T41selection} and~\ref{fig:ANA_MET_plots}, with the SM background simulations, both with respect to overall normalization and shape.

\section{Interpretation\label{sec-interpretations} }

No significant deviations from the standard model predictions are observed.
Results are interpreted as limits on SMS~\cite{bib:SMS} involving the pair production of top and bottom squarks.
Alternative decays of the top squark are studied, either $\PSQt\to\PQt\PSGczDo$ or $\PSQt\to\PQc\PSGczDo$, for a variety of top squark and LSP masses.
We also study the case when there is an intermediate chargino state between the top squark and the LSP, $\PSQt\to\PQb\PSGcpmDo\to\PQb W^\pm\PSGczDo$, where the LSP is assumed to be higgsino-like and nearly degenerate in mass with the lightest chargino: $m_{\PSGcpmDo}-m_{\PSGczDo} = 5$\GeV.
For this case, we investigate different branching fractions $\mathcal{B}(\PSQt\to\PQt\PSGczDo) = 1 - \mathcal{B}(\PSQt\to\PQb\PSGcpmDo)$ for the decay of the top squark.
Finally, we study the decay of the bottom squark via the channel $\PSQb\to\PQb\PSGczDo$ for different bottom squark and LSP masses.

The CL$_\mathrm{s}$ method~\cite{aread,tjunk} is used to estimate the lower mass exclusion limits at 95\% confidence level (CL) for third-generation squark pair production.
Signal samples are produced as discussed in Section~\ref{sec-simulation}, where
the modelling of ISR within \MADGRAPH has been re-weighted to account for observed differences between data and simulation~\cite{CMS-STOP-lepton},
and a corresponding signal uncertainty assigned.
Other sources of uncertainty arise from
the jet energy scale, the
PDFs~\cite{bib:SUSYxs,bib:PDF4LHC},
and the integrated luminosity~\cite{LUMIPAS}.
Signal cross sections include re-summation of soft-gluon emission at next-to-leading-logarithmic accuracy (NLO+NLL)
~\cite{NLO_Beenakker:1996ch,NLO_Kulesza,NLO_Kulesza:2009kq,NLO_Beenakker:2009ha,NLO_Beenakker:2011fu}.
Theoretical uncertainties are dominated by PDF uncertainties, and calculations are detailed in Ref.~\cite{bib:SUSYxs}.

The multijet t-tagged analysis and the dijet b-tagged analysis both define mutually exclusive search and control regions.
Because those two analyses are statistically independent of each other, they are combined using the CL$_\mathrm{s}$ method, assuming fully correlated systematic uncertainties as nuisance parameters.
On the other hand, when choosing between the results from the monojet analysis and the dijet b-tagged analysis, the analysis with the best {\it a priori} expected limit is selected for any particular point in the bottom squark versus neutralino mass plane.
There is no overlap between the monojet and multijet t-tagged search regions and hence no special treatment is required when displaying the results of the two analyses on the same mass plane.

\begin{figure}[htb]
  \centering
    \includegraphics[width=0.56\textwidth]{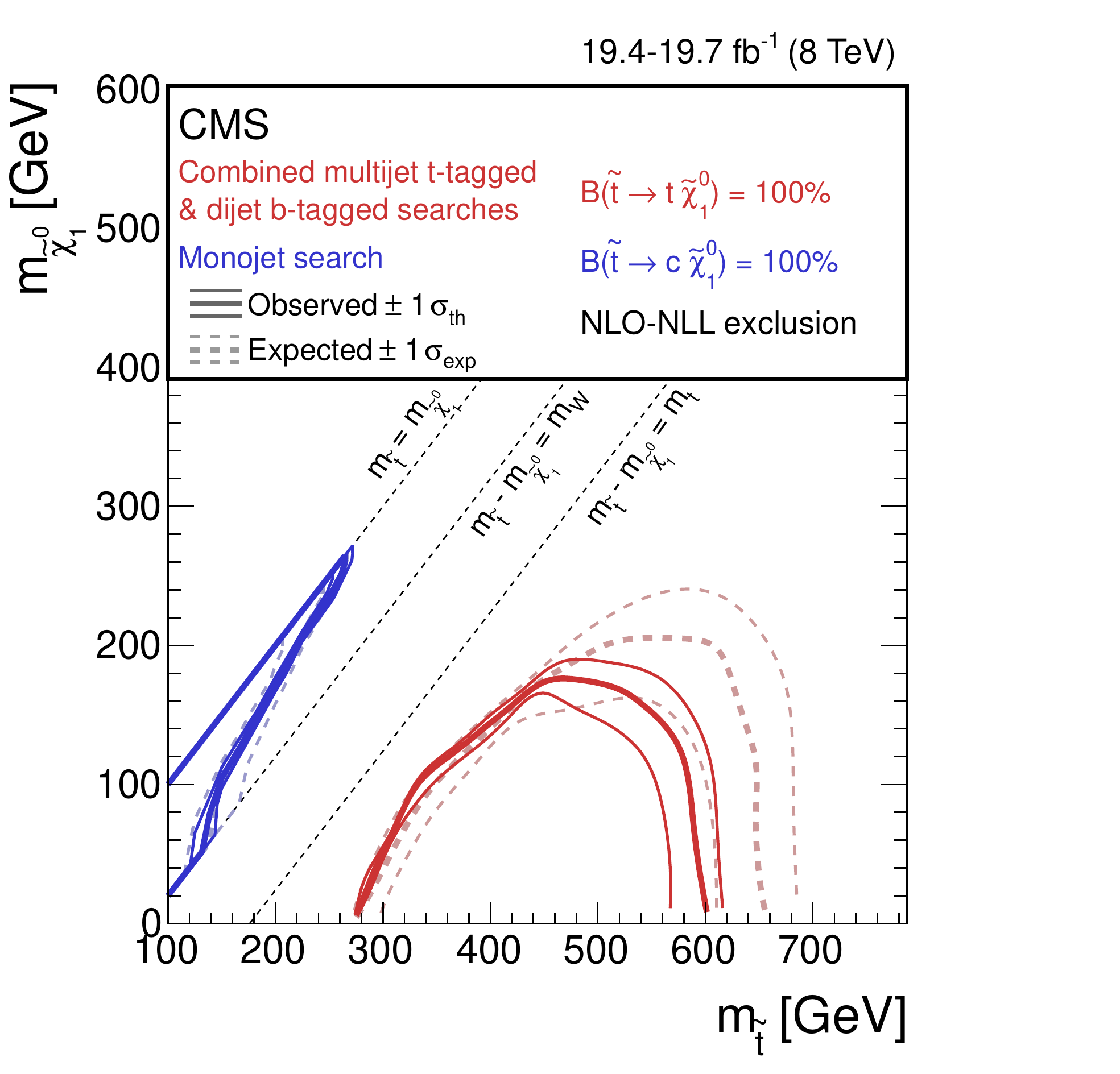}
    \caption{
    Expected and observed 95\% CL exclusion limits in the ($m_{\PSQt},m_{\PSGczDo}$) mass plane for top-squark pair production, assuming 100\% branching fraction to the decay $\PSQt\to\PQt\PSGczDo$, or, in the case of a highly compressed spectrum, to $\PSQt\to\PQc\PSGczDo$. The ${\pm}1 \sigma_{\text{exp}}$ and ${\pm}1 \sigma_{\text{th}}$ limit curves are also shown.
    The combined results from the dijet b-tagged and multijet t-tagged searches and the result from the monojet search are displayed separately.
    The dashed black diagonal lines mark the borders of the various kinematic regimes leading to different top squark decays as described in the text.
    }
    \label{fig:limitsT2tt}
\end{figure}

Figure~\ref{fig:limitsT2tt} displays the 95\% CL exclusion limits for top squark and LSP \PSGczDo masses, for either the $\ttwott$ or $\ttwocc$ simplified models, whichever is kinematically allowed.
The black diagonal dashed lines show the various kinematic regimes for top squark decay, from left to right:
 $m_{\PSQt}>m_{\PSGczDo}$ and $m_{\PSQt}-m_{\PSGczDo}<m_{\PW}$ dominated by $\ttwocc$;
 $m_{\PW} < m_{\PSQt} - m_{\PSGczDo} < m_{\PQt}$ dominated by $\ttwotb$;
 and finally $m_{\PSQt} > m_{\PQt} + m_{\PSGczDo}$, dominated by $\ttwott$.

While the multijet t-tagged search is combined with the dijet b-tagged search, the dijet b-tagged search does not contribute to the case in which the top squark decays to a top quark and the LSP with 100\% branching fraction.
This is primarily due to the jet veto requirements of the dijet b-tagged analysis, together with the high transverse momenta requirements for jets.
The observed 95\% CL exclusion limits (solid lines) are shown with the uncertainty bounds due to the uncertainty on the theoretical signal cross section (thinner, solid lines) ${\pm}1 \sigma_\text{th}$.
The expected 95\% CL exclusion limits (dashed lines) are shown with their associated uncertainty (thinner, dashed lines) ${\pm}1 \sigma_\text{exp}$.
Exclusion lines are shown in red for the combined multijet t-tagged and dijet b-tagged searches, and in blue for the monojet search.
The maximum lower limit on the top squark mass is expected to be about 620\GeV and is observed to be about 560\GeV, in the case of a massless LSP.
In the region for which $m_{\PSQt} - m_{\PSGczDo} > m_{\PW}$, the maximum lower limit on the LSP mass is expected to be just over 150\GeV
for a top squark mass of 580\GeV, and is observed to be about 180\GeV for a top squark mass of 460\GeV.
In the case of highly compressed spectra, when $m_{\PSQt}$ is close to $m_{\PSGczDo}$,
the strip below the kinematically allowed diagonal line, $m_{\PSQt}=m_{\PSGczDo}$,
and above the blue solid line is excluded, roughly up to 250\GeV in the top squark and LSP mass.

\begin{figure}[htb]
\centering
        \includegraphics[width=0.56\textwidth]{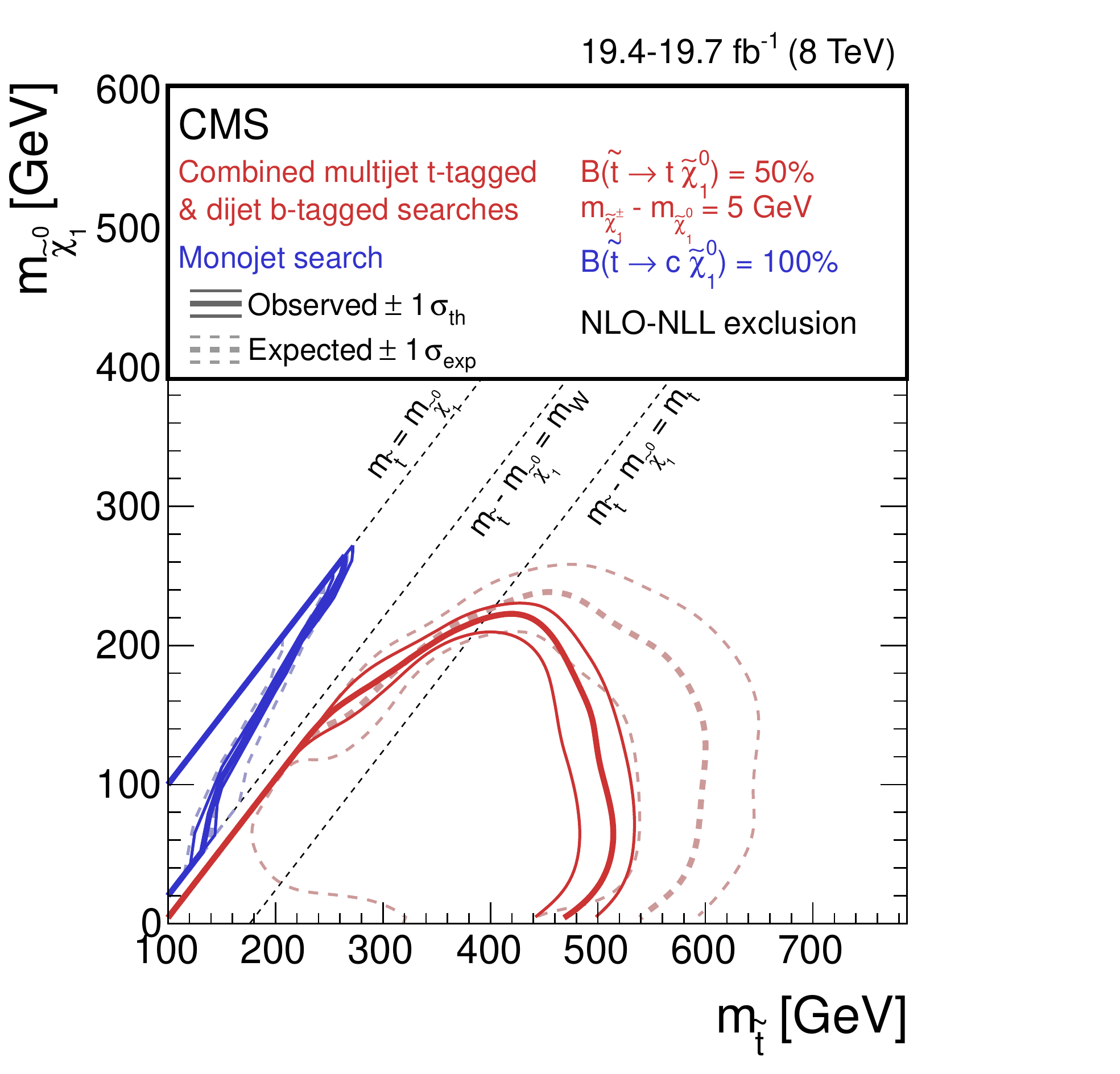}
    \caption{
    Expected and observed 95\% CL exclusion limits in the ($m_{\PSQt},m_{\PSGczDo}$)  mass plane for top-squark pair production, assuming 50\% branching fraction to the decay $\PSQt\to\PQt\PSGczDo$, with the remaining 50\% of decays proceeding via $\PSQt\to\PQb\PSGcpmDo$ and where the mass difference between the \PSGcpmDo and \PSGczDo is taken to be 5\GeV.
    In the case of a highly compressed spectrum, 100\% branching fraction to $\PSQt\to\PQc\PSGczDo$ is assumed.
    The ${\pm}1 \sigma_{\text{exp}}$ and ${\pm}1 \sigma_{\text{th}}$ limit curves are also shown.
    The combined results from the dijet b-tagged and multijet t-tagged searches and the result from the monojet search are displayed separately.
    The dashed black diagonal lines mark the borders of the various kinematic regimes leading to different top squark decays as described in the text.
    }
    \label{fig:limitsT2tb}
\end{figure}

Figure~\ref{fig:limitsT2tb} shows the same results as Fig.~\ref{fig:limitsT2tt}, except also considering a chargino \PSGcpmDo intermediate in mass to the top squark and LSP.
A 50\% branching fraction to the chargino decay channel, $\PSQt\to\PQb\PSGcpmDo$, is assumed; the other 50\% of top squarks decay via $\PSQt\to\PQt\PSGczDo$.
In this case, both the dijet b-tagged and the multijet t-tagged analyses contribute to the expected and observed limits.
The sensitivity of the dijet b-tagged analysis to this model derives from the near degeneracy of the \PSGcpmDo and \PSGczDo ($m_{\PSGcpmDo} - m_{\PSGczDo} = 5\GeV$).
The decay products of the chargino result in large missing transverse momentum together with other particles that are too soft to be reconstructed as a hard jet.
The dijet b-tagged analysis therefore primarily contributes to the moderately compressed regions, $m_{\PW} < m_{\PSQt} - m_{\PSGczDo} < m_{\PQt}$, whereas the multijet t-tagged analysis remains mainly sensitive to the bulk region.
For an LSP mass less than about 150\GeV, the lower limit on the top squark mass is expected to be about 540\GeV, and is observed to vary between about 460 and 480\GeV.
In the bulk region, the lower limit on the LSP mass is expected to be about 200\GeV for a top squark mass near 440\GeV, and is observed to be slightly lower, at about 200\GeV for a top squark mass near 400\GeV.


\begin{figure}[htb]
  \centering
    \includegraphics[width=0.56\textwidth]{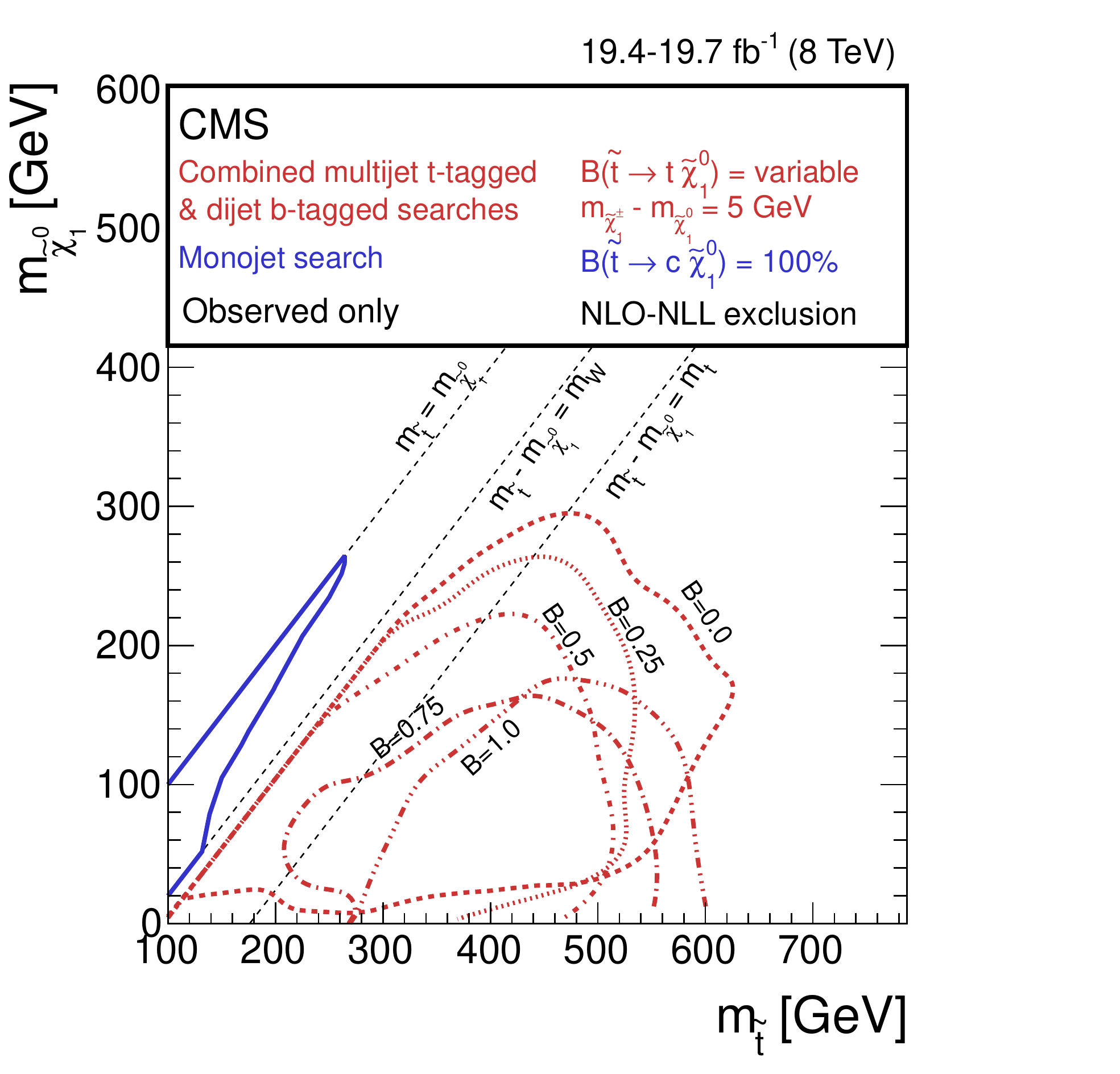}
    \caption{
Various observed 95\% CL mass exclusion limit curves for top-squark pair production, assuming different branching fractions of the two top squark decays $\PSQt\to\PQt\PSGczDo$ and $\PSQt\to\PQb\PSGcpmDo$.
The mass difference between the \PSGcpmDo and \PSGczDo is taken to be 5\GeV.
A branching fraction ($\mathcal{B}$) of 1.0 implies all decays are via $\PSQt\to\PQt\PSGczDo$, repeating the observed multijet t-tagged limit shown in Fig.~\ref{fig:limitsT2tt}, and conversely, $\mathcal{B}$ $=0.0$ implies all decays proceed through $\PSQt\to\PQb\PSGcpmDo$.
The combined results from the dijet b-tagged and multijet t-tagged searches and the result from the monojet search are displayed separately.
The dashed black diagonal lines mark the borders of the various kinematic regimes leading to different top squark decays as described in the text.
    }
    \label{fig:limitsT2tbVarBR}
\end{figure}

Figure~\ref{fig:limitsT2tbVarBR} is similar to Fig.~\ref{fig:limitsT2tb}, except that the branching fraction $\mathcal{B}(\PSQt\to\PQt\PSGczDo) = 1 - \mathcal{B}(\PSQt\to\PQb\PSGcpmDo)$ is varied between 1.0 and 0.0 in steps of 0.25.
For clarity, only the curves of the observed lower limits are displayed.
As the branching fraction $\mathcal{B}(\PSQt\to\PQt\PSGczDo)$ is reduced from 1.0 to 0.0, the dijet b-tagged analysis becomes more sensitive, excluding higher LSP higgsino masses, up to nearly 300\GeV (for a top squark mass near 480\GeV) in the case of pure $\PSQt\to\PQb\PSGcpmDo$ decays ($\mathcal{B}=0.0$).
Correspondingly, the multijet t-tagged analysis becomes less sensitive because the
events fail the $\njets \geq 5$ requirement.
For $\mathcal{B}=0.0$, the top squark mass is excluded up to 610\GeV, when the higgsino mass is about 170\GeV.

\begin{figure}[htb]
\centering
    \includegraphics[width=0.56\textwidth]{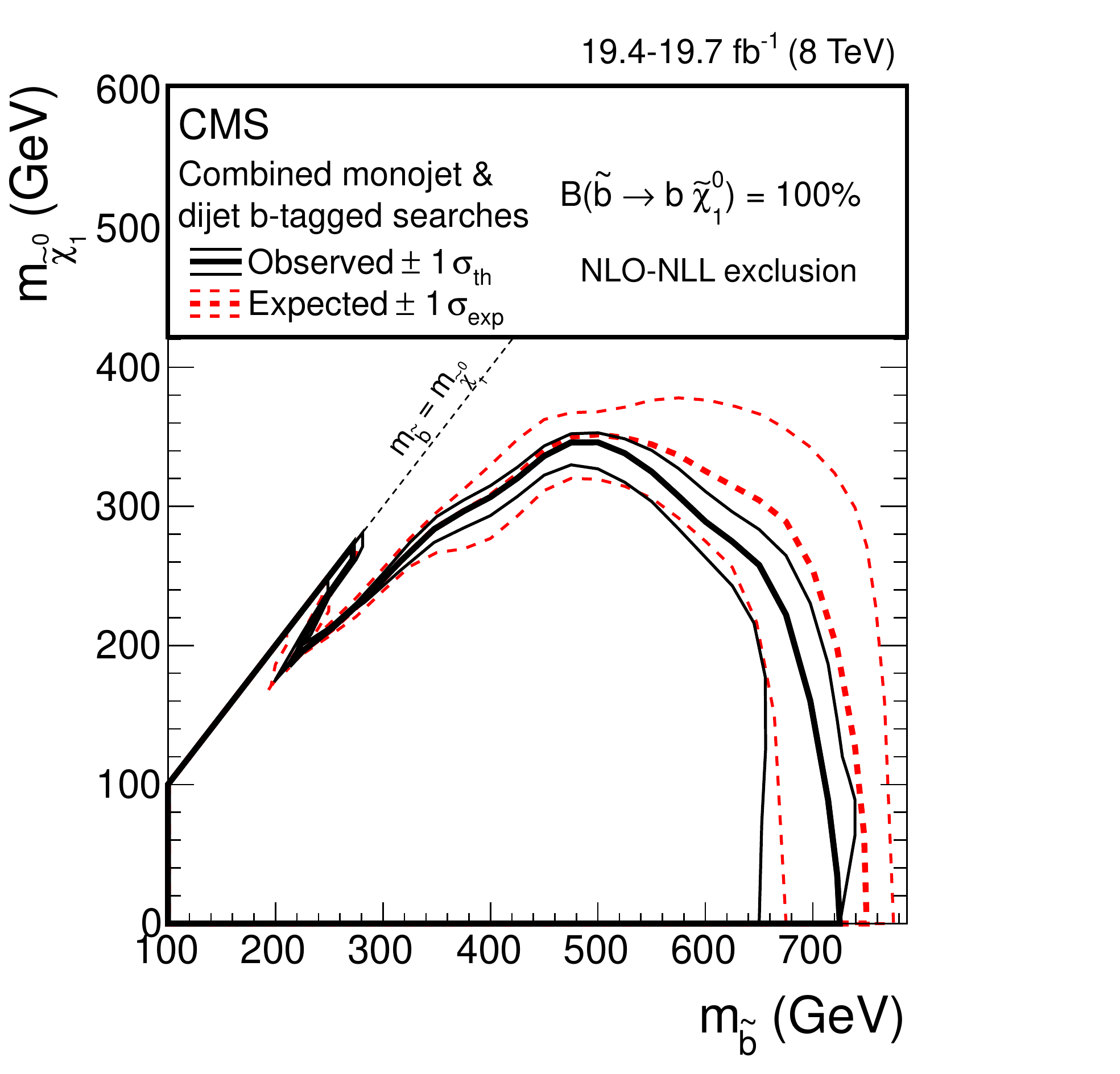}
   \caption{
       Expected and observed 95\% CL exclusion limits in the ($m_{\PSQb},m_{\PSGczDo}$) mass plane for bottom-squark pair production, assuming 100\% branching fraction to the decay $\PSQb\to \PQb \PSGczDo$.
The ${\pm}1 \sigma_{\text{exp}}$ and ${\pm}1 \sigma_{\text{th}}$ limit curves are also shown.
    The combined results from dijet b-tagged and monojet searches are displayed, taking the best expected limit from either search at each point in the mass plane.
    The black diagonal line marks the border of the kinematically allowed region.
}
    \label{fig:limitsT2bb}

\end{figure}

Finally, Fig.~\ref{fig:limitsT2bb} shows the 95\% CL exclusion limits, in the LSP mass versus bottom squark mass plane, for the simplified model $\ttwobb$ with $\mathcal{B}(\PSQb\to\PQb\PSGczDo) = 1.0$.
The black diagonal dashed line shows the allowed kinematic region for bottom squark decay, $m_{\PSQb}>m_{\PSGczDo}$.
The dijet b-tagged analysis is combined with the monojet analysis by choosing the analysis with the best expected limit at each point in the mass plane.
We expect to exclude the bottom squark up to 680\GeV for the case of a massless LSP, and are able to exclude it to 650\GeV.
In the bulk region, the four \MCT binned search regions in which $\nbjets{} = 2$ provide the best sensitivity.
We expect to exclude the LSP to 320\GeV and are able to exclude it to 330\GeV for a bottom squark mass near 480\GeV.
For mass points very close to the kinematically allowed boundary, the monojet search provides a thin strip of exclusion ranging up to about 250\GeV along the diagonal.
Otherwise, significant coverage is extended from the bulk region well into the compressed spectra region via the dijet b-tagged ISR search region with $\nbjets{} = 2$.


\section{Summary\label{sec-conclusions}}

Three complementary searches have been presented for third-generation squarks in fully
\discretionary{had-}{ronic}{hadronic} final states, corresponding to integrated luminosities of 19.4 or 19.7\fbinv
of proton-proton collision data, collected at $\sqrt{s}=8\TeV$ by the CMS experiment at the CERN LHC.
By exploiting different search techniques, the separate analyses probe similar physics processes in a variety of phase space regions, across the top/bottom squark and neutralino mass planes, including alternative SUSY scenarios in which there exists an intermediate chargino state.
No significant excesses above the predictions from the standard model are observed, and 95\% CL exclusion limits are placed on top and bottom squark masses.
A dedicated t-tagging search excludes the process $\ttwott$ with $m_{\PSQt} \le 560$\GeV for $m_{\PSGczDo}\approx 0$\GeV.
A dedicated b-tagging search excludes the process $\ttwobb$ with $m_{\PSQb} \le 650$\GeV for $m_{\PSGczDo} \approx 0$\GeV.
The process $\PSQt\PASQt\to\PQt\PSGczDo\PAQb\PSGcmDo\to\PQt\PAQb\PWm\PSGczDo\PSGczDo$ and its charge conjugate
are excluded for different branching fractions of the top
squark decay, and the two analyses are combined to exclude the
region $m_{\PSQt}\le 460$\GeV with $m_{\PSGczDo} \le 150$\GeV and
$\mathcal{B}(\PSQt\to\PQt\PSGczDo) = 0.5$ assuming the \PSGczDo and \PSGcpmDo to be
nearly mass degenerate.
A dedicated monojet search in the compressed region of
the top (bottom) squark versus LSP mass plane excludes
$\ttwocc$ ($\ttwobb$)
production for $m_{\PSQt} (m_{\PSQb}) \leq 250\GeV$ and
$m_{\PSQt} - m_{\PSGczDo} (m_{\PSQb} - m_{\PSGczDo}) <10\GeV$,
and analogously for $m_{\PSQt} (m_{\PSQb}) \leq 120\GeV$ and
$m_{\PSQt} - m_{\PSGczDo} (m_{\PSQb} - m_{\PSGczDo}) <80\GeV$.

\begin{acknowledgments}
\hyphenation{Bundes-ministerium Forschungs-gemeinschaft Forschungs-zentren} We congratulate our colleagues in the CERN accelerator departments for the excellent performance of the LHC and thank the technical and administrative staffs at CERN and at other CMS institutes for their contributions to the success of the CMS effort. In addition, we gratefully acknowledge the computing centres and personnel of the Worldwide LHC Computing Grid for delivering so effectively the computing infrastructure essential to our analyses. Finally, we acknowledge the enduring support for the construction and operation of the LHC and the CMS detector provided by the following funding agencies: the Austrian Federal Ministry of Science, Research and Economy and the Austrian Science Fund; the Belgian Fonds de la Recherche Scientifique, and Fonds voor Wetenschappelijk Onderzoek; the Brazilian Funding Agencies (CNPq, CAPES, FAPERJ, and FAPESP); the Bulgarian Ministry of Education and Science; CERN; the Chinese Academy of Sciences, Ministry of Science and Technology, and National Natural Science Foundation of China; the Colombian Funding Agency (COLCIENCIAS); the Croatian Ministry of Science, Education and Sport, and the Croatian Science Foundation; the Research Promotion Foundation, Cyprus; the Ministry of Education and Research, Estonian Research Council via IUT23-4 and IUT23-6 and European Regional Development Fund, Estonia; the Academy of Finland, Finnish Ministry of Education and Culture, and Helsinki Institute of Physics; the Institut National de Physique Nucl\'eaire et de Physique des Particules~/~CNRS, and Commissariat \`a l'\'Energie Atomique et aux \'Energies Alternatives~/~CEA, France; the Bundesministerium f\"ur Bildung und Forschung, Deutsche Forschungsgemeinschaft, and Helmholtz-Gemeinschaft Deutscher Forschungszentren, Germany; the General Secretariat for Research and Technology, Greece; the National Scientific Research Foundation, and National Innovation Office, Hungary; the Department of Atomic Energy and the Department of Science and Technology, India; the Institute for Studies in Theoretical Physics and Mathematics, Iran; the Science Foundation, Ireland; the Istituto Nazionale di Fisica Nucleare, Italy; the Ministry of Science, ICT and Future Planning, and National Research Foundation (NRF), Republic of Korea; the Lithuanian Academy of Sciences; the Ministry of Education, and University of Malaya (Malaysia); the Mexican Funding Agencies (CINVESTAV, CONACYT, SEP, and UASLP-FAI); the Ministry of Business, Innovation and Employment, New Zealand; the Pakistan Atomic Energy Commission; the Ministry of Science and Higher Education and the National Science Centre, Poland; the Funda\c{c}\~ao para a Ci\^encia e a Tecnologia, Portugal; JINR, Dubna; the Ministry of Education and Science of the Russian Federation, the Federal Agency of Atomic Energy of the Russian Federation, Russian Academy of Sciences, and the Russian Foundation for Basic Research; the Ministry of Education, Science and Technological Development of Serbia; the Secretar\'{\i}a de Estado de Investigaci\'on, Desarrollo e Innovaci\'on and Programa Consolider-Ingenio 2010, Spain; the Swiss Funding Agencies (ETH Board, ETH Zurich, PSI, SNF, UniZH, Canton Zurich, and SER); the Ministry of Science and Technology, Taipei; the Thailand Center of Excellence in Physics, the Institute for the Promotion of Teaching Science and Technology of Thailand, Special Task Force for Activating Research and the National Science and Technology Development Agency of Thailand; the Scientific and Technical Research Council of Turkey, and Turkish Atomic Energy Authority; the National Academy of Sciences of Ukraine, and State Fund for Fundamental Researches, Ukraine; the Science and Technology Facilities Council, UK; the US Department of Energy, and the US National Science Foundation.

Individuals have received support from the Marie-Curie programme and the European Research Council and EPLANET (European Union); the Leventis Foundation; the A. P. Sloan Foundation; the Alexander von Humboldt Foundation; the Belgian Federal Science Policy Office; the Fonds pour la Formation \`a la Recherche dans l'Industrie et dans l'Agriculture (FRIA-Belgium); the Agentschap voor Innovatie door Wetenschap en Technologie (IWT-Belgium); the Ministry of Education, Youth and Sports (MEYS) of the Czech Republic; the Council of Science and Industrial Research, India; the HOMING PLUS programme of the Foundation for Polish Science, cofinanced from European Union, Regional Development Fund; the Compagnia di San Paolo (Torino); the Consorzio per la Fisica (Trieste); MIUR project 20108T4XTM (Italy); the Thalis and Aristeia programmes cofinanced by EU-ESF and the Greek NSRF; and the National Priorities Research Program by Qatar National Research Fund.

 \end{acknowledgments}
\bibliography{auto_generated}

\newpage
\appendix

\section{Information for additional model testing \label{infoAdd}}

Information needed to enable additional model testing is provided here.

\begin{figure}[hbt]
  \centering
    \includegraphics[width=0.49\textwidth]{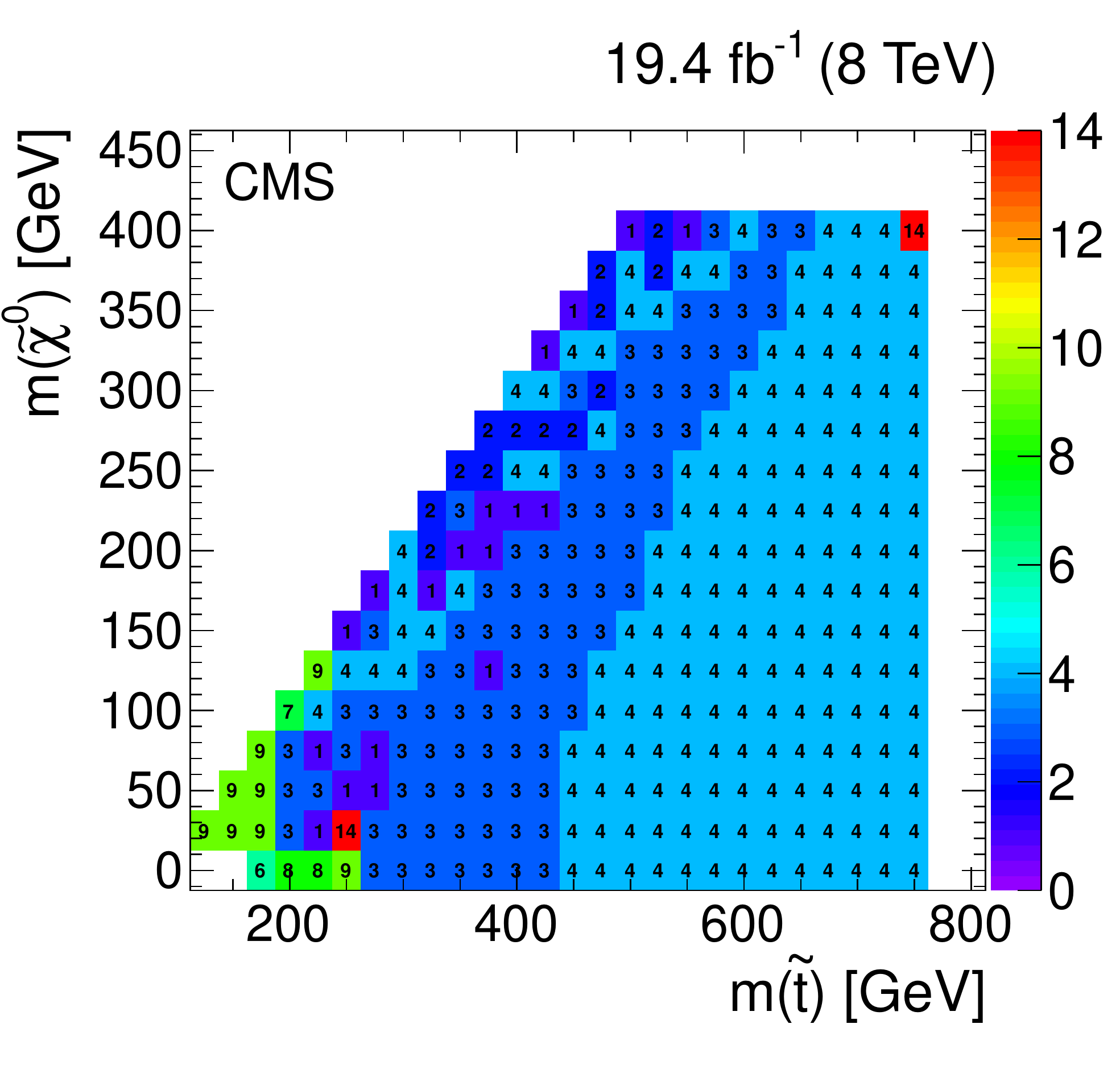}
    \includegraphics[width=0.49\textwidth]{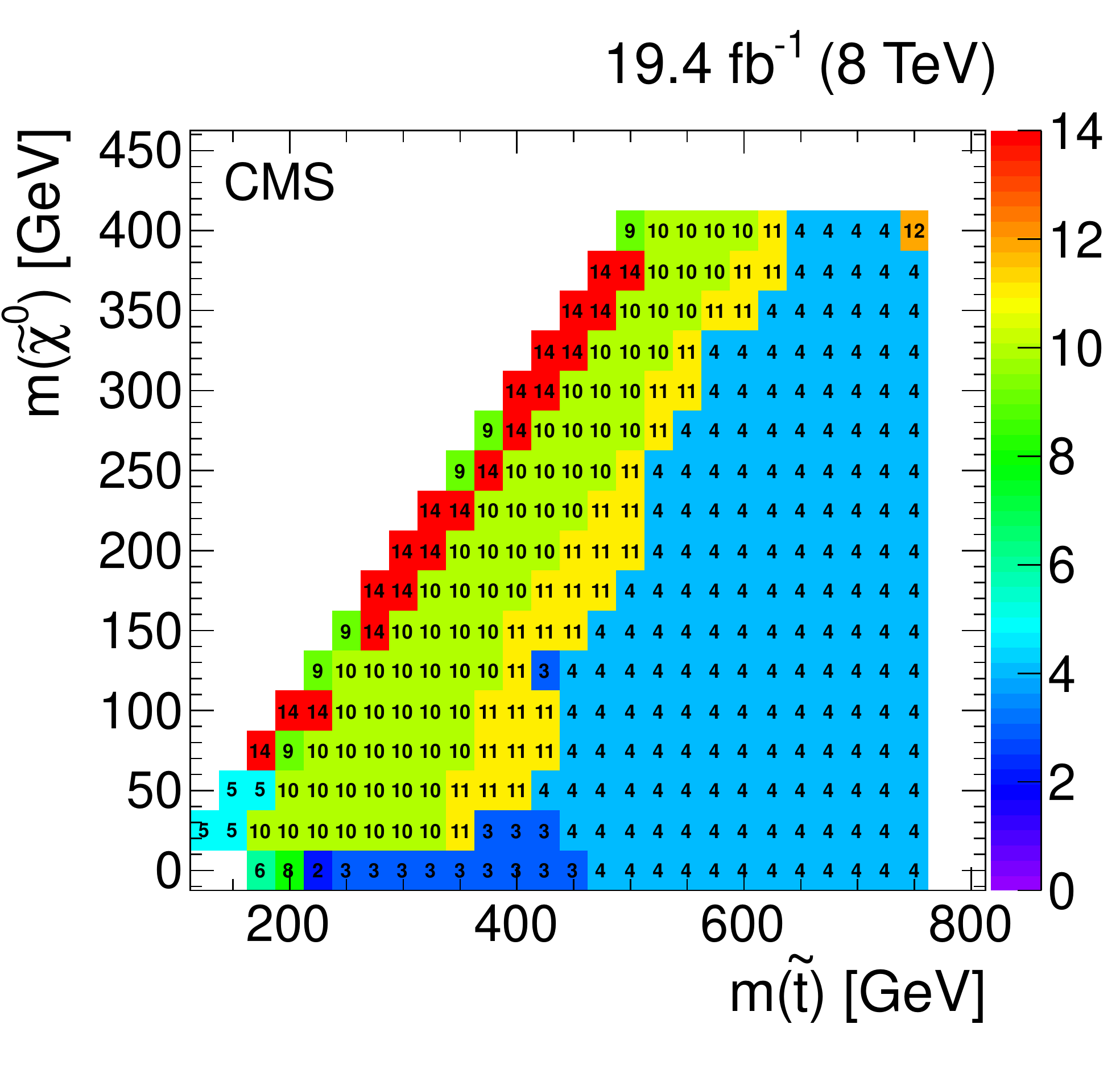}
    \includegraphics[width=0.49\textwidth]{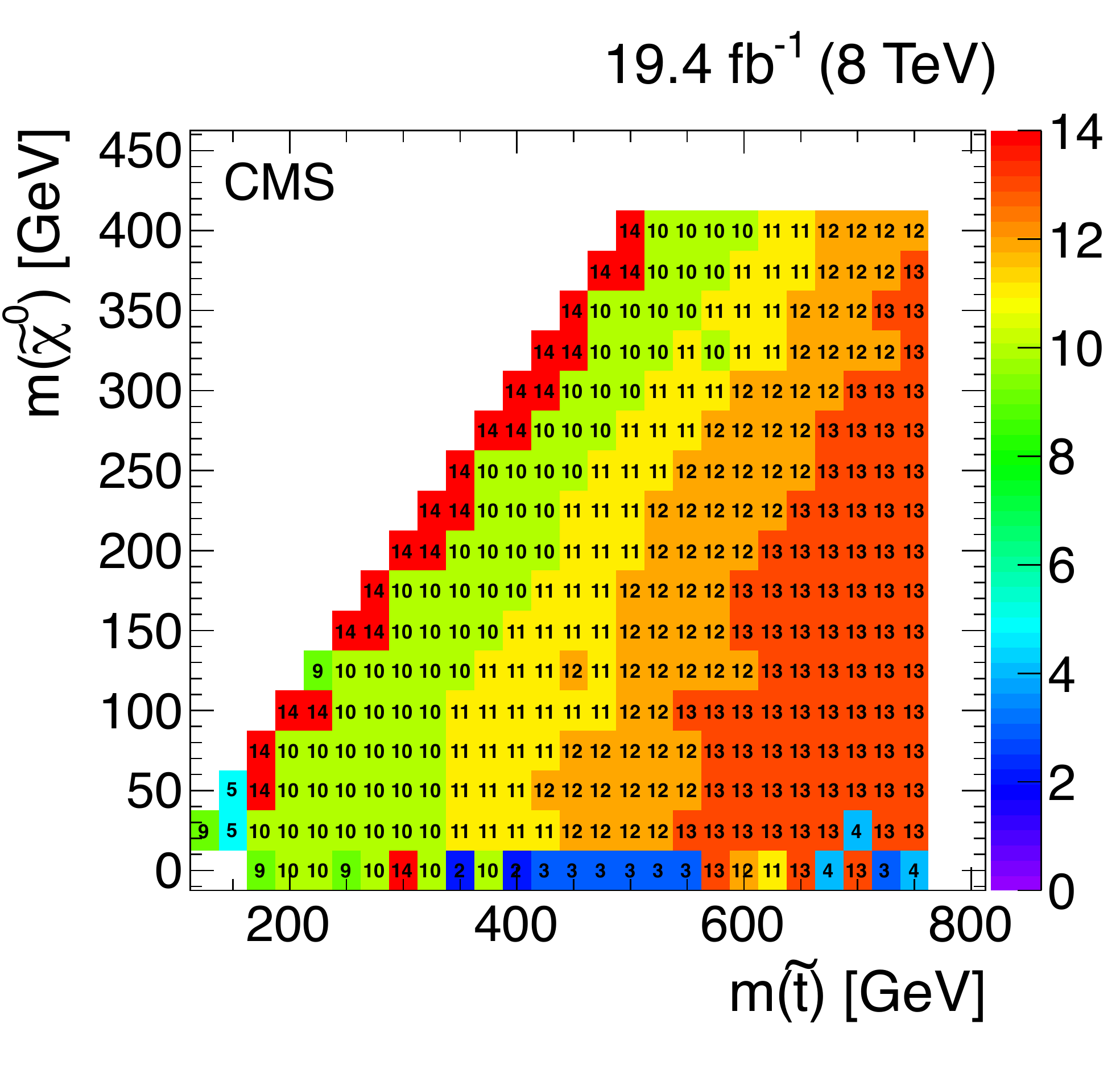}
   \caption{
   The search region from the combined multijet t-tagged and dijet b-tagged analyses resulting in the best 95\% CL expected limit is indicated by a number and shown on the colour scale in the ($m_{\PSQt},m_{\PSGczDo}$) mass plane.
   Search regions are numbered as follows.
Regions labelled 1--4 are for the multijet t-tagged analysis:
(1) $1 \PQb + 200\le \ptmiss \le 350$; (2) $1 \PQb + \ptmiss > 350$; (3) $2 \PQb+200 < \ptmiss < 350$; and (4) $2 \PQb + \ptmiss > 350$\GeV.
Numbers 5-14 are from the dijet b-tagged analysis.
Regions labelled 5-9 are for those search regions with 1 b tag:
(5) $M_\mathrm{CT} < 250$; (6) $250 < M_\mathrm{CT} < 350$; (7) $350 < M_\mathrm{CT} < 450$; (8) $M_\mathrm{CT} > 450$\GeV and (9) the ISR region.
Regions labelled 10--14 are for similar regions with 2 b tags.
In the top left-hand plot,
the optimal search regions are shown for the
best expected limit curve in Fig.~\ref{fig:limitsT2tt} in which
$\mathcal{B}(\PSQt\to\PQt\PSGczDo)=1.0$ is assumed,
i.e. all top squarks decay via $\PSQt\to\PQt\PSGczDo$.
In the top right-hand plot, the optimal search regions are shown for
the expected limit curve in Fig.~\ref{fig:limitsT2tb} in which
$\mathcal{B}(\PSQt\to\PQt\PSGczDo)=0.5$ is assumed.
In the bottom plot,
the optimal search regions are shown for
the expected limit curve shown in Fig.~\ref{fig:limitsT2tbVarBR},
i.e. all top squarks decay via $\PSQt\to\PQb\PSGcpmDo$.
}
    \label{fig:bestExplimitsT2tb}

\end{figure}

Figure~\ref{fig:bestExplimitsT2tb} shows the search regions that give the best expected 95\% CL limit for top-squark pair production assuming different branching fractions to the top squark decays $\PSQt\to\PQt\PSGczDo$ and $\PSQt\to\PQb\PSGcpmDo$ in the ($m_{\PSQt}, m_{\PSGczDo}$) mass plane.
The top-left hand plot of Fig.~\ref{fig:bestExplimitsT2tb} illustrates the optimal search region assuming $\mathcal{B}(\PSQt\to\PQt\PSGczDo)=1.0$.
Contributions only come from the multijet t-tagged analysis.
The top-right hand plot illustrates the optimal search region assuming $\mathcal{B}(\PSQt\to\PQt\PSGczDo)=0.5$.
Here, the multijet t-tagged analysis has no sensitivity when the top squark mass is less than the top mass,
so the dijet b-tagged search dominates.
At large top squark mass the multijet t-tagged analysis dominates.
The bottom plot of Fig.~\ref{fig:bestExplimitsT2tb} illustrates the optimal search region assuming
$\mathcal{B}(\PSQt\to\PQt\PSGczDo)=0.0$ (all top squarks decay via $\PSQt\to\PQb\PSGcpmDo$), in which contributions only come from the dijet b-tagged analysis.

\begin{figure}[hbt]
  \centering
    \includegraphics[width=0.45\textwidth]{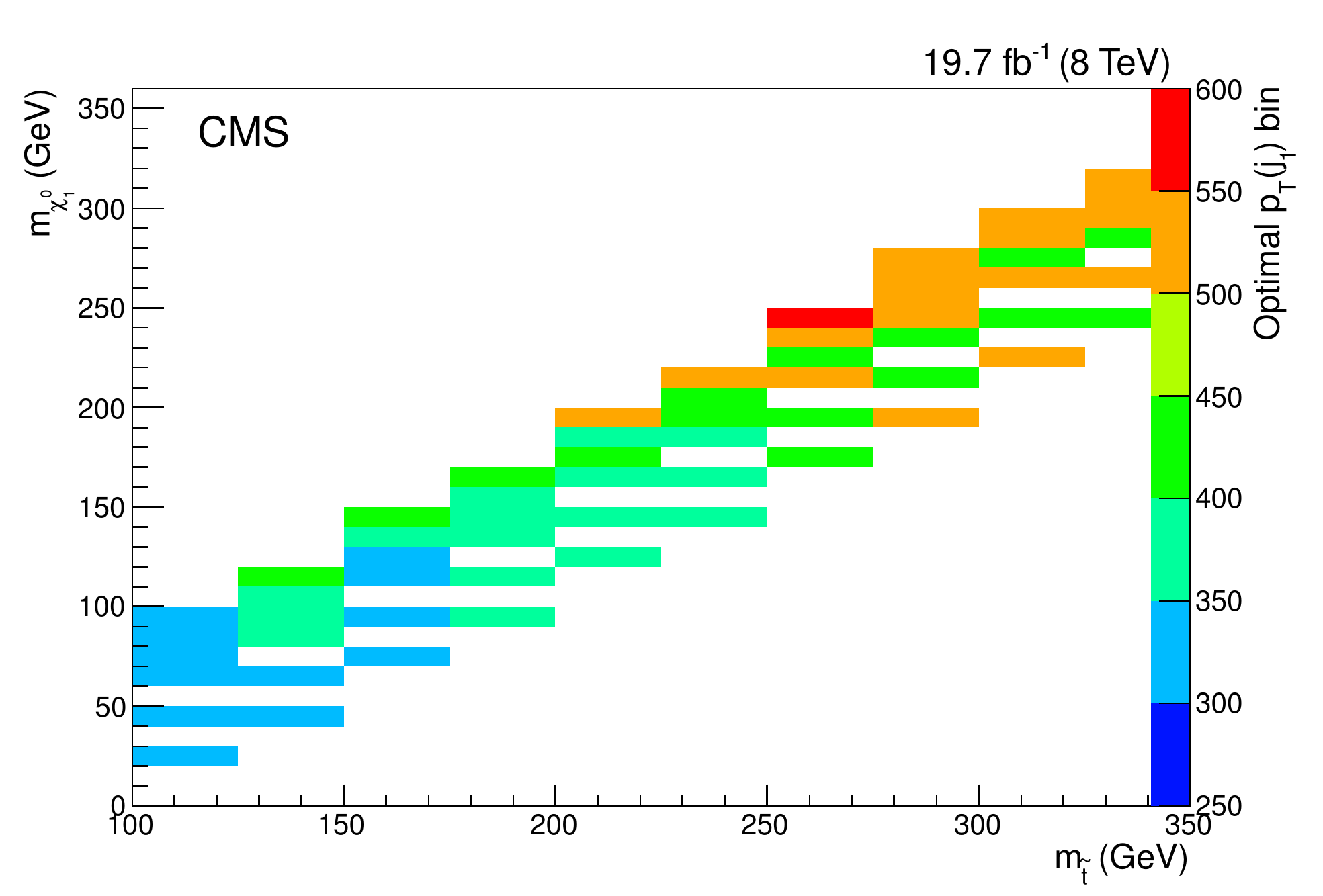}
    \includegraphics[width=0.45\textwidth]{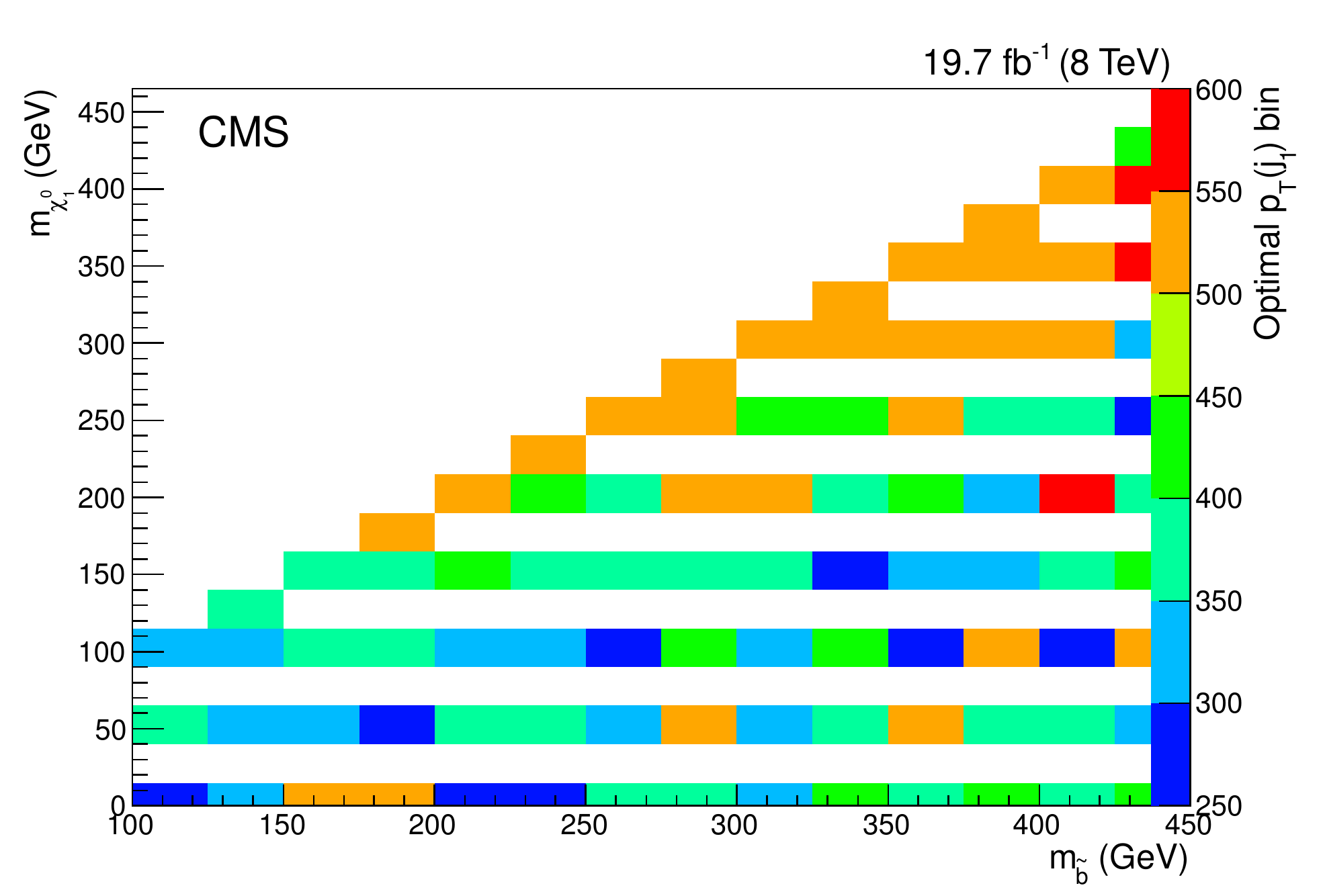}
   \caption{
   The search region resulting in the best 95\% CL expected limit for the monojet search, in the ($m_{\PSQt}, m_{\PSGczDo}$) mass plane for $\ttwocc$ (left) and the ($m_{\PSQb}, m_{\PSGczDo}$) mass plane for $\ttwobb$ (right).
}
    \label{fig:bestExplimitsMono}

\end{figure}

Figure~\ref{fig:bestExplimitsMono}
similarly shows the search regions in the monojet analysis
that give the best expected 95\% CL limit for the top squark decay $\PSQt\to\PQc\PSGczDo$ in the ($m_{\PSQt}, m_{\PSGczDo}$) mass plane, and bottom squark decay $\PSQb\to\PQb\PSGczDo$ in the ($m_{\PSQb}, m_{\PSGczDo}$) mass plane.
   Typically, harder jet thresholds are found to give better expected limits close to the diagonal, and softer jet thresholds are better for lower $\PSQt$ and $\PSQb$ masses.

\begin{figure}[hbt]
  \centering
    \includegraphics[width=0.45\textwidth]{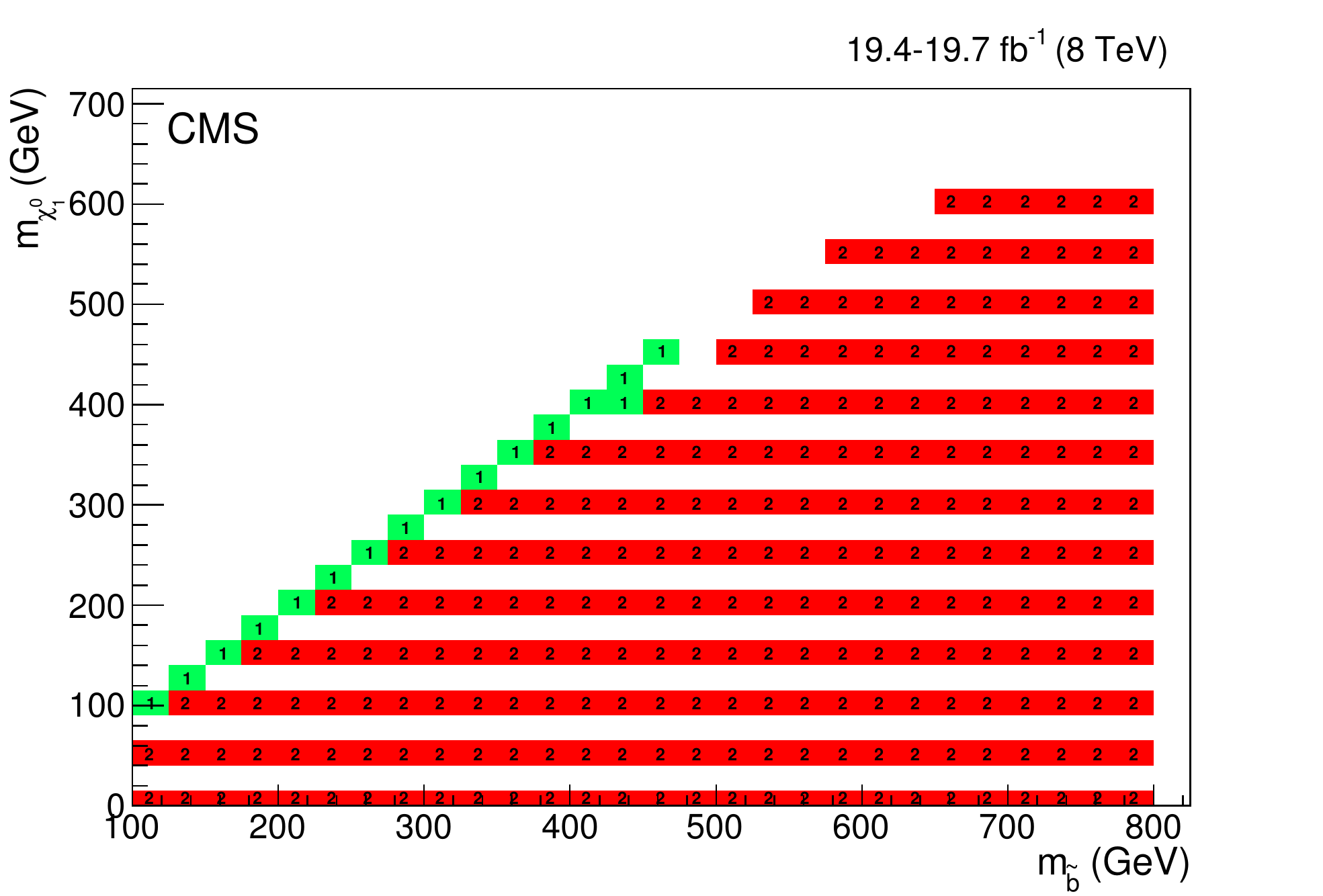}
   \caption{
   The search resulting in the best 95\% CL expected limit, for different sparticle mass hypotheses for the process $\ttwobb$.
   Bins shown in green (`1') are where the monojet search gives the best expected limit, and those shown in red (`2') are where the dijet b-tagged search gives the best expected limit.
}
    \label{fig:bestExplimitsT2bb}

\end{figure}

\begin{figure}[hbt]
  \centering
    \includegraphics[width=0.55\textwidth]{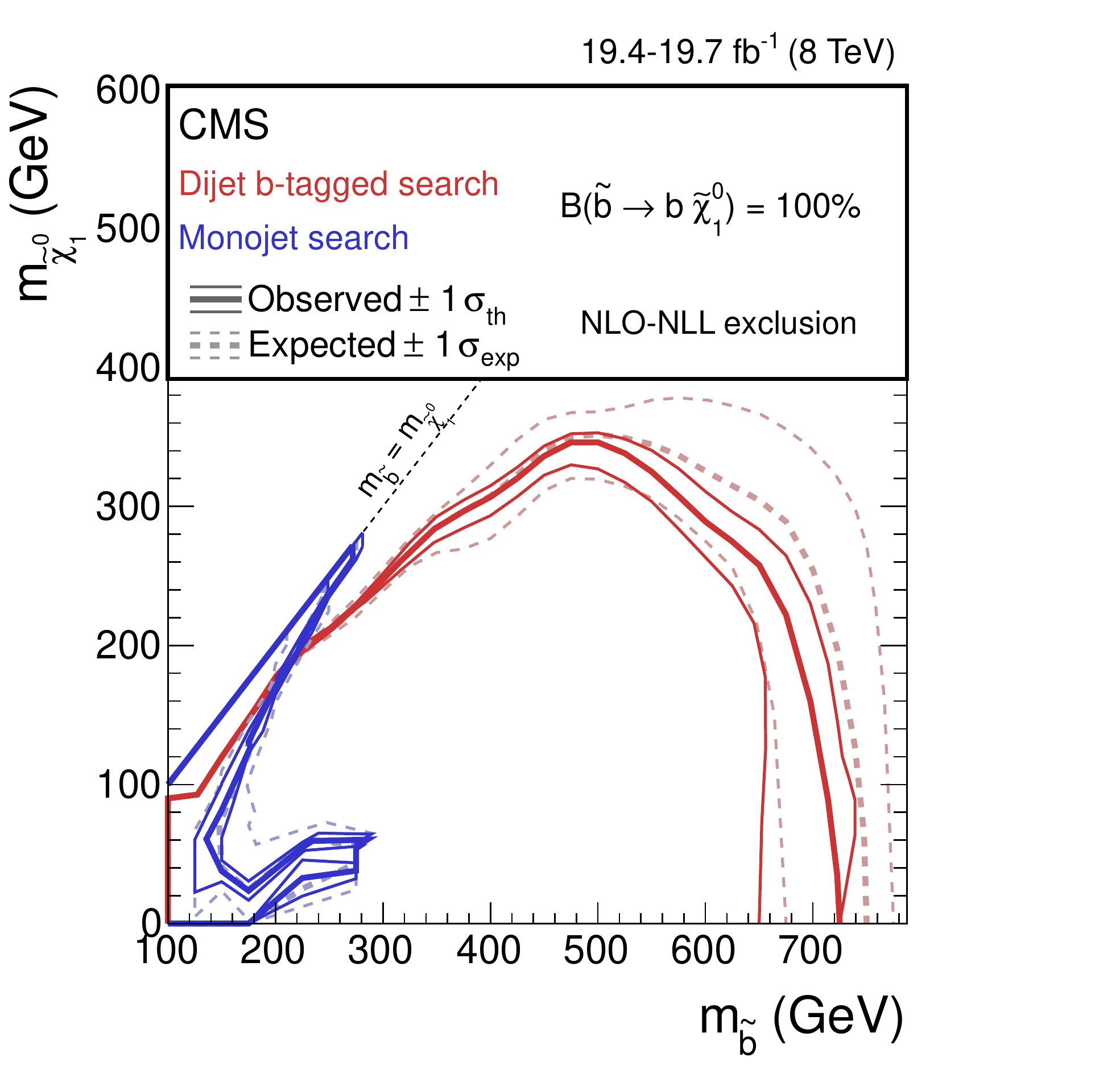}
   \caption{
Expected and observed  95\% CL exclusion limits in the mass plane ($m_{\PSQb},m_{\PSGczDo}$), for bottom-squark pair production, assuming 100\% branching fraction to the decay $\PSQb\to\PQb\PSGczDo$. The ${\pm}1\, \sigma_{\text{exp}}$ and ${\pm}1\, \sigma_{\text{th}}$ limit curves are also shown. Limits for the dijet b-tagged search and monojet search are superimposed, to illustrate where in the parameter space each search dominates.
    The black diagonal line shows the region of parameter space for bottom squark decay; $m_{\PSQb} > m_{\PSGczDo}$.
}
    \label{fig:limitsT2bb_app}

\end{figure}

Figure~\ref{fig:bestExplimitsT2bb} shows the analysis giving the best expected 95\% CL limit for bottom squark decay $\PSQb\to\PQb\PSGczDo$ in the ($m_{\PSQb}, m_{\PSGczDo}$) mass plane when the results from the monojet and dijet b-tagged analyses are combined.
Figure~\ref{fig:limitsT2bb_app} shows the individual 95\% CL exclusion limits for the dijet b-tagged and monojet searches for $\ttwobb$.
The monojet search gives the better exclusion close to the diagonal, showing the kinematic limit of mass degeneracy between the bottom squark and LSP. The dijet b-tagged search, including ``ISR'' regions, dominate in the rest of the parameter space.

Cut flow tables detailing the fraction of total events passing event selections at each step are also shown.
Table~\ref{tab:ttag_sigAcc} shows the signal acceptance $\times$ efficiency for different top squark and LSP mass hypotheses at each stage event selection in the multijet t-tagged search.
Similarly, Tables~\ref{tab:dijetbtag_sigAcc},~\ref{tab:ISR_sigAcc}, and~\ref{tab:mono_sigAcc} show the signal acceptance $\times$ efficiency for different third-generation squark and LSP mass hypotheses at each stage of the event selection in the dijet b-tagged and monojet searches.
In these tables, ``Event cleaning'' (the first of the cuts applied to events) are the requirements used to remove events with badly measured \ptmiss, beam halo, detector noise, etc.

\begin{table}[ht]
        \centering
        \topcaption{\label{tab:ttag_sigAcc} Signal acceptance $\times$ efficiency, shown in \%, for each step of the event selection of the multijet t-tagged analysis.
        Two representative mass points are shown: ($m_{\PSQt}$, $m_{\PSGczDo}$) = (500,125) and (650,25)\GeV, for $\ttwotb$ with $\mathcal{B}(\PSQt\to\PQt\PSGczDo)$ = 0.0, 0.5 and 1.0. Only statistical uncertainties are shown.}
\resizebox{\textwidth}{!}{
\begin{tabular}{l|cc|cc|cc}
        \hline
  \multirow{3}{*}{ t-tagged event selection } & \multicolumn{6}{c}{\rule{0pt}{2.5ex}$\ttwotb$} \\[0.5ex]
  \cline{2-7}
   & \multicolumn{2}{c|}{\rule{0pt}{2.5ex} $\mathcal{B}(\PSQt\to\PQt\PSGczDo)$ = 0.0 } & \multicolumn{2}{c|}{ $\mathcal{B}(\PSQt\to\PQt\PSGczDo)$ = 0.5 } & \multicolumn{2}{c}{ $\mathcal{B}(\PSQt\to\PQt\PSGczDo)$ = 1.0 } \\[0.5ex]
  \cline{2-7  }
        & (500,125)\GeV & (650,25)\GeV & (500,125)\GeV & (650,25)\GeV & (500,125)\GeV & (650,25)\GeV  \\[0.5ex]
  \hline
  Event cleaning  & $98.02 \pm 0.09$ &  $97.35 \pm 0.10$  &  $98.05 \pm 0.04$  &  $97.29 \pm 0.05$  &  $98.13 \pm 0.08$  &  $97.44 \pm 0.10$\\
  $\mu$ veto      & $87.16 \pm 0.21$ &  $76.60 \pm 0.27$  &  $79.58 \pm 0.13$  &  $74.36 \pm 0.14$  &  $72.16 \pm 0.28$  &  $72.50 \pm 0.29$\\
  $\Pe$ veto        & $83.60 \pm 0.23$ &  $64.21 \pm 0.31$  &  $68.77 \pm 0.14$  &  $59.72 \pm 0.16$  &  $55.41 \pm 0.31$  &  $55.55 \pm 0.32$\\
  $\njets(\pt > 70\GeV) \ge 2 $     &  $74.49 \pm 0.27$  &  $61.72 \pm 0.31$  &  $61.02 \pm 0.15$  &  $57.06 \pm 0.16$  &  $49.55 \pm 0.31$  &  $52.72 \pm 0.32$\\
  $\njets(\pt > 50\GeV) \ge 4 $    &  $14.99 \pm 0.22$  &  $30.17 \pm 0.29$  &  $23.86 \pm 0.13$  &  $32.76 \pm 0.15$  &  $31.16 \pm 0.29$  &  $34.55 \pm 0.31$\\
  $\njets(\pt > 30\GeV) \ge 5 $    &  $9.11 \pm 0.18$  &  $22.06 \pm 0.27$  &  $17.54 \pm 0.12$  &  $25.63 \pm 0.14$  &  $26.25 \pm 0.28$  &  $28.49 \pm 0.29$\\
  $\Delta\phi(\ptvec{}^{j},\ptvecmiss)(j=1,2,3)$  &  \multirow{2}{*}{$7.47 \pm 0.17$}  &  \multirow{2}{*}{$17.39 \pm 0.24$}  &  \multirow{2}{*}{$14.77 \pm 0.11$}  &  \multirow{2}{*}{$21.54 \pm 0.13$}  &  \multirow{2}{*}{$22.46 \pm 0.26$}  &  \multirow{2}{*}{$24.98 \pm 0.28$}\\
\hspace{0.5cm}$ \ge 0.5, 0.5, 0.3$ &&&&&&\\

  $\nbjets\ge1$  &  $6.77 \pm 0.16$  &  $15.48 \pm 0.23$  &  $13.09\pm 0.11$  &  $19.12 \pm 0.13$  &  $19.63 \pm 0.25$  &  $21.81 \pm 0.26$\\
  $\ptmiss > 200\GeV$  &  $4.79 \pm 0.13$  &  $11.84 \pm 0.21$  &  $8.54\pm 0.09$  &  $15.02 \pm 0.11$  &  $12.21 \pm 0.20$  &  $17.60 \pm 0.24$\\

Top tagger, $\MTTwo \ge 300$\GeV , & \multirow{3}{*}{$0.96 \pm 0.06$}  &  \multirow{3}{*}{$1.89 \pm 0.09$}  &  \multirow{3}{*}{$3.01 \pm 0.05$}  &  \multirow{3}{*}{$5.15 \pm 0.07$}  &  \multirow{3}{*}{$4.87 \pm 0.13$}  &  \multirow{3}{*}{$8.37 \pm 0.18$}  \\
       $(0.5\MTt+\MTb)$  &&&&&& \\
       \hspace{0.5cm}$\ge500$\GeV  &&&&&& \\
         \hline
  $ \nbjets=1$,   &  \multirow{2}{*}{$0.18 \pm 0.03$}  &  \multirow{2}{*}{$0.31 \pm 0.04$}  &  \multirow{2}{*}{$0.63 \pm 0.02$}  &  \multirow{2}{*}{$0.67 \pm 0.03$}  &  \multirow{2}{*}{$1.19 \pm 0.07$}  &  \multirow{2}{*}{$1.06 \pm 0.07$}\\
  \hspace{0.5cm}$\ptmiss\in[200, 350]\GeV$ &&&&&& \\
  $ \nbjets=1$,  &  \multirow{2}{*}{$0.23 \pm 0.03$}  &  \multirow{2}{*}{$0.52 \pm 0.05$}  &  \multirow{2}{*}{$0.58 \pm 0.02$}  &  \multirow{2}{*}{$1.38 \pm 0.04$}  &  \multirow{2}{*}{$0.93 \pm 0.06$}  &  \multirow{2}{*}{$2.49 \pm 0.10$}\\
  \hspace{0.5cm}$\ptmiss \ge 350\GeV$ &&&&&& \\
  $ \nbjets\ge2$,  &  \multirow{2}{*}{$0.32 \pm 0.04$}  &  \multirow{2}{*}{$0.46 \pm 0.04$} &  \multirow{2}{*}{$0.99 \pm 0.03$}  &  \multirow{2}{*}{$0.99 \pm 0.03$}  &  \multirow{2}{*}{$1.64 \pm 0.08$}  &  \multirow{2}{*}{$1.34 \pm 0.07$}\\
  \hspace{0.5cm}$\ptmiss\in[200, 350]\GeV$ &&&&&& \\
  $ \nbjets\ge2$,   &  \multirow{2}{*}{$0.23 \pm 0.03$}  &  \multirow{2}{*}{$0.61 \pm 0.05$}  &  \multirow{2}{*}{$0.80 \pm 0.03$}  &  \multirow{2}{*}{$2.11 \pm 0.05$}  &  \multirow{2}{*}{$1.11 \pm 0.07$}  &  \multirow{2}{*}{$3.48 \pm 0.12$}\\
  \hspace{0.5cm}$\ptmiss \ge 350\GeV$ &&&&&& \\
\hline
\end{tabular}
}
\end{table}

\begin{table}[htb]
\centering
\topcaption{\label{tab:dijetbtag_sigAcc} Signal acceptance $\times$ efficiency, shown in \%, for each step of the event selection of the dijet b-tagged analysis, for the \MCT search regions.
        Two representative mass points are shown; ($m_{\PSQb}$, $m_{\PSGczDo}$) = (275,250) and (700,100)\GeV for $\ttwobb$ with $\mathcal{B}(\PSQb\to\PQb\PSGczDo)$ = 1.0, and ($m_{\PSQt}$, $m_{\PSGczDo}$) = (250,125) and (500,125)\GeV for $\ttwotb$ with $\mathcal{B}(\PSQt\to\PQt\PSGczDo)$ = 0.0, 0.5. The dijet b-tagged analysis has no sensitivity to the $\mathcal{B}(\PSQt\to\PQt\PSGczDo)$ = 1.0 case so it is not shown.
        Only statistical uncertainties are shown.}
\resizebox{\textwidth}{!}{
\begin{tabular}{l|cc|cc|cc}
        \hline
\multirow{3}{*}{\MCT event selection } & \multicolumn{2}{c|}{\rule{0pt}{3ex}$\ttwobb$} & \multicolumn{4}{c}{$\ttwotb$} \\[0.5ex] \cline{2-7}
 & \multicolumn{2}{c|}{\rule{0pt}{2.5ex} $\mathcal{B}(\PSQb\to\PQb\PSGczDo)$ = 1.0 } & \multicolumn{2}{c|}{ $\mathcal{B}(\PSQt\to\PQt\PSGczDo)$ = 0.0 } & \multicolumn{2}{c}{ $\mathcal{B}(\PSQt\to\PQt\PSGczDo)$ = 0.5 } \\[0.5ex]\cline{2-7  }
                                         & (275,250)\GeV & (700,100)\GeV & (250,125)\GeV & (500,125)\GeV & (250,125)\GeV & (500,125)\GeV  \\[0.5ex]\hline

 Event cleaning                          & 91.50 $\pm$0.11 & 92.92 $\pm$0.13  &  91.63$\pm$0.08   & 93.54$\pm$0.18 & 91.63$\pm$0.06 & 92.87$\pm$0.07  \\
 $\njets(\pt > 70\GeV)=2$             & 2.01$\pm$0.30 & 37.62 $\pm$0.38  &  3.22$\pm$0.17    & 31.71$\pm$0.29 & 15.86$\pm$0.11 & 23.73 $\pm$0.17 \\
 $\Pe,\mu$ veto                            & 1.97$\pm$0.28 & 36.01$\pm$0.36   &  3.19$\pm$0.17    & 31.60$\pm$0.29 & 13.62$\pm$0.12 & 19.09 $\pm$0.15 \\
 IsoTrk Veto                             & 1.87$\pm$0.26 & 33.64$\pm$0.34   &  2.92$\pm$0.15    & 30.63$\pm$0.28 & 12.08$\pm$0.12 & 16.62$\pm$ 0.14 \\
 $3^{rd}$ Jet Veto                       & 1.24$\pm$0.21 & 23.1$\pm$0.22    &  2.34$\pm$0.13    & 21.24$\pm$0.33 & 6.57 $\pm$0.08 & 7.32$\pm$0.09   \\
 $\HT >250\GeV$                         & 0.56$\pm$0.14 & 22.13$\pm$0.20    &  1.13$\pm$0.10    & 20.91$\pm$0.32 & 2.76 $\pm$0.04& 6.41$\pm$0.08    \\
 $\ptmiss >175\GeV$                        & 0.13$\pm$0.10 & 16.41$\pm$0.18    &  0.92$\pm$0.09    & 17.83$\pm$0.26 & 0.55$\pm$0.03 & 4.81$\pm$0.07    \\
 $\MT >200\GeV$                         & 0.074$\pm$0.009 & 14.4 $\pm$0.15    &  0.86$\pm$0.08    & 16.24$\pm$0.25 & 0.25$\pm$0.03 & 3.9  $\pm$0.05   \\
 $\Delta\phi(\PQb_{1},\PQb_{2}) <$ 2.5         & 0.071$\pm$0.009 & 13.0$\pm$0.14     &  0.86$\pm$0.08    & 16.21$\pm$0.25 & 0.86$\pm$0.04 & 3.49$\pm$0.06    \\
 $\nbjets=$ 1                            & 0.039$\pm$0.007 & 6.42 $\pm$0.09    &  0.033$\pm$0.001  & 5.92$\pm$0.13  & 0.19$\pm$0.02 & 2.89$\pm$0.04    \\
 $\nbjets=$ 2                            & 0.002$\pm$0.001 & 5.2$\pm$0.08      &  0.002$\pm$0.001  & 6.49$\pm$0.16  & 0.02$\pm$0.001 & 1.56 $\pm$0.03  \\
 \hline
 $\nbjets=$ 1,        & \multirow{2}{*}{0.027$\pm$0.005 } & \multirow{2}{*}{1.74$\pm$0.08}     &  \multirow{2}{*}{0.017$\pm$0.001 } & \multirow{2}{*}{0.68$\pm$0.03}  & \multirow{2}{*}{0.14$\pm$0.01  } & \multirow{2}{*}{0.92$\pm$ 0.04} \\
  \hspace{0.5cm}$\MCT< 250\GeV$ &&&&&& \\
 $\nbjets=$ 1, & \multirow{2}{*}{0.011$\pm$0.004 } & \multirow{2}{*}{2.82$\pm$0.09}     &  \multirow{2}{*}{0.015$\pm$0.001 } & \multirow{2}{*}{1.12$\pm$0.07}  & \multirow{2}{*}{0.04$\pm$0.003 } & \multirow{2}{*}{1.28$\pm$0.05 } \\
  \hspace{0.5cm}$\MCT\in[250,350]\GeV$  &&&&&& \\
 $\nbjets=$ 1,  & \multirow{2}{*}{0.001$\pm$0.0006} & \multirow{2}{*}{1.73$\pm$0.08}     &  \multirow{2}{*}{0.001$\pm$0.0006} & \multirow{2}{*}{1.56$\pm$0.08}  & \multirow{2}{*}{0.01$\pm$0.001 } & \multirow{2}{*}{0.61$\pm$0.03 } \\
  \hspace{0.5cm}$\MCT\in[250,350]\GeV$ &&&&&& \\
 $\nbjets=$ 1,       & \multirow{2}{*}{0.00$\pm$0.00 } & \multirow{2}{*}{0.13$\pm$0.06}     &  \multirow{2}{*}{0.00$\pm$0.00   } & \multirow{2}{*}{2.54$\pm$0.09}  & \multirow{2}{*}{0.00$\pm$0.00  } & \multirow{2}{*}{0.06$\pm$0.004} \\
   \hspace{0.5cm}$\MCT > 450\GeV$ &&&&&& \\
 $\nbjets=$ 2,        & \multirow{2}{*}{0.001$\pm$0.005} & \multirow{2}{*}{1.29$\pm$0.06}     &  \multirow{2}{*}{0.002$\pm$0.0006} & \multirow{2}{*}{0.44$\pm$0.02}  & \multirow{2}{*}{0.02$\pm$0.001 } & \multirow{2}{*}{0.50$\pm$0.03 } \\
  \hspace{0.5cm}$\MCT < 250\GeV$ &&&&&& \\
 $\nbjets=$ 2,  & \multirow{2}{*}{0.001$\pm$0.005 } & \multirow{2}{*}{1.98$\pm$0.09}     &  \multirow{2}{*}{0.001$\pm$0006  } & \multirow{2}{*}{1.21$\pm$0.07}  & \multirow{2}{*}{0.002$\pm$0.001} & \multirow{2}{*}{0.72$\pm$0.04 } \\
   \hspace{0.5cm}$\MCT\in[250,350]\GeV$ &&&&&& \\
 $\nbjets=$ 2,  & \multirow{2}{*}{0.00$\pm$0.00 } & \multirow{2}{*}{1.52$\pm$0.08}     &  \multirow{2}{*}{0.00$\pm$0.00   } & \multirow{2}{*}{1.58$\pm$0.08}  & \multirow{2}{*}{0.0$\pm$0.00  }  & \multirow{2}{*}{0.38$\pm$0.04 } \\
  \hspace{0.5cm}$\MCT\in[250,350]\GeV$ &&&&&& \\
 $\nbjets=$ 2,         & \multirow{2}{*}{0.00$\pm$0.00 } & \multirow{2}{*}{0.15$\pm$0.03}     &  \multirow{2}{*}{0.00$\pm$0.00   } & \multirow{2}{*}{3.21$\pm$0.11}  & \multirow{2}{*}{0.00$\pm$0.00 }  & \multirow{2}{*}{0.00$\pm$0.00 } \\
  \hspace{0.5cm}$\MCT >450\GeV$  &&&&&& \\ \hline
\end{tabular}
}
\end{table}
\begin{table}[ht]
\centering
\topcaption{\label{tab:ISR_sigAcc} Signal acceptance $\times$ efficiency, shown in \%, for each step of the event selection of the dijet b-tagged analysis, for the ISR search regions.
        Two representative mass points are shown; ($m_{\PSQb}$, $m_{\PSGczDo}$) = (275,250) and (700,100)\GeV, for $\ttwobb$ with $\mathcal{B}(\PSQb\to\PQb\PSGczDo)$ = 1.0, and ($m_{\PSQt}$, $m_{\PSGczDo}$) = (250,125) and (500,125)\GeV, for $\ttwotb$ with $\mathcal{B}(\PSQt\to\PQt\PSGczDo)$ = 0.0, 0.5. The dijet b-tagged analysis has no sensitivity to the $\mathcal{B}(\PSQt\to\PQt\PSGczDo)$ = 1.0 case so it is not shown.
        Only statistical uncertainties are shown.}
\resizebox{\textwidth}{!}{
\begin{tabular}{l|cc|cc|cc}
        \hline
\multirow{3}{*}{ISR event selection } & \multicolumn{2}{c|}{\rule{0pt}{3ex} $\ttwobb$} & \multicolumn{4}{c}{$\ttwotb$} \\[0.5ex] \cline{2-7}
 & \multicolumn{2}{c|}{\rule{0pt}{2.5ex}  $\mathcal{B}(\PSQb\to\PQb\PSGczDo)$ = 1.0 } & \multicolumn{2}{c|}{ $\mathcal{B}(\PSQt\to\PQt\PSGczDo)$ = 0.0 } & \multicolumn{2}{c}{ $\mathcal{B}(\PSQt\to\PQt\PSGczDo)$ = 0.5 } \\[0.5ex]\cline{2-7  }
                            & (275,250)\GeV & (700,100)\GeV &(250,125)\GeV & (500,125)\GeV & (250,125)\GeV & (500,125)\GeV  \\[0.5ex]\hline
Event cleaning                                    & 94.40$\pm$0.07 & 96.50$\pm$0.17    & 94.62$\pm$0.14 & 95.28$\pm$0.15   & 94.61$\pm$0.08 & 95.34$\pm$0.08   \\
$\njets(\pt > 30\GeV) ==3 $        & 9.98$\pm$0.31 & 28.90$\pm$0.29    & 28.57$\pm$0.26 &  31.52$\pm$0.29  & 36.72$\pm$0.25 &  29.61 $\pm$0.21 \\
$1^{st},2^{nd}$ jets $(\pt > 70\GeV)$    & 2.50$\pm$0.14 & 27.60$\pm$0.28     & 17.21$\pm$0.21 & 27.90$\pm$ 0.26   & 14.91$\pm$0.14 & 21.63 $\pm$0.19  \\
$\Pe,\mu$ and IsoTrk veto                         & 2.40$\pm$0.14 & 27.50$\pm$0.28     & 15.84$\pm$0.20 & 24.08$\pm$0.23   & 9.42$\pm$0.11 & 14.48$\pm$ 0.16   \\
$\HT >250\GeV$                           & 1.40$\pm$0.09 & 27.20$\pm$0.28     & 9.46$\pm$0.12 & 23.13 $\pm$0.23   & 7.58$\pm$0.09 & 10.21$\pm$0.12    \\
$\ptmiss >175\GeV$                                & 0.90$\pm$0.07 & 22.10$\pm$0.24     & 0.73$\pm$0.10 & 12.10$\pm$0.17     & 0.43$\pm$0.04 & 4.77$\pm$0.07     \\
Min($\Delta\phi(j_{1,2,3},\ptmiss))>0.5$           & 0.36$\pm$0.04 & 17.30$\pm$0.19    & 0.58$\pm$0.08 & 9.72$\pm$0.15      & 0.36$\pm$0.04 & 3.1$\pm$0.06     \\
\hline
$\nbjets=$ 1 ($2^{nd}$ or $3^{rd}$ jets)        & 0.13$\pm$0.01 & 3.4$\pm$0.12     & 0.29$\pm$0.06 & 2.10 $\pm$0.09     & 0.090$\pm$0.008 & 1.26$\pm$0.02    \\
$\nbjets=$ 2 ($2^{nd}$ and $ 3^{rd}$ jets)      & 0.0080$\pm$0.0002 & 0.43$\pm$0.02 & 0.09$\pm$0.04 & 0.29$\pm$0.07     & 0.015$\pm$0.001 & 0.43$\pm$0.03   \\
$\nbjets=$ 1, $\pt^{\text{non-b}}> 250\GeV$         & 0.087$\pm$0.003 & 2.60$\pm$0.09   & 0.14$\pm$0.09 & 1.44$\pm$0.08     & 0.070$\pm$0.008& 0.87$\pm$0.06     \\
$\nbjets=$ 2, $\pt^{\text{non-b}}> 250\GeV$         & 0.0060$\pm$0002 & 0.37$\pm$0.02   & 0.050$\pm$0.007 & 0.24$\pm$0.04    & 0.012$\pm$0.001 & 0.16$\pm$0.02   \\\hline
\end{tabular}
}
\end{table}

\begin{table}[h]
\centering
\topcaption{\label{tab:mono_sigAcc} Signal acceptance $\times$ efficiency, shown in \%, for each step of the event selection of the monojet analysis.
        Two representative mass points are shown; ($m_{\PSQb}$, $m_{\PSGczDo}$) = (250,240) and (150,50)\GeV for $\ttwobb$ where $\mathcal{B}(\PSQb\to\PQb\PSGczDo)$ = 1.0,
        and ($m_{\PSQt}$, $m_{\PSGczDo}$) = (250,240)\GeV and (200,120) for $\ttwocc$, where $\mathcal{B}(\PSQt\to\PQc\PSGczDo)$ = 1.0.
        Only statistical uncertainties are shown.}
{
\begin{tabular}{l|cc|cc}
        \hline

\multirow{3}{*}{Monojet event selection } & \multicolumn{2}{c|}{\rule{0pt}{3ex} $\ttwobb$} & \multicolumn{2}{c}{$\ttwocc$} \\[0.5ex] \cline{2-5}
 & \multicolumn{2}{c|}{\rule{0pt}{2.5ex}  $\mathcal{B}(\PSQb\to\PQb\PSGczDo)$ = 1.0 } & \multicolumn{2}{c}{ $\mathcal{B}(\PSQt\to\PQc\PSGczDo)$ = 1.0 }\\[0.5ex]\cline{2-5}
 & (250, 240)\GeV & (150, 50)\GeV& (250, 240)\GeV & (200, 120)\GeV   \\ [0.5ex]\hline

Event cleaning                     &   98.61 $\pm$   0.24 &   98.79 $\pm$  0.02 &   97.54 $\pm$   0.14  &   99.21 $\pm$  0.03 \\
$\ptmiss >$~200\GeV               &   7.41 $\pm$   0.49  &   2.37 $\pm$  0.02  &    7.17 $\pm$   0.25  &    4.29 $\pm$  0.06 \\
Noisy events                    &   6.90 $\pm$   0.47  &   2.22 $\pm$  0.02  &    6.68 $\pm$   0.24  &    4.01 $\pm$  0.06 \\
$\pt(\jet_1)>110$\GeV          &   6.58 $\pm$   0.46  &   2.08 $\pm$  0.02  &    6.35 $\pm$   0.23  &    3.71 $\pm$  0.06 \\
$\njets<3$                     &   5.78 $\pm$   0.44  &   1.39 $\pm$  0.02  &    5.56 $\pm$   0.22  &    2.30 $\pm$  0.04 \\
$\Delta \phi(\jet_1,\jet_2)<2.5$&   5.58 $\pm$   0.43  &   1.170 $\pm$  0.015  &    5.36 $\pm$   0.21  &    1.96 $\pm$  0.04 \\
$\mu$ veto                      &   5.57 $\pm$   0.43  &   1.170 $\pm$  0.015  &    5.36 $\pm$   0.21  &    1.96 $\pm$  0.04 \\
$\Pe$ veto                        &   5.57 $\pm$   0.43  &   1.160 $\pm$  0.015  &    5.36 $\pm$   0.21  &    1.96 $\pm$  0.04 \\
$\tauh$ veto                 &   5.52 $\pm$   0.43  &   1.14 $\pm$  0.015  &    5.30 $\pm$   0.21  &    1.93 $\pm$  0.04 \\
\hline
\ptmiss \& $\pt(\jet_1) > 250$\GeV&   2.08 $\pm$   0.27  &  0.222 $\pm$ 0.006  &    2.04 $\pm$   0.13  &    0.42 $\pm$  0.02 \\
$\pt(\jet_1) > 300$\GeV        &   1.32 $\pm$   0.21  &  0.122 $\pm$ 0.005  &    1.32 $\pm$   0.11  &    0.25 $\pm$  0.01 \\
$\pt(\jet_1) > 350$\GeV        &   0.80 $\pm$   0.17  &  0.058 $\pm$ 0.003  &    0.81 $\pm$  0.08  &    0.13 $\pm$  0.01 \\
$\pt(\jet_1) > 400$\GeV        &   0.49 $\pm$   0.13  &  0.027 $\pm$ 0.002  &    0.50 $\pm$  0.07  &   0.072 $\pm$ 0.008 \\
$\pt(\jet_1) > 450$\GeV        &   0.31 $\pm$   0.11  &  0.016 $\pm$ 0.002  &    0.32 $\pm$  0.05  &   0.041 $\pm$ 0.006 \\
$\pt(\jet_1) > 500$\GeV        &   0.19 $\pm$  0.08  &  0.009 $\pm$ 0.001  &    0.19 $\pm$  0.04  &   0.023 $\pm$ 0.005 \\
$\pt(\jet_1) > 550$\GeV        &   0.12 $\pm$  0.07  &  0.006 $\pm$ 0.001  &    0.12 $\pm$  0.03  &   0.013 $\pm$ 0.003 \\ \hline
\end{tabular}
}
\end{table}

\cleardoublepage \section{The CMS Collaboration \label{app:collab}}\begin{sloppypar}\hyphenpenalty=5000\widowpenalty=500\clubpenalty=5000\textbf{Yerevan Physics Institute,  Yerevan,  Armenia}\\*[0pt]
V.~Khachatryan, A.M.~Sirunyan, A.~Tumasyan
\vskip\cmsinstskip
\textbf{Institut f\"{u}r Hochenergiephysik der OeAW,  Wien,  Austria}\\*[0pt]
W.~Adam, E.~Asilar, T.~Bergauer, J.~Brandstetter, E.~Brondolin, M.~Dragicevic, J.~Er\"{o}, M.~Flechl, M.~Friedl, R.~Fr\"{u}hwirth\cmsAuthorMark{1}, V.M.~Ghete, C.~Hartl, N.~H\"{o}rmann, J.~Hrubec, M.~Jeitler\cmsAuthorMark{1}, V.~Kn\"{u}nz, A.~K\"{o}nig, M.~Krammer\cmsAuthorMark{1}, I.~Kr\"{a}tschmer, D.~Liko, I.~Mikulec, D.~Rabady\cmsAuthorMark{2}, B.~Rahbaran, H.~Rohringer, J.~Schieck\cmsAuthorMark{1}, R.~Sch\"{o}fbeck, J.~Strauss, W.~Treberer-Treberspurg, W.~Waltenberger, C.-E.~Wulz\cmsAuthorMark{1}
% ...
\vskip\cmsinstskip
% ...
1:~~Also at Vienna University of Technology, Vienna, Austria\\
2:~~Also at CERN, European Organization for Nuclear Research, Geneva, Switzerland\\
% ...

\end{sloppypar}
\end{document}